\documentclass{bmcart}

\usepackage[utf8]{inputenc} 
\usepackage{booktabs} 
\usepackage{subcaption}
\usepackage{xspace}
\usepackage{hyphenat}
\usepackage{makecell}
\usepackage{footmisc}
\usepackage{url}
\usepackage{graphicx}
\usepackage{amsmath}

\newcommand{\trackedChange}[1]{\textcolor{blue}{#1}}

\newcommand{\est}{\mathrm{est}}
\newcommand{\true}{\mathrm{true}}
\newcommand{\err}{\mathrm{err}}
\newcommand{\reference}{\mathrm{ref}}

\newcommand{\ie}{{i.e.}\xspace}
\newcommand{\eg}{{e.g.}\xspace}
\newcommand{\cf}{{cf.}\xspace}
\newcommand{\vs}{{vs.}\xspace}

\renewcommand{\etal}{{et al.}\xspace}

\newcommand{\hide}[1]{}

\newcommand{\xhdr}[1]{\vspace{1.7mm}\noindent{{\bf #1.}}}

\newcommand{\Secref}[1]{Sec.~\ref{#1}}
\newcommand{\Eqnref}[1]{Eq.~\ref{#1}}

\newcommand{\Tabref}[1]{Table~\ref{#1}}
\newcommand{\Figref}[1]{Fig.~\ref{#1}}

\newenvironment{squishlist}
{\begin{list}{$\bullet$}
  { \setlength{\itemsep}{0pt}
     \setlength{\parsep}{1pt}
     \setlength{\topsep}{1pt}
     \setlength{\partopsep}{0pt}
     \setlength{\leftmargin}{1.5em}
     \setlength{\labelwidth}{1em}
     \setlength{\labelsep}{0.5em} } }
{\end{list}}

\startlocaldefs
\endlocaldefs

\begin{document}

\begin{frontmatter}

\begin{fmbox}
\dochead{Research}


\title{Human biases in body measurement estimation}


\author[
   addressref={aff1},                   
   email={kirmartynov@gmail.com}   
]{\inits{KM}\fnm{Kirill} \snm{Martynov}}
\author[
   addressref={aff2},
   email={garimell@mit.edu}
]{\inits{KG}\fnm{Kiran} \snm{Garimella}}
\author[
   addressref={aff3},
   corref={aff3},
   email={robert.west@epfl.ch}
]{\inits{RW}\fnm{Robert} \snm{West}}


\address[id=aff1]{
  \orgname{Google}, 
  \city{Munich},                              
  \cny{Germany}                                    
}
\address[id=aff2]{%
  \orgname{Massachusetts Institute of Technology},
  \city{Cambridge, MA},
  \cny{USA}
}
\address[id=aff3]{%
  \orgname{EPFL},
  \city{Lausanne},
  \cny{Switzerland}
}


\begin{artnotes}
\note[id=aff1]{Work performed at EPFL and TUM.} 
\note[id=aff2]{Work performed  at EPFL.} 
\end{artnotes}

\end{fmbox}


\begin{abstractbox}

\begin{abstract}
Body measurements, including weight and height, are key indicators of health.
Being able to visually assess body measurements reliably is a step towards
increased awareness of overweight and obesity and is thus important for
public health. Nevertheless it is currently not well understood how accurately
humans can assess weight and height from images, and when and how they
fail. To bridge this gap, we start from 1,682 images of persons collected
from the Web, each annotated with the true weight and height, and ask crowd
workers to estimate the weight and height for each image. We conduct a
faceted analysis taking into account characteristics of the images as well
as the crowd workers assessing the images, revealing several novel findings:
(1)~Even after aggregation, the crowd's accuracy is overall low.
(2)~We find strong evidence of contraction bias toward a reference value, such that
the weight of light people and the height of short people
are overestimated, whereas
the weight of heavy people and the height of tall people
are underestimated.
(3)~We estimate workers'
individual reference values using a Bayesian model, finding that reference
values strongly correlate with workers' own height and weight, indicating
that workers are better at estimating people similar to themselves. (4)~The
weight of tall people is underestimated more than that of short people;
yet, knowing the height decreases the weight error only mildly. (5)~Accuracy
is higher on images of females than of males, but female and male workers
are no different in terms of accuracy. (6)~Crowd workers improve over time
if given feedback on previous guesses. Finally, we explore various bias
correction models for improving the crowd's accuracy, but find that this
only leads to modest gains. Overall, this work provides important insights
on biases in body measurement estimation as obesity-related conditions
are on the rise.
\end{abstract}




\end{abstractbox}
%

\end{frontmatter}


\section{Introduction} \label{sec:intro}

Height and weight play a key role as indicators of health~\cite{who2006child}.
Population\hyp level height and weight statistics open a window onto the general health of a region.
With obesity and other lifestyle diseases on the rise, it is important to understand how humans perceive height and weight.
For instance, research has shown that exposure to obesity changes the perception of the weight status of others~\cite{robinson2014he}, and that perceived deviations of one's own weight from an ``ideal'' weight often lead to issues such as social anxiety, depression, peer victimization, and lower self-worth~\cite{lanza2012deviating}.
In terms of action, understanding human perception of body measurements should therefore play a critical role for public health agencies when striving to design effective strategies for raising awareness about the prevalence and risks of overweight and obesity.
For all these reasons, the estimation of human body measurements is an active area of interest in medical research \cite{coetzee2009facial,burton2013judgments,robinson2014he,robinson2015visual,robinson2017overweight}.

While a thorough understanding of the perception biases in height and weight estimation is important in and of itself, it is also important for several potential downstream applications.
For instance, crowdsourcing systems for inferring body measurements from social\hyp media images \cite{weber2016crowdsourcing} could be leveraged to estimate regional weight and height distributions across the world, but such systems require accurate crowd labels.
Understanding humans' accuracy and biases is thus essential for establishing the bounds of what one might hope to be achieved by crowd\hyp based weight and height sensing systems.
Crowdsourced estimates could also be useful for training machine learning models that could ultimately produce height and weight estimates in an automated fashion~\cite{benabdelkader2008statistical,dey2014estimating,gunel2018face,DBLP:journals/tifs/JiangG19,kocabey2017face,wen2013computational}.
Such methods require large numbers of labeled training images, and crowdsourcing could potentially provide an efficient way of collecting such labeled samples.
Machine learning models can only be as good as the data they are trained on, so here too, quantifying how well crowds perform at height and weight estimation establishes upper bounds for machine learning systems that use crowdsourced training data, and understanding human biases opens up avenues for addressing the crowd's specific shortcomings.

Existing studies of body measurement estimation \cite{coe1999,coetzee2009facial,burton2013judgments,robinson2014he,robinson2015visual,cornelissen2016visual,robinson2017overweight} are typically based on surveys that reach only small numbers of people and thus tend to lack the scale to provide generalizable insights.
The present paper contributes to bridging this gap through a large-scale, systematic analysis of how humans estimate others' weight and height.
We designed and conducted a large study on an online crowdsourcing platform, where we asked participants (``workers'') to estimate the height and weight of people in 1,682 images that had been collected from an online weight\hyp loss forum and labeled with ground-truth height and weight values by the persons in the images themselves.
In addition to a total of over 100,000 guesses, our pool of 1,767 crowd workers also provided their own height, weight, and demographic variables.
This data allows us to conduct a faceted analysis of the accuracy of crowd workers as well as their biases---the systematic ways in which they succeed and fail.

Our main contributions are as follows. In a
\textbf{large-scale crowdsourced study} (\Secref{sec:data}), we collected
over 100,000 height and weight estimates from a diverse set of crowd workers,
showing that the \textbf{crowd's accuracy} is low (\Secref{sec:gen accuracy}),
with an overall mean absolute error of 15.5~kg (6.3~cm), and 8.8~kg (5.2~cm)
on images subsampled to match the weight (height) distribution of workers.
A small number of workers (20 or 30) sufficed to achieve low variance,
but the error remained large due to contraction bias toward a reference
value, such that
the weight of light people
and the height of short people
were systematically
overestimated, whereas
the weight of heavy people and the height of tall people
were underestimated. Also,
the taller a person, the more their weight was underestimated.

We then analyzed the
\textbf{dependency of accuracy on worker characteristics} (\Secref{sec:worker spec}).
Although the height and weight of females were found to overall be easier
to guess than those of males, female and male workers were no different
in terms of accuracy. Workers were, however, better at estimating bodies
similar to their own in terms of height and weight. We explain this finding
with a Bayesian model that assumes every worker to be biased toward a worker-specific
reference value. By inferring a worker's reference values from the worker's
estimates and from ground-truth image labels alone (without using the worker's
own measurements), we found that the inferred reference values correlated
strongly with workers' own measurements.

Based on these insights, we explored paths
\textbf{towards more accurate crowdsourcing} (\Secref{sec:Towards more accurate crowdsourcing}).
We found that various post-hoc bias correction models only led to modest
gains, whereas varying the setup of the crowdsourcing task itself is more
promising: the accuracy could be improved by giving workers feedback on the
quality of their guesses, effectively teaching them to make better guesses
in the future. Another setup, however, where we gave workers access to
the height of the person to be estimated, decreased the weight error only
mildly.

We conclude the paper (\Secref{sec:discussion}) by discussing limitations
of our work and by revisiting related work (cf. \Secref{sec:relwork})
in the context of our results.
Finally, hoping to inform future research in this area, we digest our
findings into a set of implications and best practices for crowdsourcing and machine learning
tasks involving body measurements.

\section{Related work}
\label{sec:relwork}

The study of how humans estimate body measurements is an active area of research spanning multiple diverse domains, including psychology, medicine, and computer science.
The most relevant pieces of prior work pertain to human perception biases and to the usage of crowdsourcing.
We review these two directions in turn.

\xhdr{Perception biases in body measurement estimation}
Biases in weight and height estimates have been studied in medicine and behavioral psychology \cite{coe1999,cornelissen2016visual,poulton1989bias}.
When asked to judge their own weight, people tend to systematically under- or overestimate.
For instance, it has been established through lab studies that people show a progressive underestimation of overweight and obese bodies~\cite{cornelissen2016visual}, attributed to a psychological phenomenon known as contraction bias~\cite{poulton1989bias}.
According to these studies, such underestimation may compromise people's ability to recognize weight gain and undertake compensatory weight control behaviors.

Other psychological experiments in lab settings have tried to explain the aforementioned biases by constructing mental models for how people judge weight and height.
Facial adiposity~\cite{coetzee2009facial} was found to constitute an important signal used in weight estimation, whereas features such as dominance and facial maturity \cite{burton2013judgments}, as well as head\hyp to\hyp shoulder ratio~\cite{mather2010head}, are prominent in height estimation.
Robinson \cite{robinson2017overweight} proposed visual normalization theory, a framework for explaining why people under- or overestimate certain body shapes.
This theory is based on the notion that weight is judged relative to the body\hyp size norms prevalent in the respective society. As larger body sizes are common in the United States, participants in Robinson's \cite{robinson2017overweight} study assessed overweight individuals as normal.
Robinson \etal\ \cite{robinson2015visual} studied a similar phenomenon for young adults from the United States (high obesity prevalence), the United Kingdom, and Sweden (lower obesity prevalence) to understand cultural differences between perceptions.
They did not find cross-cultural differences in judging overweight males.

Based on questions and answers from Yahoo!\ Answers, Yom-Tov \cite{yom2016crowdsourced} found that men estimate their weight status fairly well, whereas women do not. This could potentially be explained by visual normalization theory (see above), where, due to societal norms on body shapes of women, being slightly over the median might be considered ``overweight'' for women but not for men.
Apart from gender, weight perception was also shown to be subject to generational shifts \cite{hansen2014generational}.

Humans tend to update their perceptions when shown evidence that contradicts their mental model. For instance, Robinson and Kirkham \cite{robinson2014he} showed that people exposed to photographs of obese young males changed their visual judgments of whether an overweight young male was of healthy weight.

Researchers have also worked on statistical models for correcting human estimation biases, usually based on simple regression techniques \cite{brettschneider2015development,perez2015measuring}.
Finally, Martynov et al.~\cite{martynov2020darks} have investigated the impact of clothing on body-size perception,
showing that dark clothes have a small but statistically significant slimming effect.

\xhdr{Crowdsourcing for body measurement estimation}
Some pieces of previous work have leveraged crowdsourcing to estimate height and weight.
Indeed, historically speaking, body measurement and crowdsourcing are tightly intertwined.
In his 1907 paper \textit{Vox populi} \cite{galton1907vox}---probably the first paper on the ``wisdom of crowds''---Francis Galton evaluated data from a weight\hyp judging competition held at a Plymouth county fair, where 787 participants guessed the weight of an ox.
Galton observed that the median guess was surprisingly accurate, only 9 pounds (0.8\%) above the true weight of the ox (1,198 pounds).
One might hope that, if humans excel at estimating the weight of an ox, they would perform even better when the ox is replaced with a fellow human.
As we will see, though, this is unfortunately not the case.

In the context of humans, rather than oxen, crowdsourcing has been shown to be a useful tool for assessing childhood predictors of adult obesity~\cite{bevelander2014crowdsourcing}
and for estimating weight from social media profile images
\cite{weber2016crowdsourcing}.

It has been established that, when using crowds as estimators, perception biases~\cite{eickhoff2018cognitive} and social influence~\cite{lorenz2011social} can lead to severe biases.
In the present work, we turn the vice into a virtue: our primary goal is not to build accurate weight and height estimators based on crowds, but rather to assess the very biases inherent in crowds as estimators.

\xhdr{Present work}
Our work is situated at the intersection of the above two strands of research.
On the one hand, we replicate findings about biases in weight and height perception from small\hyp scale lab studies in a much larger setting with thousands of crowd workers, revealing similar biases.
On the other hand, by recruiting a global and diverse pool of crowd workers, we can go beyond the scope of previous studies and investigate how biases vary across workers' countries of residence and demographic strata: by collecting workers' own weight, height, age, gender, and country, we can systematically study the impact of those factors and their interplay with the image whose weight is to be estimated.

\section{Data}
\label{sec:data}

\subsection{Height- and weight-labeled body images}
\label{sec:data main}

The original data was provided by Kocabey \etal\ \cite{kocabey2017face} and includes about 10k samples
collected from a Reddit forum (``subreddit'') called
\textit{r/progresspics},\footnote{\url{https://www.reddit.com/r/progresspics}}
where users who intend to lose (or sometimes gain) weight post pictures of themselves before and after their weight transformation. Each sample contains the ID of the Reddit post, the height and gender of the user, their weight before and after the transformation, as well as one or several images. Note that the weight and height labels were provided by the Reddit-users themselves and might not always be exact or accurate. However, throughout the paper, we refer to these labels as the \textit{(weak) ground-truth} values, because we are not aware of a method that could check or improve their quality.

Several preprocessing steps were required to prepare the data for our weight and height estimation task.
In many cases, users combined the ``before'' and ``after'' pictures into a single image file, so our first processing step consisted in automatically splitting such collages into its component images, such that a single weight could be associated with each image.
To avoid ambiguous assignments, collages with more than two images were excluded from the dataset. Furthermore, we dropped images that contain weight or height labels as text or that depict multiple people. Finally, low-quality photos, as well as images that contain only faces, were omitted. 

\begin{figure}[t]
    \centering  
    \begin{subfigure}{.49\hsize}
        \centering
        \includegraphics[width=\hsize]{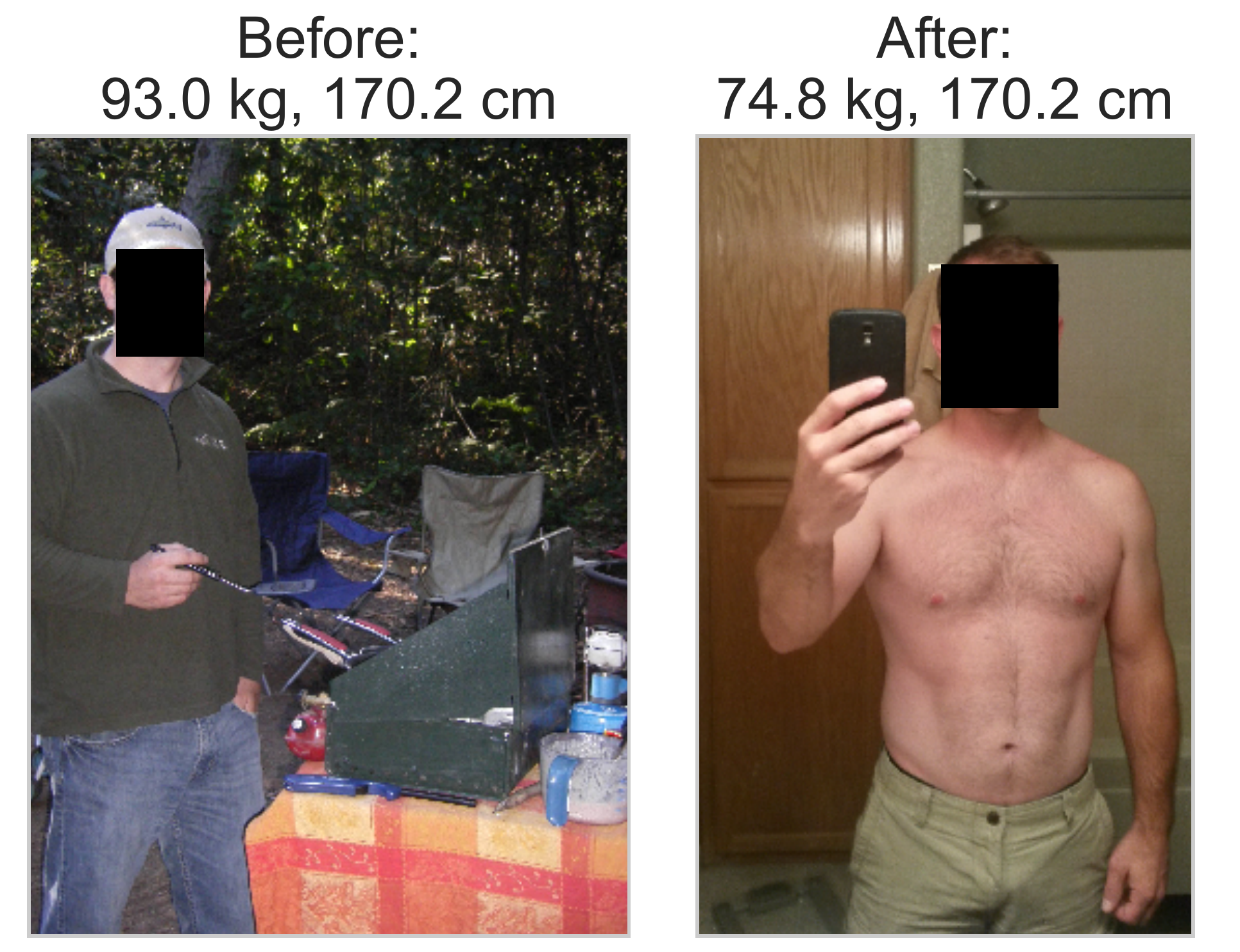}
        \caption{}
        \label{fig:sample1}
    \end{subfigure}%
    \begin{subfigure}{.49\hsize}
        \centering
        \includegraphics[width=\hsize]{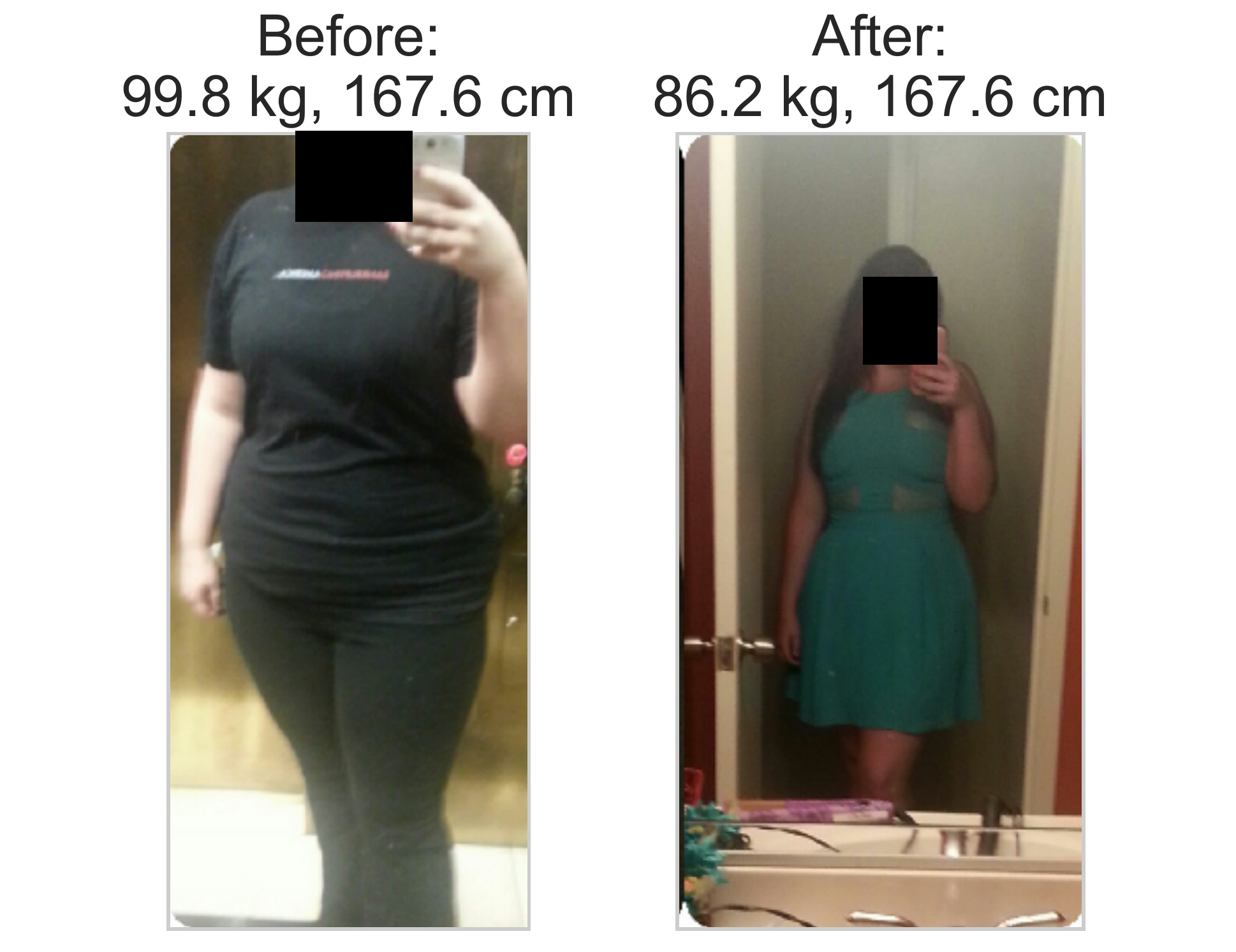}
        \caption{}
        \label{fig:sample2}
    \end{subfigure}
    \captionsetup{width=0.9\hsize} \caption{Two samples from the dataset. Each sample contains a ``before'' and an ``after'' picture. The height and weight before and after are shown on the top.}
\label{fig:samples}
\end{figure}

As a result, we obtained 841 samples (431 and 410 posts from male and female users, respectively), for a total of 1,682 images (two images per person, each with a weight and a height label).
Two samples from the dataset are shown in \Figref{fig:samples}.
We manually inspected the dataset and characterized the different types of images present in our data.
Only around 10\% of the images are full-body pictures, the rest lack part of the body (\eg, head or legs);
80\% depict fully dressed people;
around 50\% are selfies taken in front of a mirror;
and in around 75\% we can see the head.
\Figref{fig:men image char} (men) and \Figref{fig:women image char} (women) show distributions of weight, height, and body-mass index (BMI)
for the collected images.


\subsection{Crowdsourced height and weight estimates}
\label{sec:data crowd}

We used Amazon Mechanical Turk to collect weight and height estimates from diverse groups of workers. For this purpose, the dataset described in \Secref{sec:data main} was split into 168 tasks with 10 images in each task. For each image, crowd workers were asked to guess the weight and height of the depicted person. Both input fields were located under the image, with the field for weight estimates coming first. The workers could choose between ``kg'' and ``lbs'' units for weight estimation, and ``cm'' and ``ft/in'' for height estimation. Conversions between different units were automatically performed on the fly, i.e., the other field was updated at the same time as workers were typing the guesses in their preferred format. 
To encourage high-quality estimates, 25\% of the most accurate (on a per-task basis) workers were awarded with a bonus that doubled their usual reward.

Each task (a collection of 10 images) was independently performed by 45 workers who estimated both weight and height.
In this way, we gathered 45 weight and height labels per image, or 75,600 estimates from 1,767 unique workers overall. \Secref{sec:gen accuracy} and \ref{sec:worker spec} focus mostly on this main dataset. 

Apart from the main experiment described above, we explored two additional setups, where workers estimated
(1)~the weight while being shown the true height (\Secref{sec:known height}) or
(2)~the weight and height while given the accuracy of their previous guess (\Secref{sec:feedback}).
In each of these experiments, we collected another 20k estimates.

\begin{figure}[t]
\centering  
\begin{subfigure}{.24\hsize}
    \centering
    \includegraphics[width=\hsize]{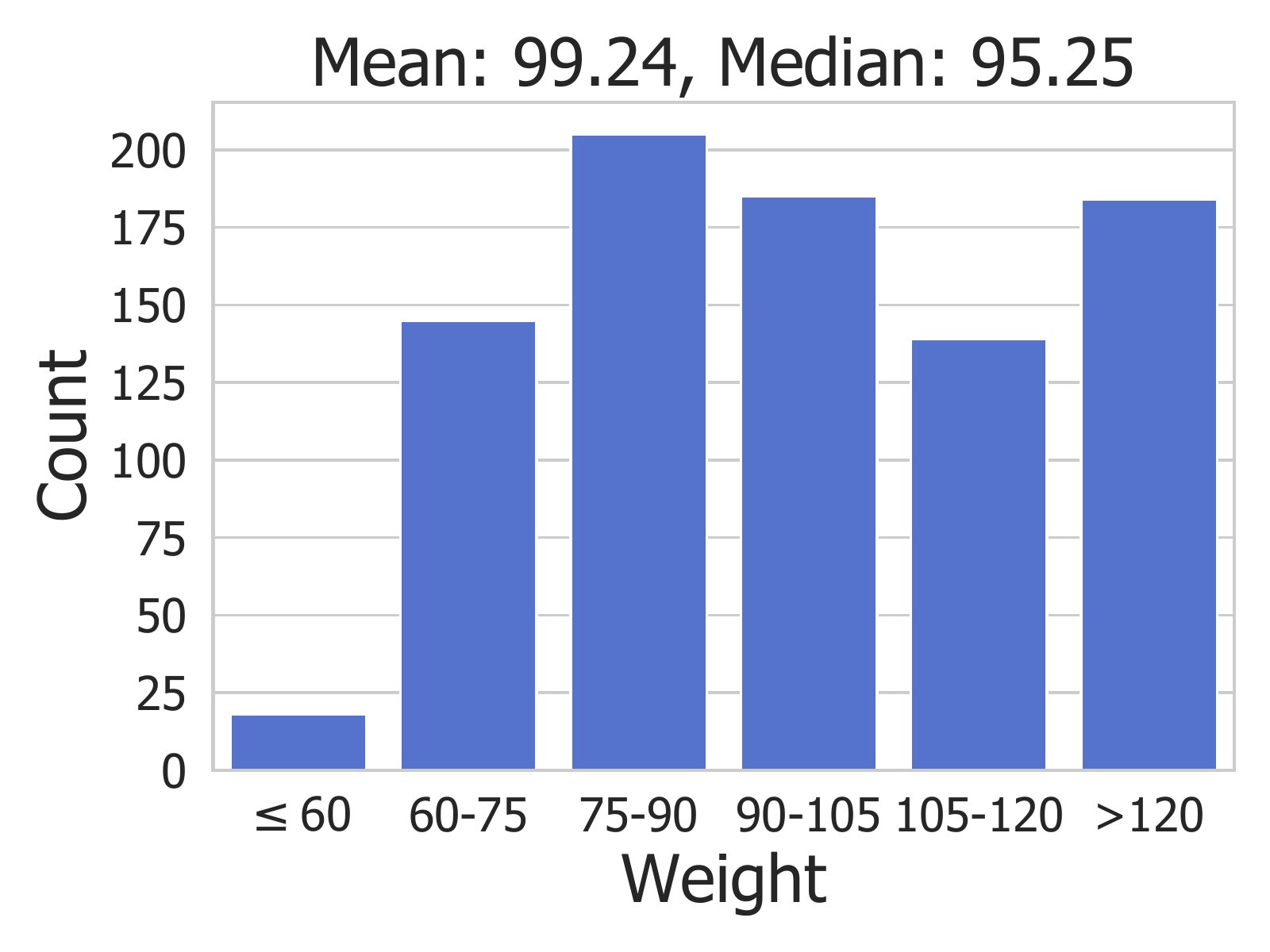}
\end{subfigure}%
\begin{subfigure}{.24\hsize}
    \centering
    \includegraphics[width=\hsize]{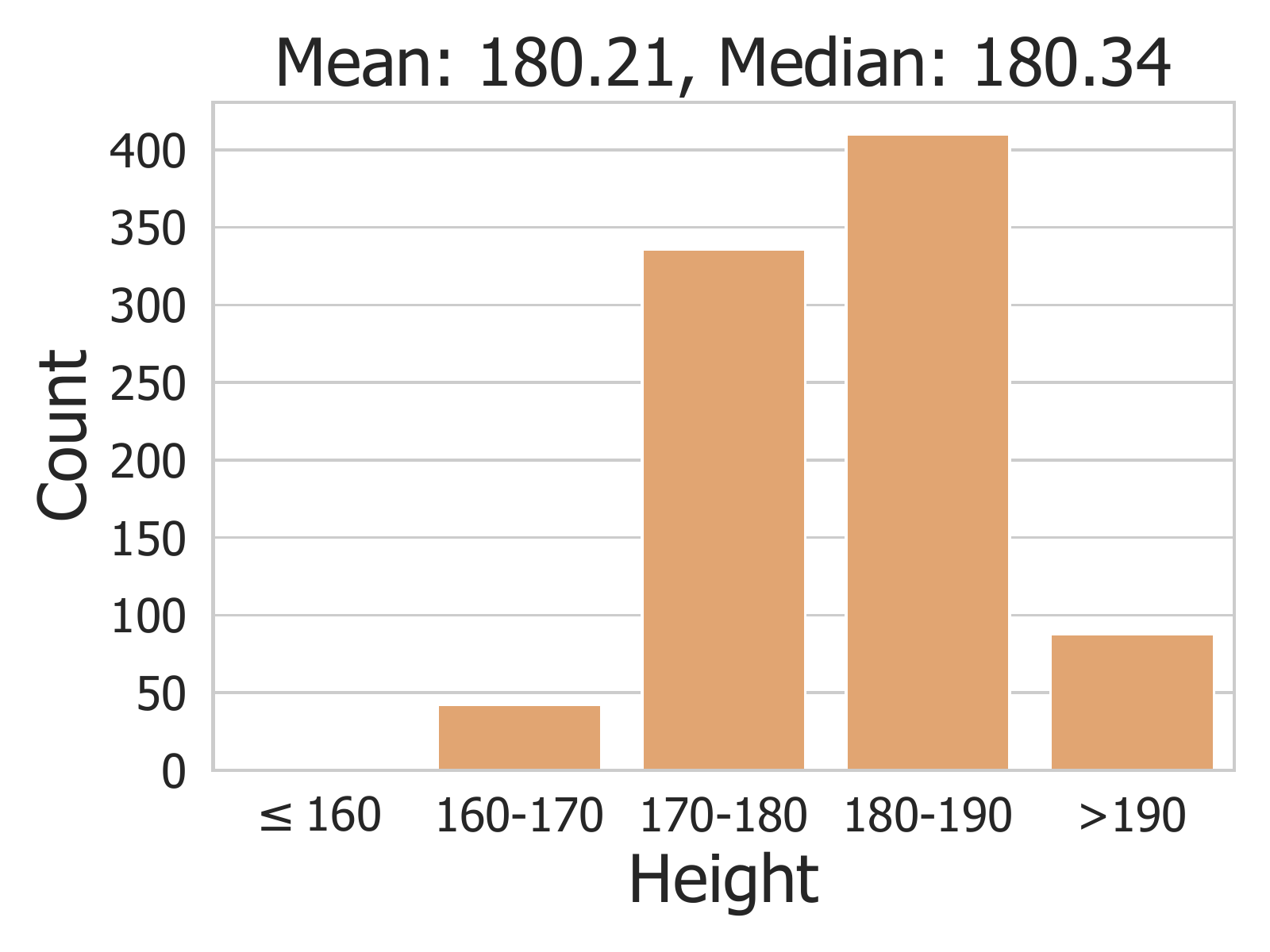}
\end{subfigure}
\begin{subfigure}{.24\hsize}
    \centering
    \includegraphics[width=\hsize]{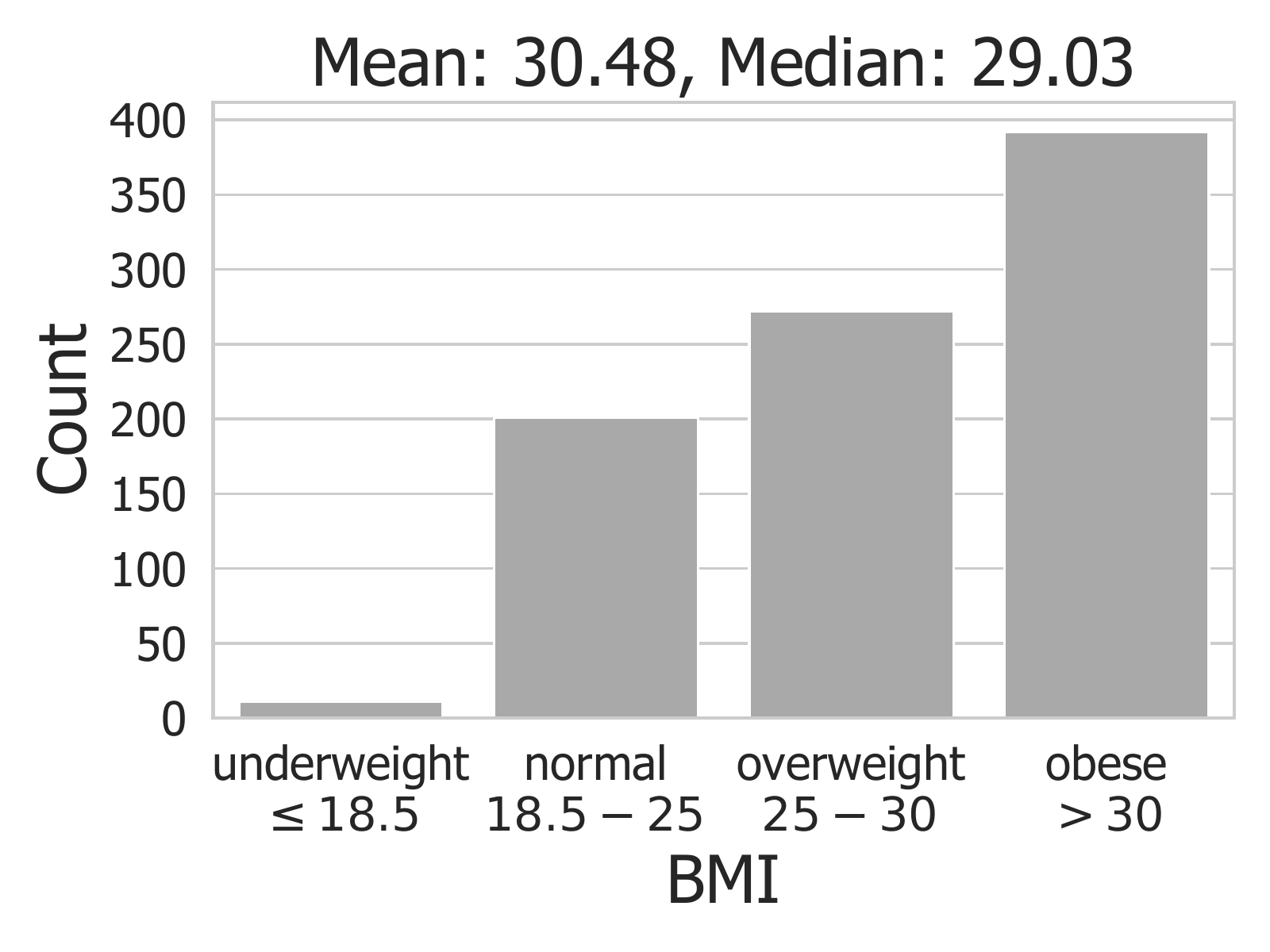}
\end{subfigure}
\begin{subfigure}{.24\hsize}
    \centering
    \includegraphics[width=\hsize]{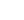}
\end{subfigure}
\captionsetup{width=0.9\hsize}
\vspace*{-0.25cm}
\caption{Male images in Reddit data (N=876, 52.3\% of all images).}
\label{fig:men image char}

\vspace*{\floatsep}
\centering  
\begin{subfigure}{.24\hsize}
    \centering
    \includegraphics[width=\hsize]{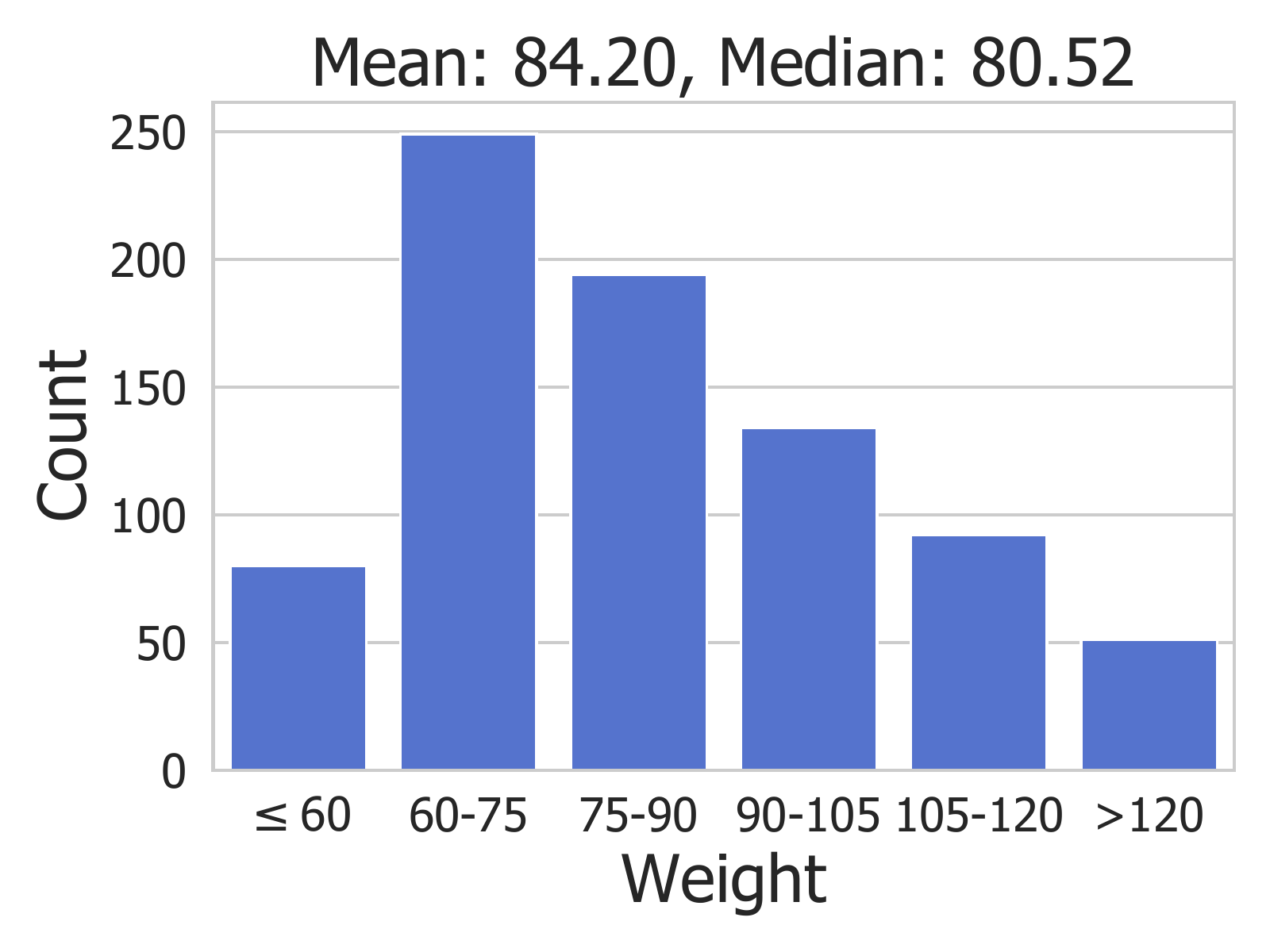}
\end{subfigure}%
\begin{subfigure}{.24\hsize}
    \centering
    \includegraphics[width=\hsize]{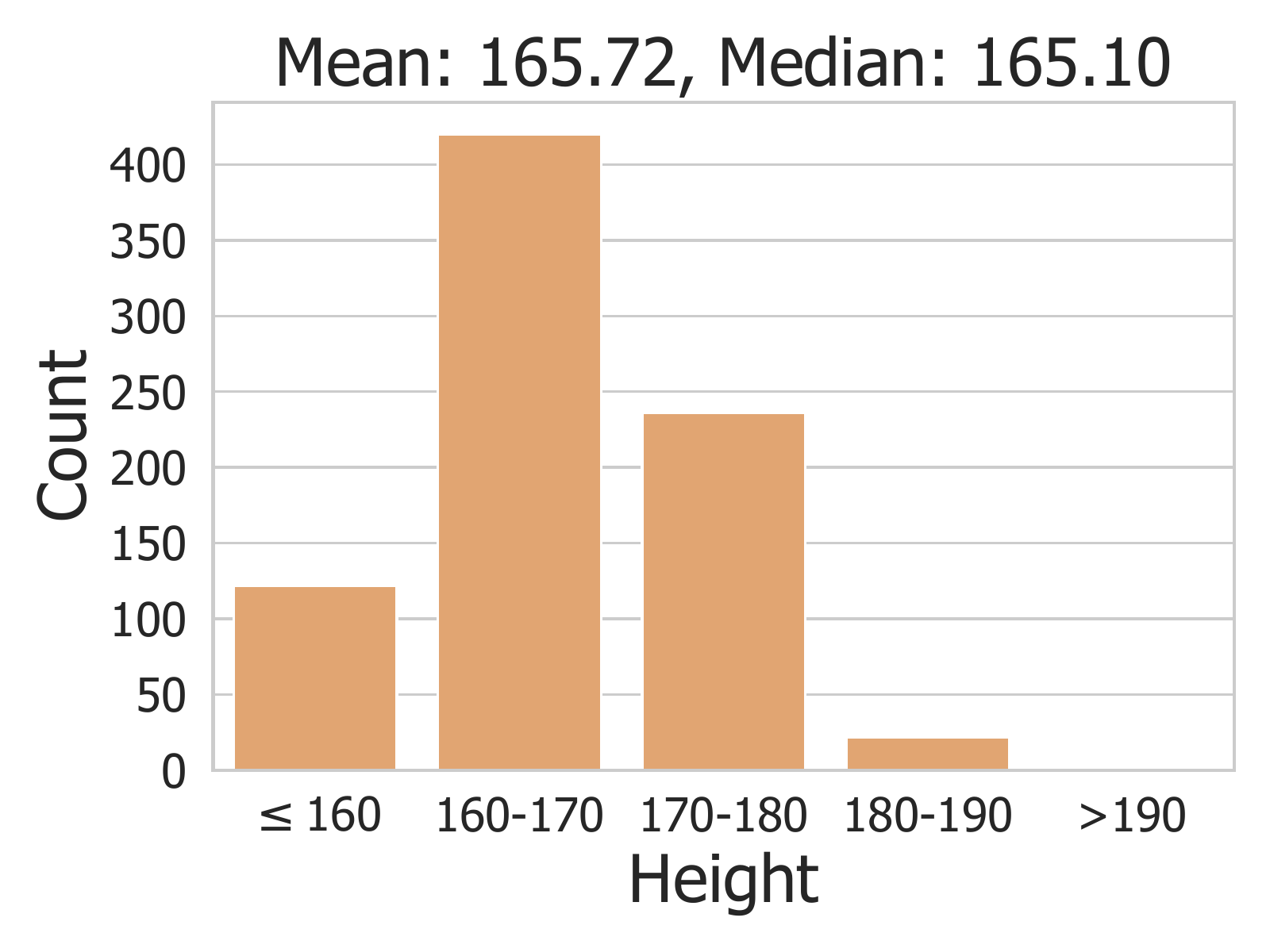}
\end{subfigure}
\begin{subfigure}{.24\hsize}
    \centering
    \includegraphics[width=\hsize]{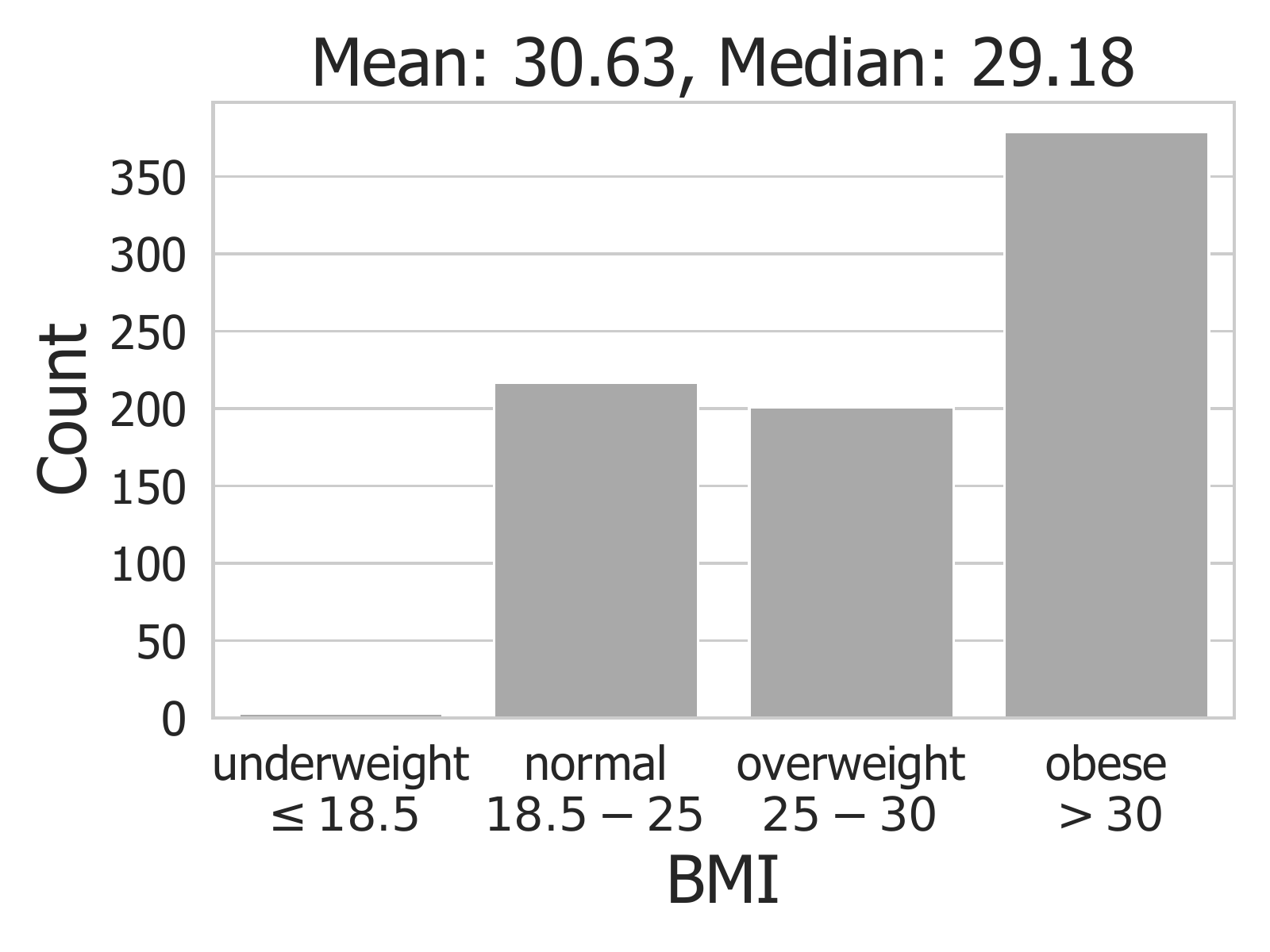}
\end{subfigure}
\begin{subfigure}{.24\hsize}
    \centering
    \includegraphics[width=\hsize]{blank.png}
\end{subfigure}
\captionsetup{width=0.9\hsize}
\vspace*{-0.25cm}
\caption{Female images in Reddit data (N=800, 47.7\% of all images).}
\label{fig:women image char}
\centering 
\vspace*{\floatsep}

\begin{subfigure}{.24\hsize}
    \centering
    \includegraphics[width=\hsize]{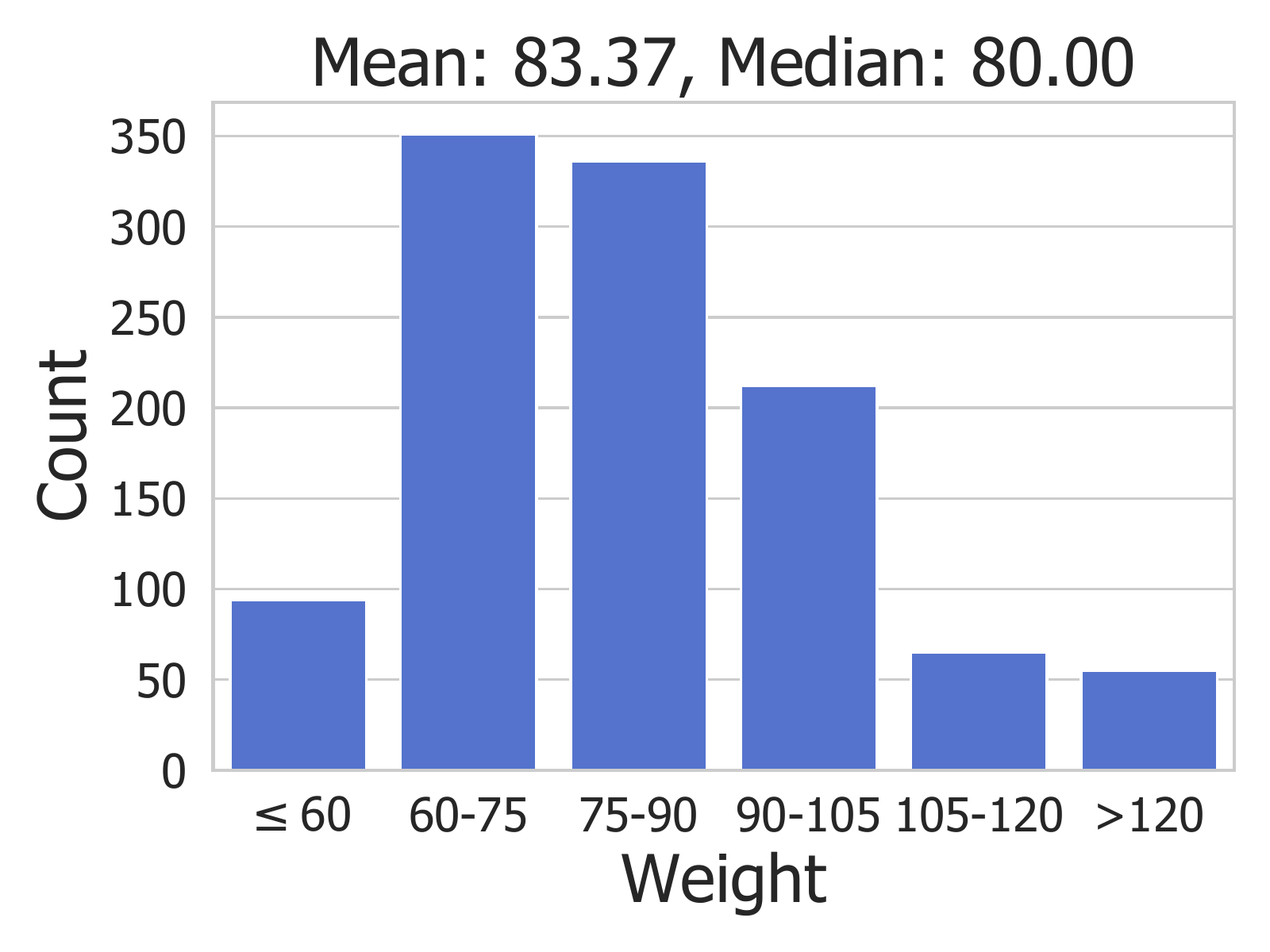}
\end{subfigure}%
\begin{subfigure}{.24\hsize}
    \centering
    \includegraphics[width=\hsize]{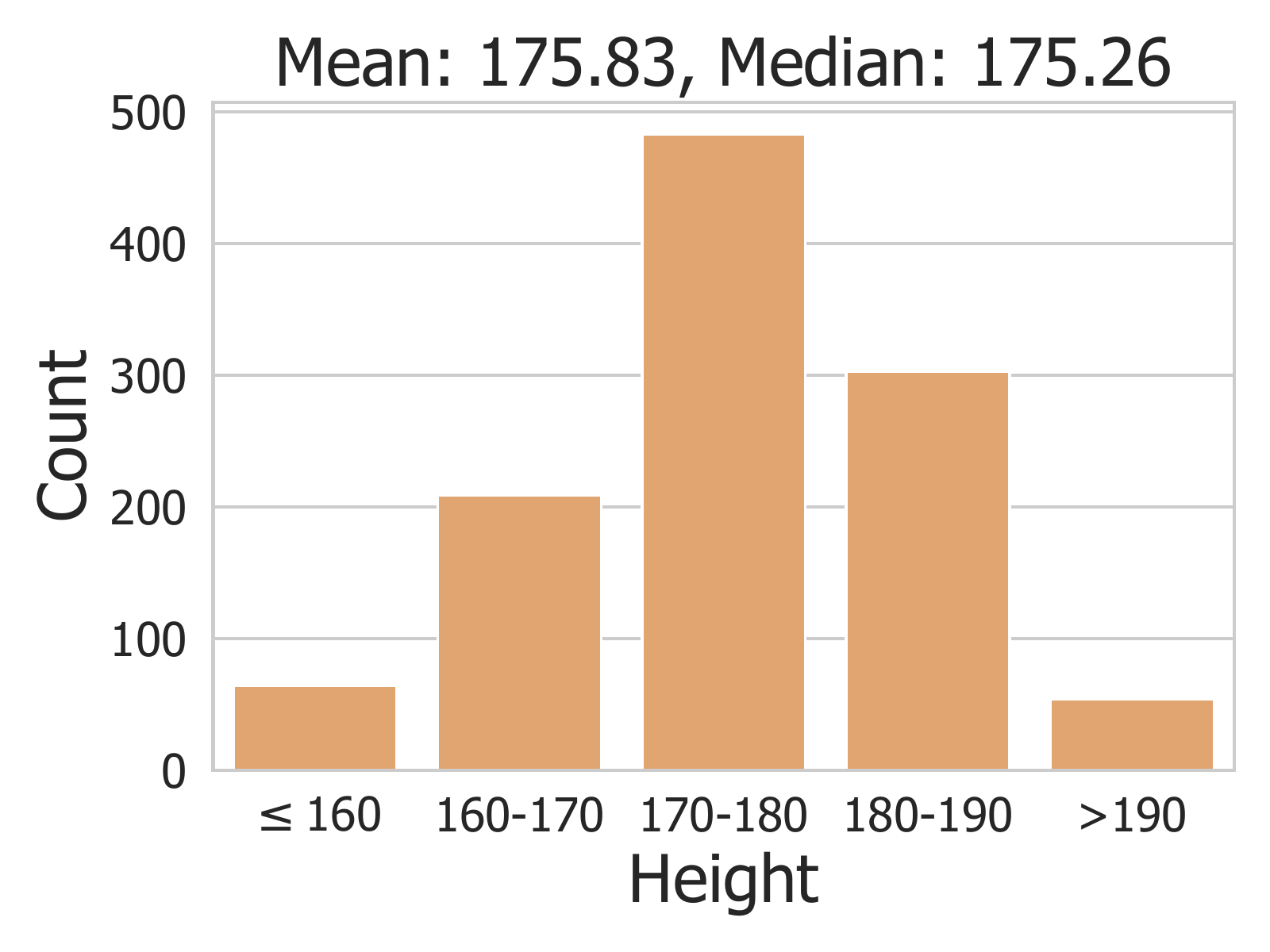}
\end{subfigure}
\begin{subfigure}{.24\hsize}
    \centering
    \includegraphics[width=\hsize]{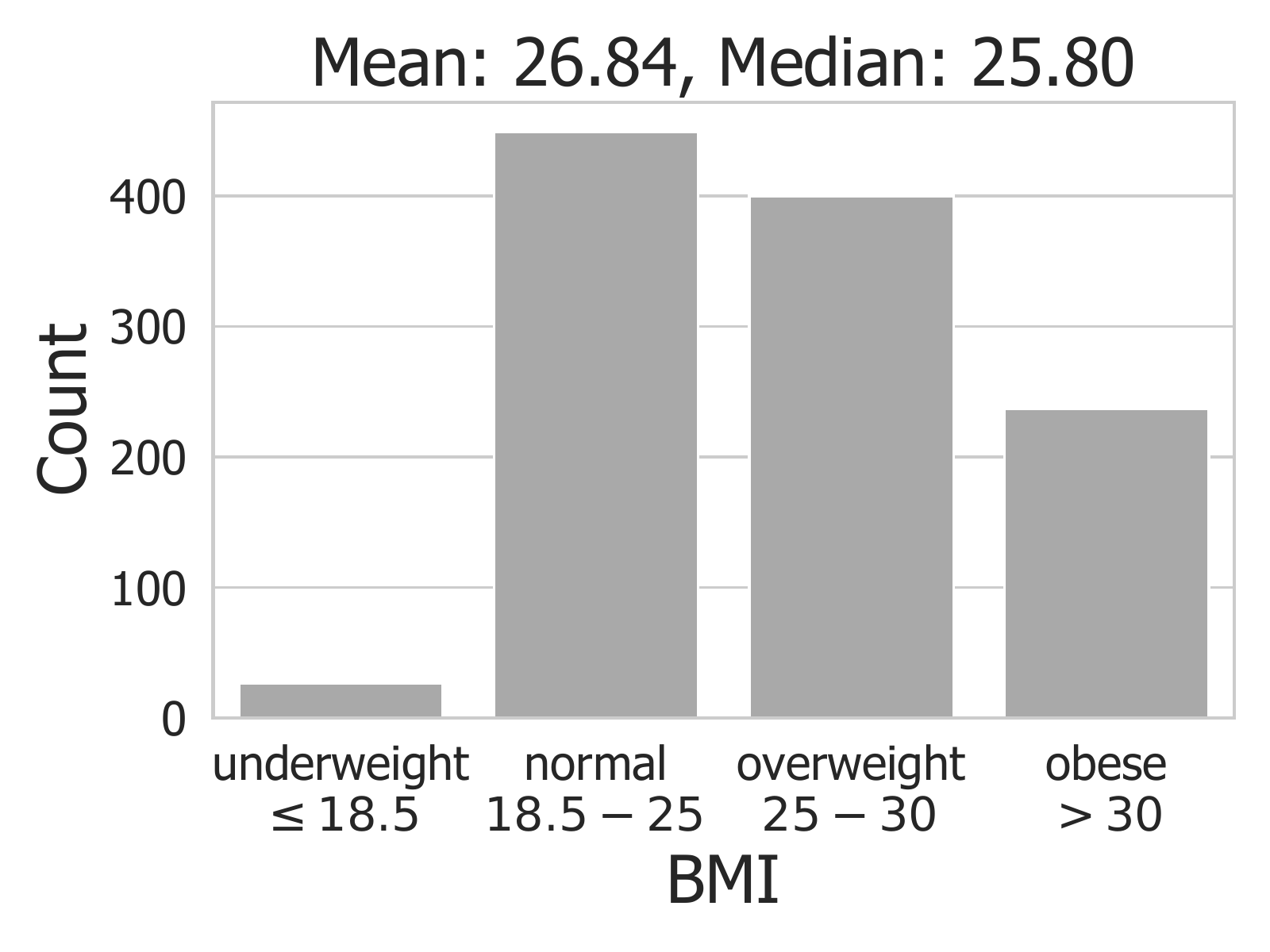}
\end{subfigure}
\begin{subfigure}{.24\hsize}
    \centering
    \includegraphics[width=\hsize]{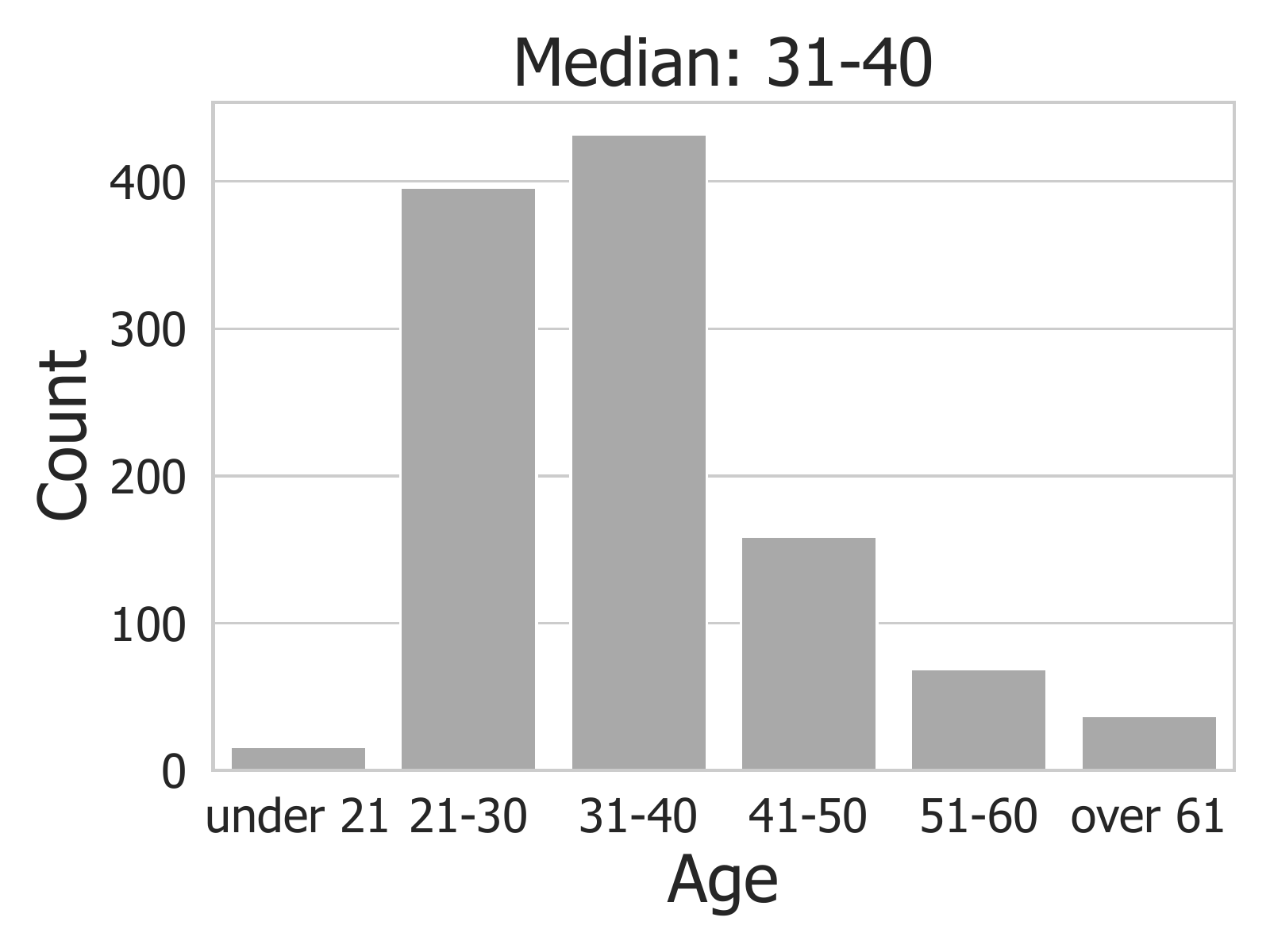}
\end{subfigure}
\captionsetup{width=0.9\hsize}
\vspace*{-0.25cm}
\caption{Men crowd workers (N=1,113, 45.9\% of all workers).}
\label{fig:men worker char}
\vspace*{\floatsep}

\centering  
\begin{subfigure}{.24\hsize}
    \centering
    \includegraphics[width=\hsize]{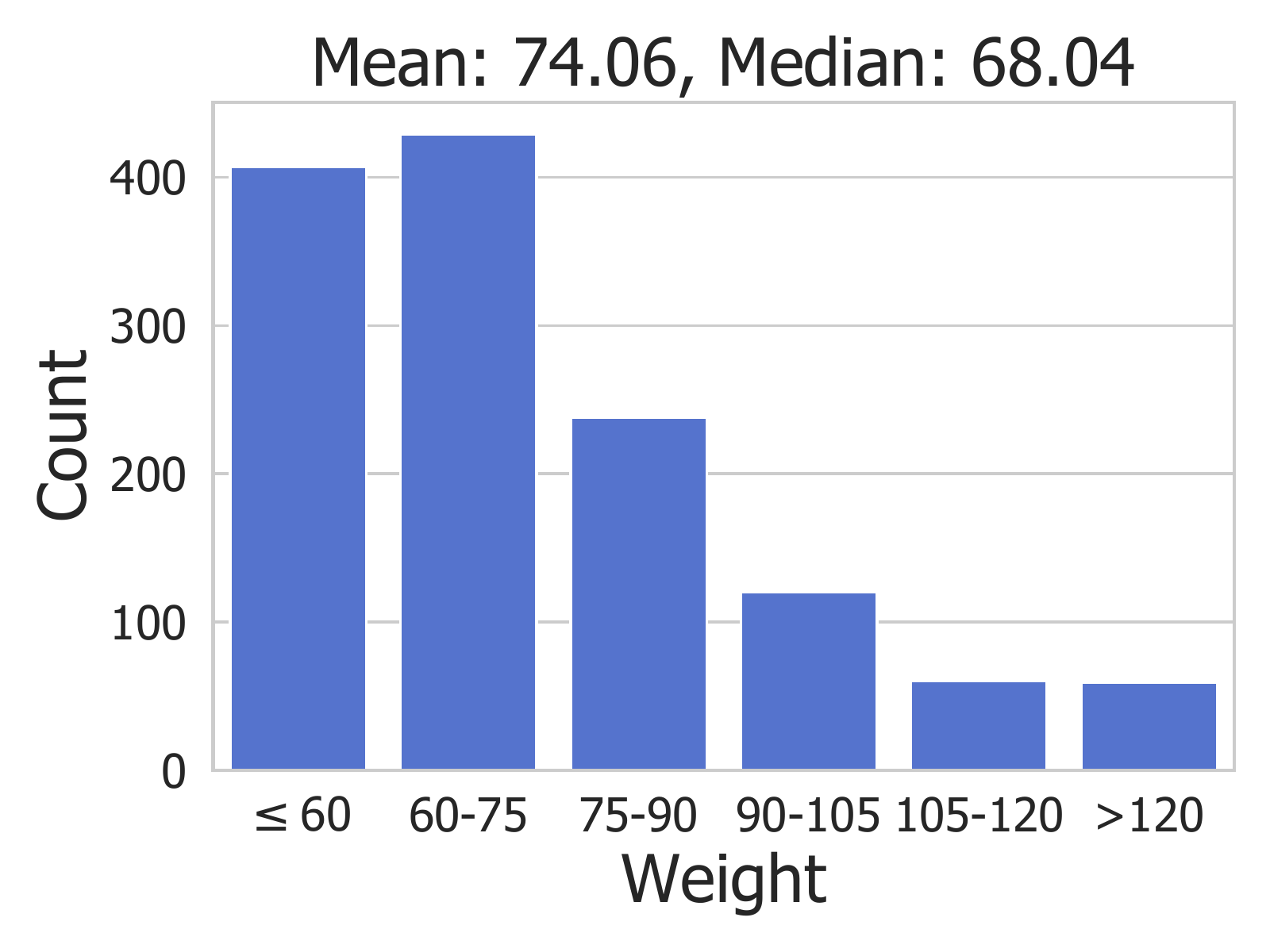}
\end{subfigure}%
\begin{subfigure}{.24\hsize}
    \centering
    \includegraphics[width=\hsize]{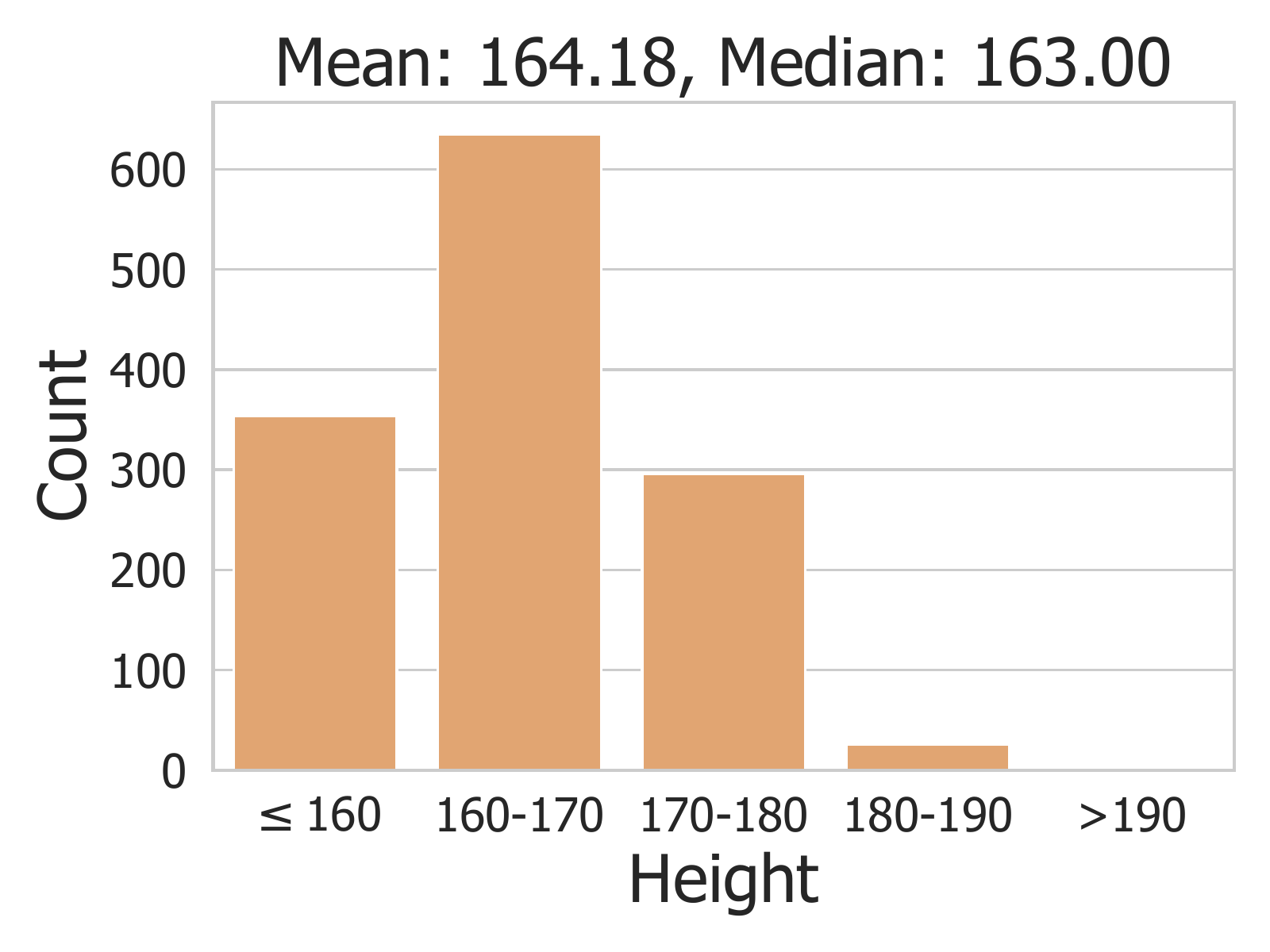}
\end{subfigure}
\begin{subfigure}{.24\hsize}
    \centering
    \includegraphics[width=\hsize]{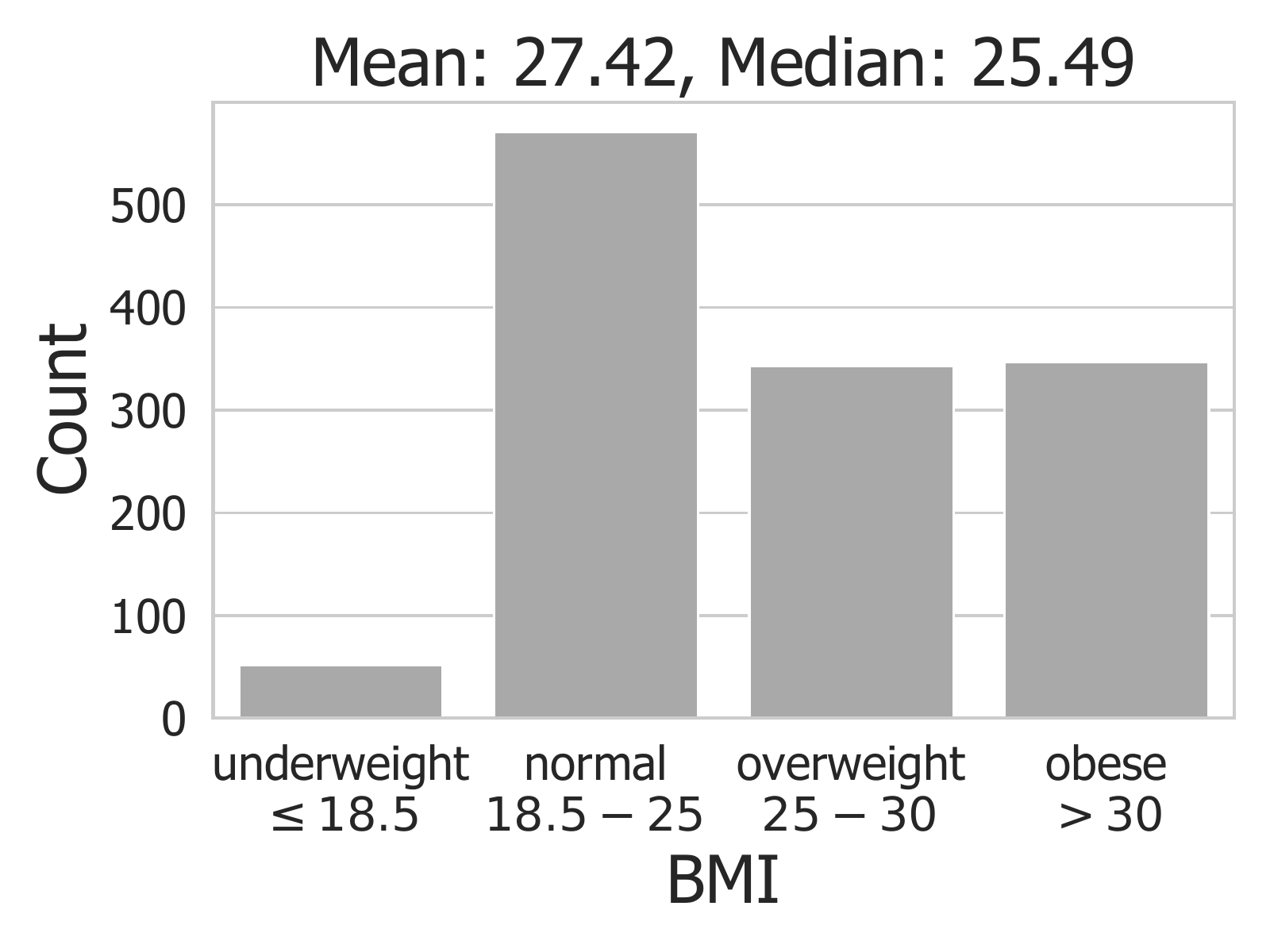}
\end{subfigure}
\begin{subfigure}{.24\hsize}
    \centering
    \includegraphics[width=\hsize]{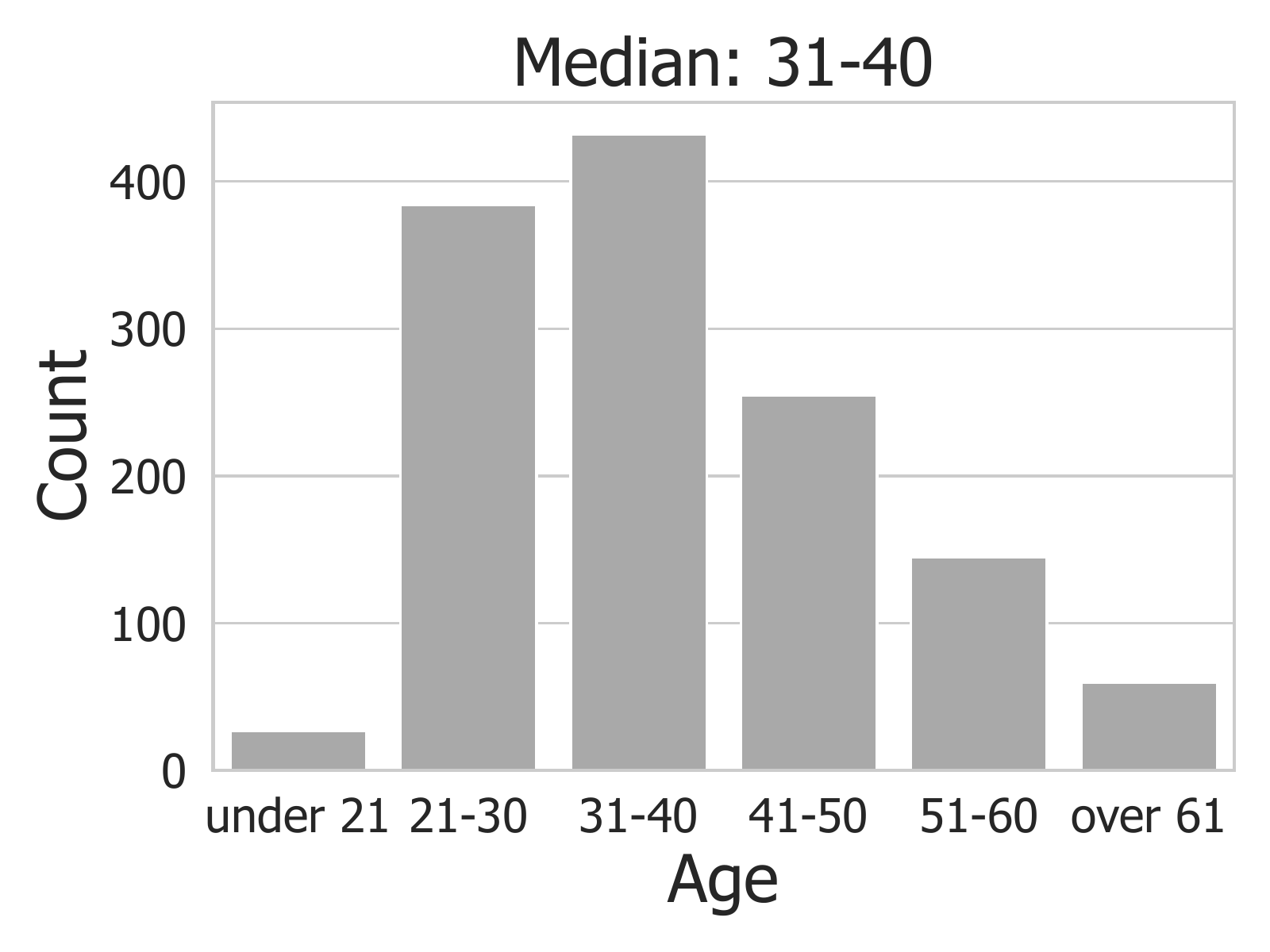}
\end{subfigure}
\captionsetup{width=0.9\hsize}
\vspace*{-0.25cm}
\caption{Women crowd workers (N=1,313, 54.1\% of all workers).}
\label{fig:women worker char}
\end{figure}

Finally, for each experimental setup, we collected the following personal information from all participating workers: their own height and weight, age, gender,
and country of residence.
The numerical characteristics are summarized in \Figref{fig:men worker char} (men) and \Figref{fig:women worker char} (women) for the 2,426 crowd workers who performed at least one of our tasks. Regarding the country of residence, most of the crowd workers are residents of the United States (74\%) or India (17\%).

The raw weight and height estimates obtained from the crowd workers were filtered and preprocessed in several ways. In particular, we removed data from workers who appear to have used scripts to generate random guesses, as well as
estimates from those who did not provide their personal information. Furthermore, several kinds of erroneous guesses, such as obvious typos or usage of wrong measurement units (e.g., a worker accidentally chose lbs instead of kg or vice versa), were detected.
After these preprocessing steps, roughly 75\% of the initial estimates remained in each of the experiments.
 

\section{Crowd-worker accuracy}
\label{sec:gen accuracy}

The basic idea underlying the ``wisdom of crowds'' is that, for certain estimation tasks, aggregating local estimates from multiple people can lead to a much more accurate global estimate.
Aggregation can be performed in multiple ways, \eg, via the mean or the median.
As we will see later (\Tabref{tab:mean vs median}), the two yield similar results, so in the rest of the paper, we will use the mean of all individual guesses collected for an image $i$ as an estimate of the true weight of $i$: 
\begin{equation}
	w_{\est}^i =  \frac{1}{n} \sum_{j=1}^{n} w_{j}^i,
\end{equation}
where $w_1^i,\dots,w_n^i$ are the weight guesses collected for image $i$.
The error for image $i$ can then be defined as
\begin{equation}
	w^i_\err = w^i_\true - w^i_\est,
\end{equation}
where $w^i_{\true}$ is the true weight for image $i$.
That is, \textit{positive (negative) errors signify underestimation (overestimation).}
Similarly, we define the true height $h^i_{\true}$, the estimated height $h^i_{\est}$, and the height error $h^i_{\err}$. 

Remember that, for each sample of the dataset from \Secref{sec:data main}, two weights and two images (``before'' and ``after'') need to be matched. This is done in an intuitive manner that also minimizes the error: the image with the higher value of $w^i_\est$ also is assigned the higher ground-truth label $w^i_{\true}$. The number of wrong assignments is very low with this strategy: among 70 samples (140 images) that we manually inspected, only two were assigned incorrectly. Thus, the resulting accuracy of the ground truth weight labels is expected to be above 95\%.


\subsection{Height and weight estimation errors}

\begin{figure}[t]
\centering  
\begin{subfigure}{.49\hsize}
  \centering
      \includegraphics[width=\hsize]{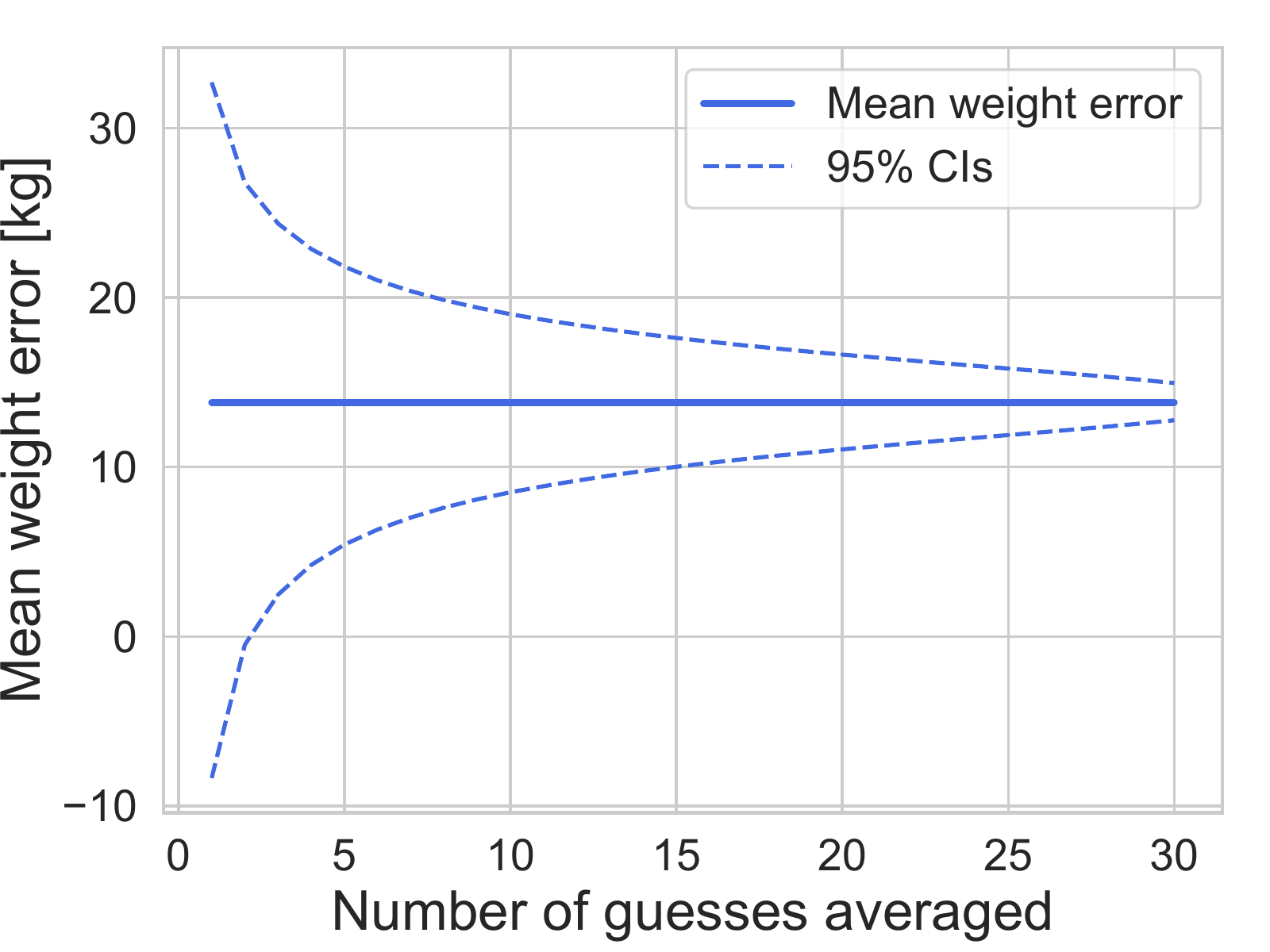}
\end{subfigure}%
\begin{subfigure}{.49\hsize}
  \centering
      \includegraphics[width=\hsize]{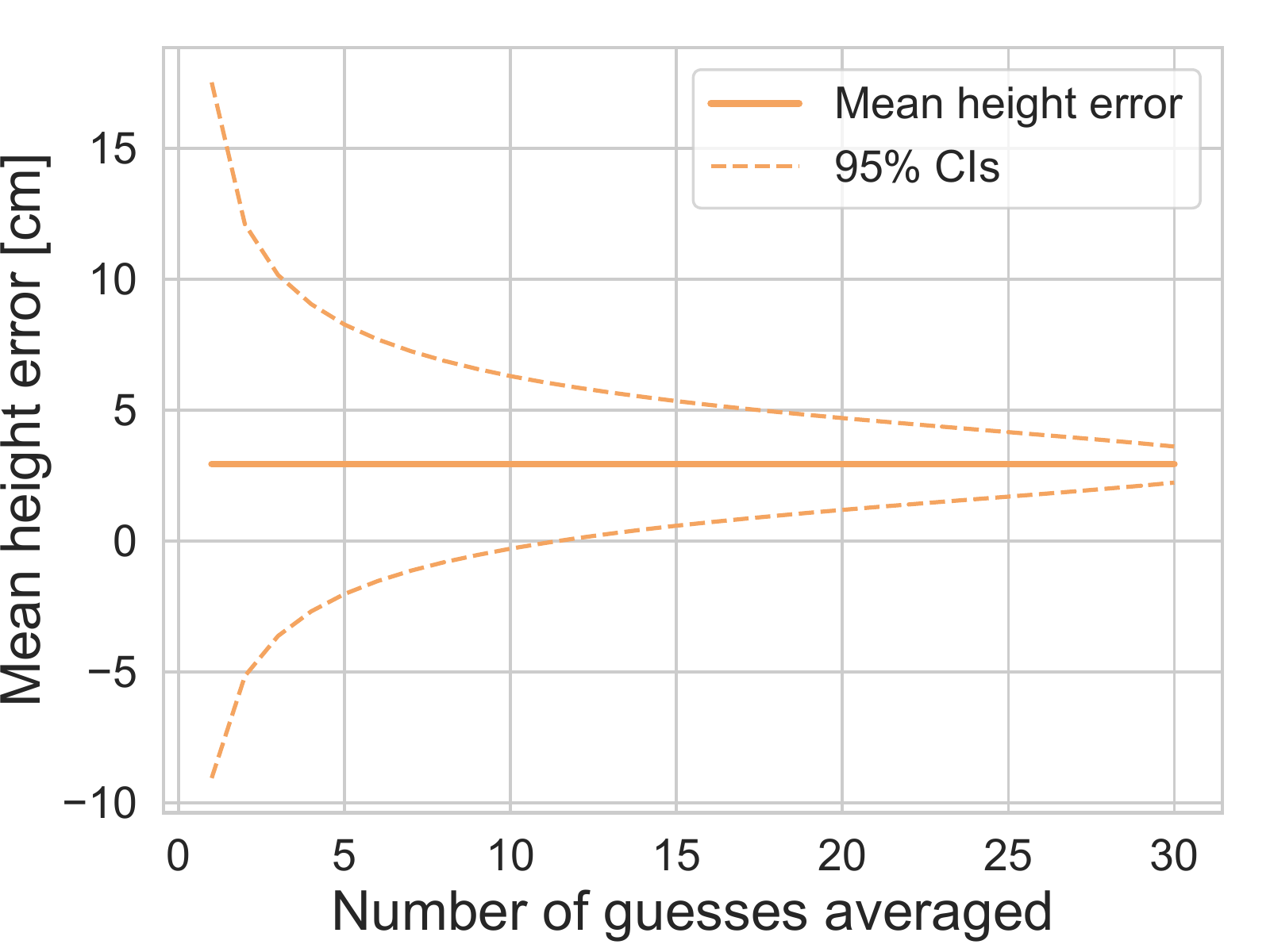}
\end{subfigure}%
\captionsetup{width=0.9\hsize} \caption{Convergence of mean error on all 1,682 images.}
\label{fig:convergence}
\end{figure}

\Figref{fig:convergence} shows the dependence of the mean height and weight error on the number of collected guesses. We see that the 95\% confidence intervals around the mean error become very small after around 30 guesses both for height and weight. As a side note, this fact justifies, \textit{post hoc,} our choice of collecting 45 guesses per image.

\begin{figure}[t]
\centering  
\begin{subfigure}{.5\hsize}
  \centering
      \includegraphics[width=\hsize]{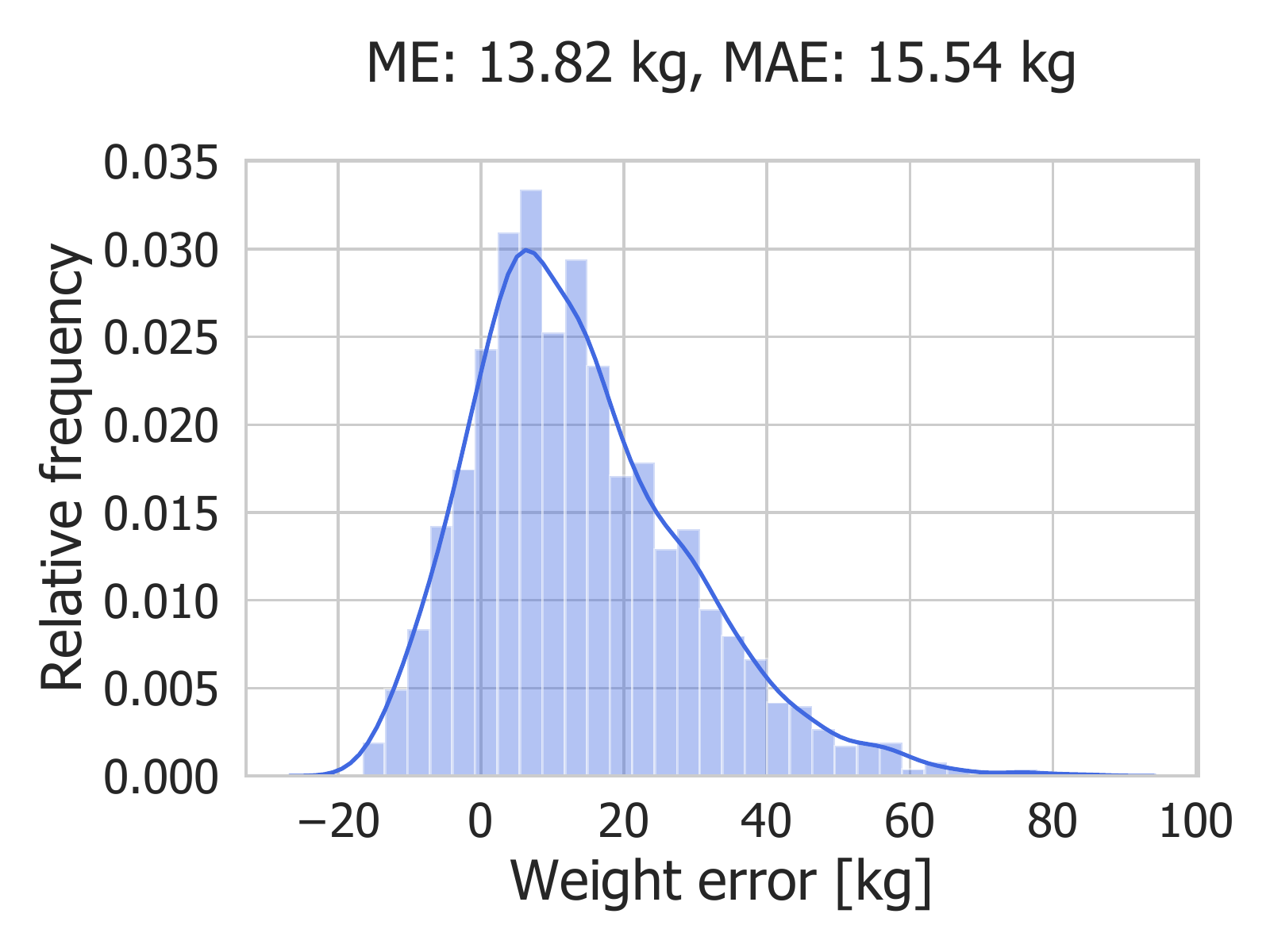}
\end{subfigure}%
\begin{subfigure}{.5\hsize}
  \centering
  \includegraphics[width=\hsize]{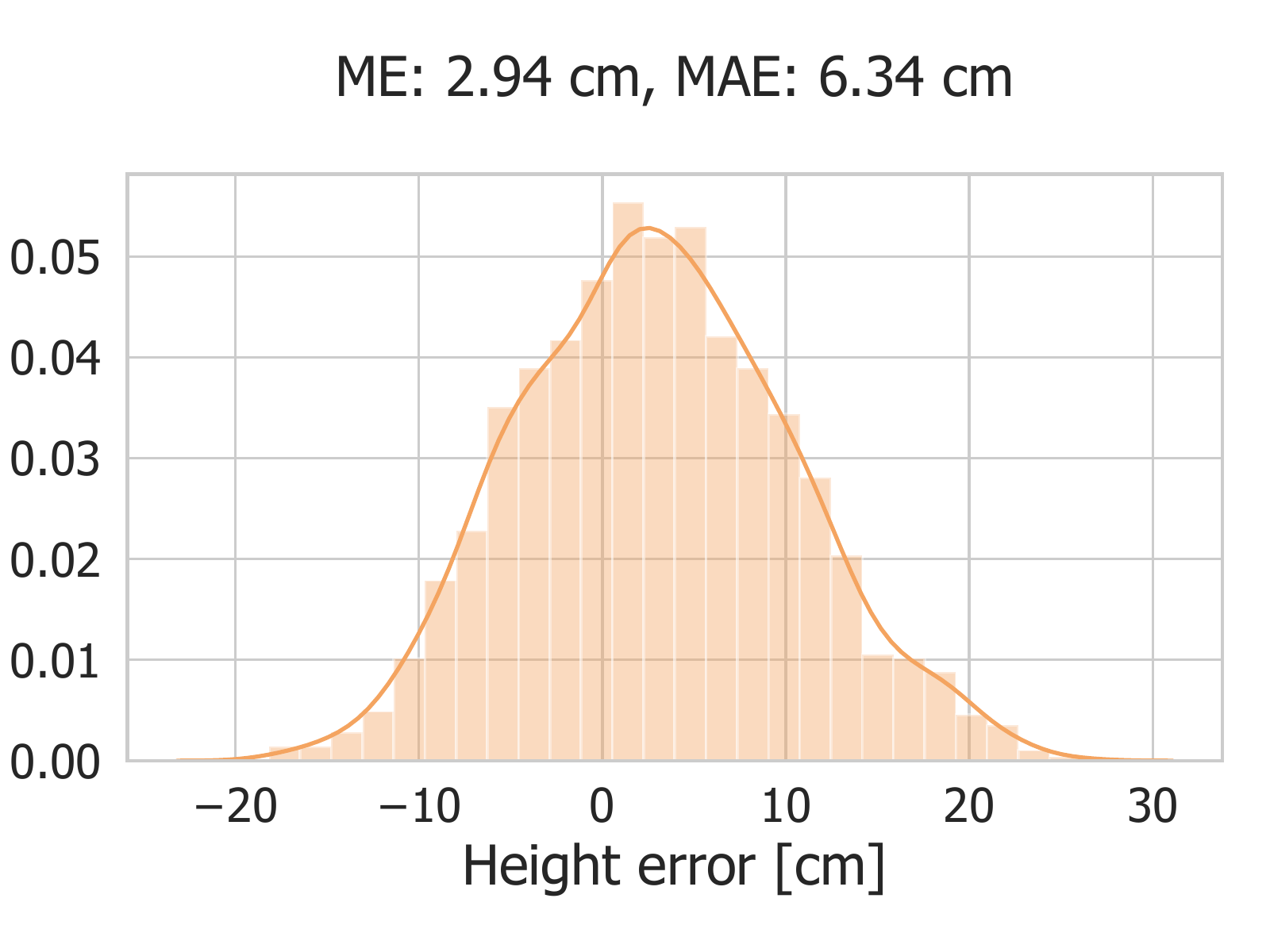}
\end{subfigure}
\captionsetup{width=0.9\hsize} \caption{Distributions of weight and height errors.}
\label{fig:error}
\end{figure}

The distributions of errors in terms of mean error (\ie, the mean of $w_{\err}$ and $h_{\err}$, respectively; ME) and mean absolute error (\ie, the mean of the absolute value of $w_{\err}$ and $h_{\err}$, respectively; MAE) are shown in \Figref{fig:error}. 
We see that the quality of the obtained estimates is low and that the errors of these ``raw'' crowdsourced estimates are too large to be of immediate use for any practical applications. 

According to the original idea of the ``wisdom of crowds'' as introduced by Galton \cite{galton1907vox}, we could also use the median of all collected guesses as the aggregate estimate $w^i_\est$. \Tabref{tab:mean vs median} shows that the difference between both methods is marginal.
Given the similar accuracy, we use the mean value as the aggregate estimate in the rest of the paper.

\begin{table}[]
    \centering
    \caption{Mean error (ME) and mean absolute error (MAE) when aggregating individual estimates via mean and median, with 95\% confidence intervals.}
    \begin{tabular}{l|c|c}
        \hline
        &  Mean & Median \\
        \hline
        Weight ME [kg] & $13.82 \pm 0.7$ & $14.47 \pm 0.7$ \\
        Weight MAE [kg] & $15.54 \pm 0.65$ & $16.09 \pm 0.7$ \\
        Height ME [cm] & $2.94 \pm 0.35$ & $2.55 \pm 0.35$ \\
        Height MAE [cm] & $6.34 \pm 0.25$ & $6.20 \pm 0.2$ \\
        \hline
    \end{tabular}

    \label{tab:mean vs median}
\end{table}

\subsection{Dependence of errors on true measurements}
\label{sec:Dependence of errors on true measurements}

\begin{figure}[t]
\centering  
\begin{subfigure}{.5\hsize}
    \centering
    \includegraphics[width=\hsize]{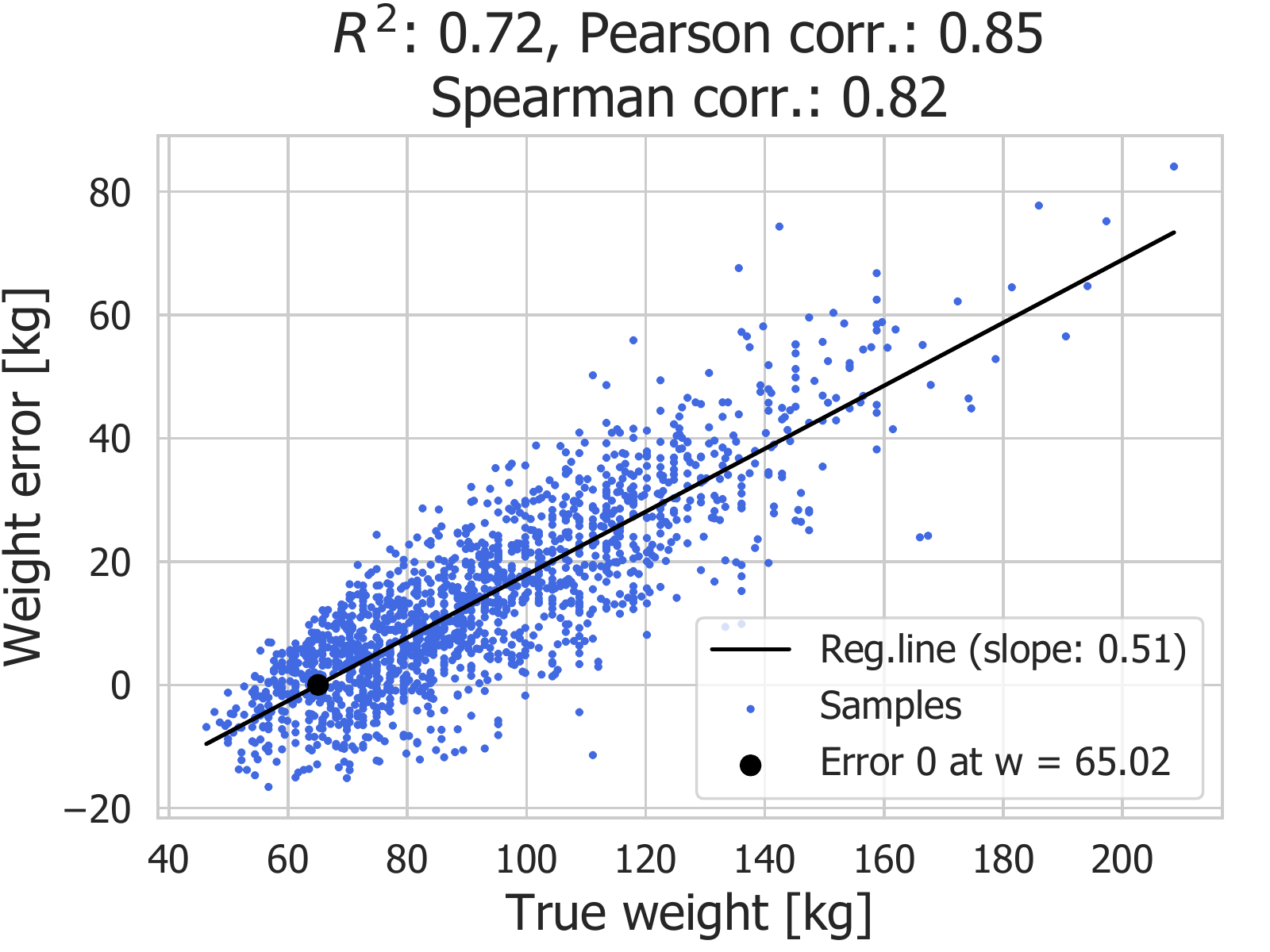}
\end{subfigure}%
\begin{subfigure}{.5\hsize}
    \centering
    \includegraphics[width=\hsize]{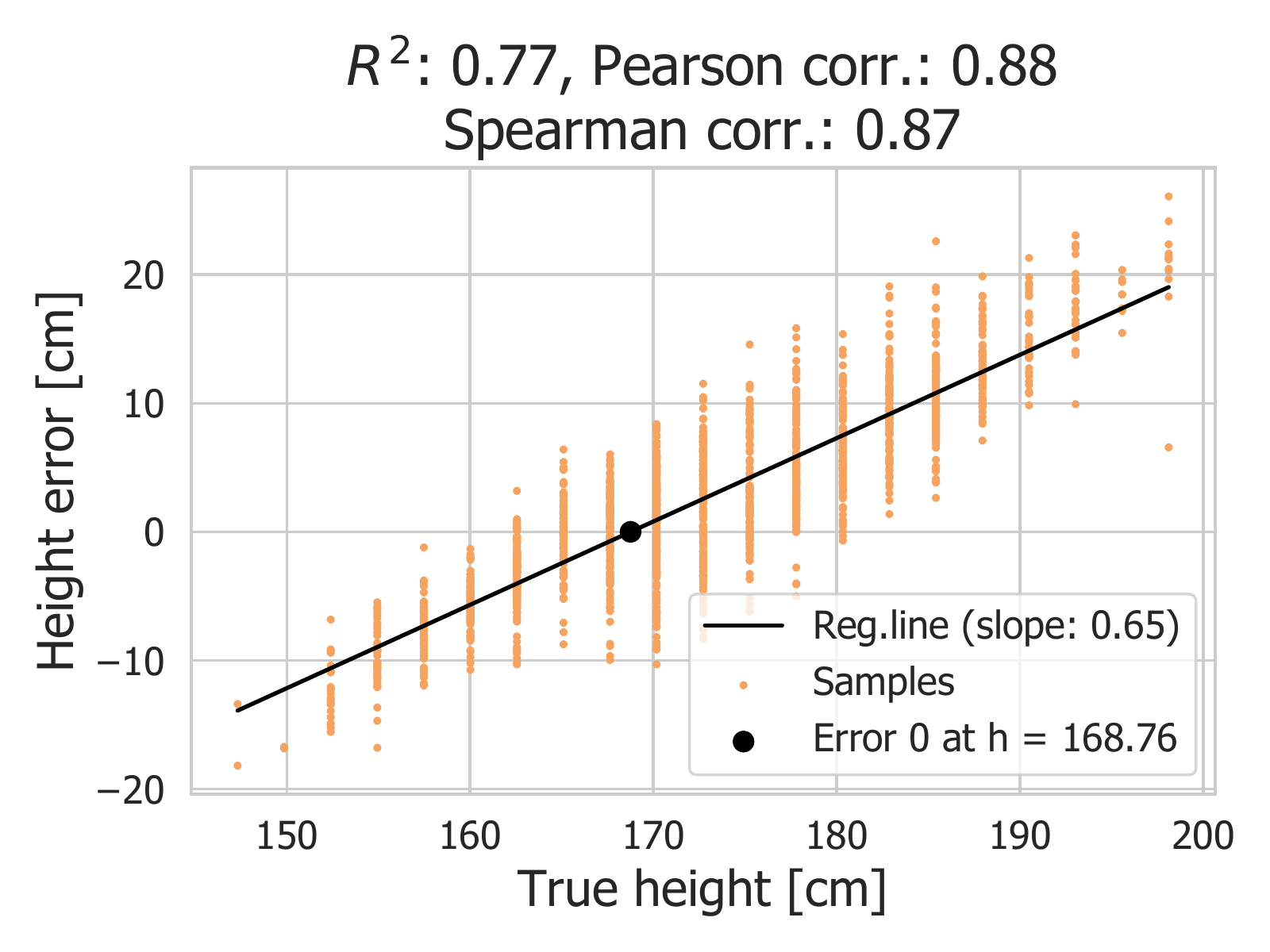}
\end{subfigure}
\captionsetup{width=0.9\hsize} \caption{Dependence of the crowd's estimation errors on the true weight/height of the images.
The $x$-axes appear quantized because the ground-truth data consists of integer numbers (pounds for weight, inches for height).}
\label{fig:error vs true value}
\end{figure}

\begin{figure}[t]
\centering  
\includegraphics[width=.5\hsize]{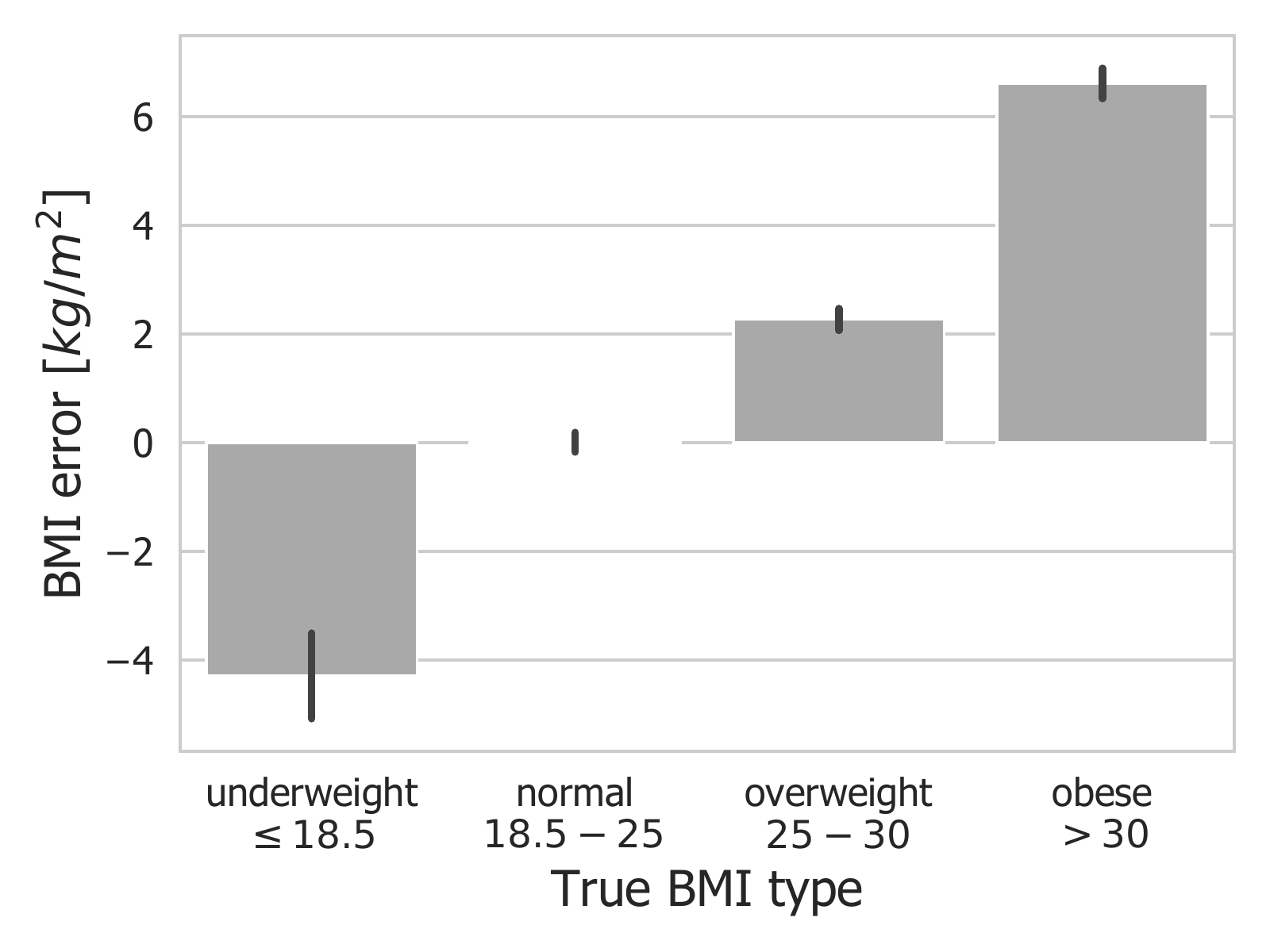}
\captionsetup{width=0.9\hsize} \caption{Crowd's estimation error \vs true BMI in images. Crowd workers are not asked to estimate BMI directly; instead, we compute the estimated BMI from the collected weight and height guesses.}
\label{fig:bmi error}
\end{figure}

From \Figref{fig:convergence} and~\ref{fig:error}, it becomes clear that, while the estimates converge, the resulting $w_\est$ differ strongly from the ground truth labels $w_\true$. 
Such behavior indicates that there is a systematic error associated with visual perception of weight and height. 
\Figref{fig:error vs true value} shows how the errors 
depend on the true measurements. We can see that deviations from average weight or height are largely underestimated, such that, \eg, the weight of light people is overestimated (negative errors), whereas the weight of heavy people is underestimated (positive errors). This observation explains the large positive mean error that we observe in \Figref{fig:error}: most images in the dataset represent obese people whose weight is clearly above average (cf.\ \Figref{fig:men image char} and \ref{fig:women image char}). This is an important finding, as it implies that the BMI of overweight and obese people is strongly underestimated, as also confirmed by \Figref{fig:bmi error}, which demonstrates that we cannot rely on humans' visual perception of weight in order to recognize obesity. 

These findings are in agreement with the literature. The results of Winkler and Rhodes \cite{winkler2005} suggest that people adapt their estimate to a \textit{reference value} of weight and height determined by all the bodies that people have seen, with a particularly strong influence of the most recent observations. 
In simplified terms, we can view crowd workers' reference values as their idea of a normal human weight or height. Cornelissen et al.~\cite{cornelissen2016visual} suggest that systematic visual biases in the estimation of weight can be explained by combining the assumption about reference values with a phenomenon known as ``contraction bias'' \cite{poulton1989bias}, which shifts the estimated value towards the guesser's reference value.
Thus, the trend in \Figref{fig:error vs true value} can be explained by contraction bias if we assume that workers' reference values are close to the average human weight and height.
In \Secref{sec:ref values derivation} and \ref{sec:ref values analysis}, we formulate a statistical, Bayesian model to derive individual reference values in a data-driven fashion, confirming this intuition.

\begin{figure}[t]
\centering  
\begin{subfigure}{.5\hsize}
    \centering
    \includegraphics[width=\hsize]{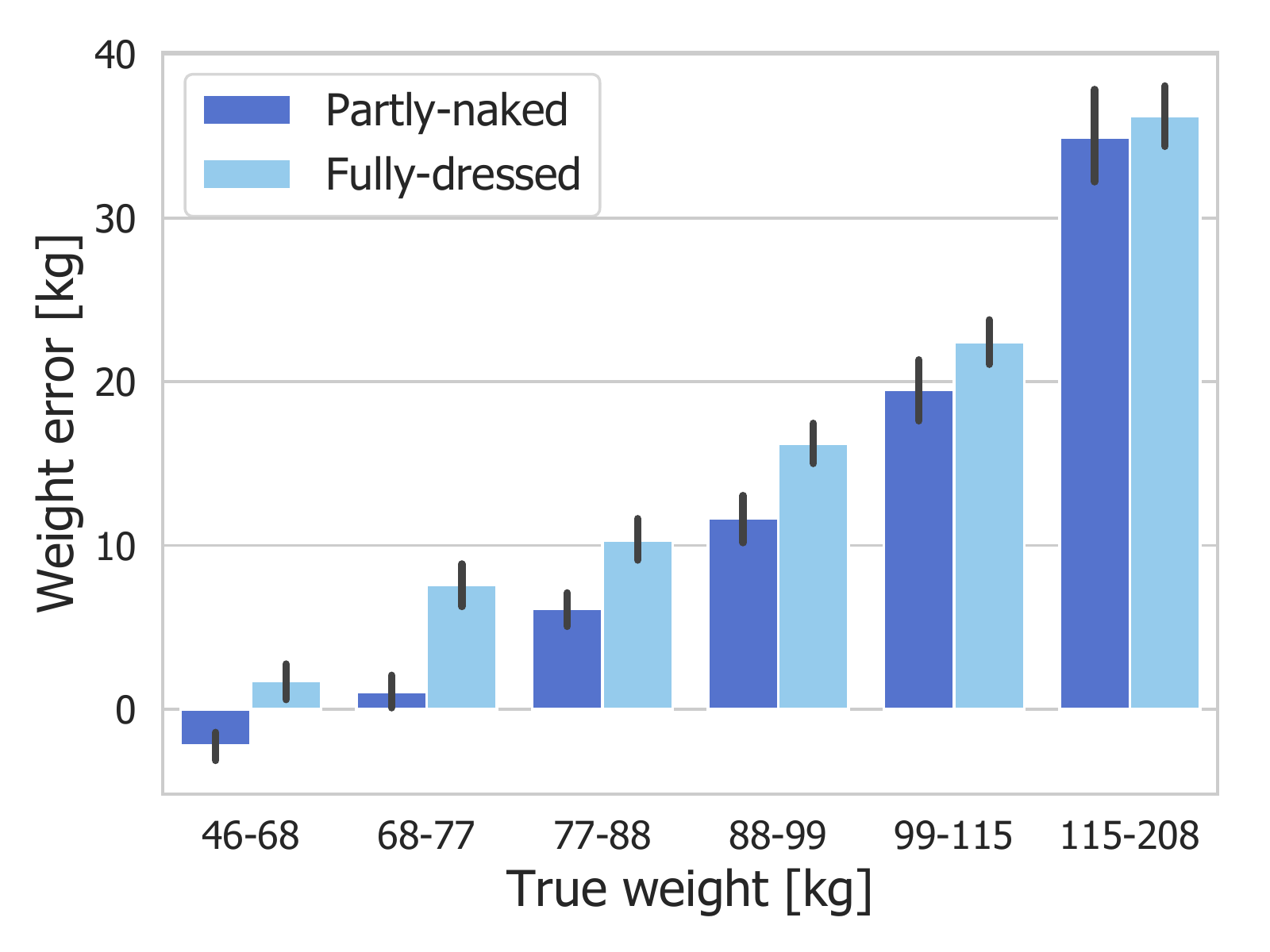}
\end{subfigure}%
\begin{subfigure}{.5\hsize}
    \centering
    \includegraphics[width=\hsize]{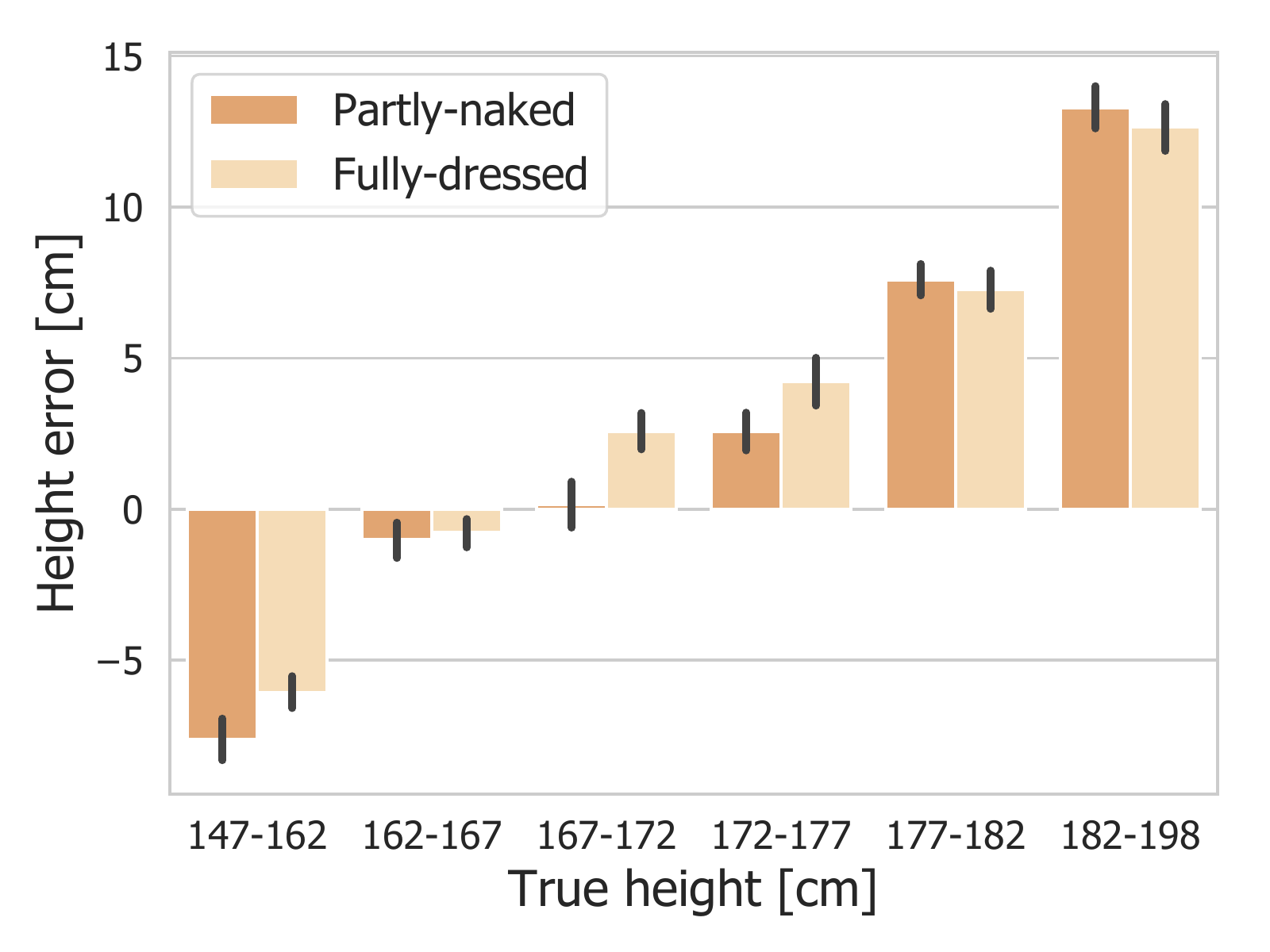}
\end{subfigure}
\captionsetup{width=0.9\hsize} \caption{Crowd's estimation errors by clothing type. Dressed people are estimated as lighter.}
\label{fig:error vs clothes}
\end{figure}

Clearly, the estimates of the crowd workers are influenced by other factors apart from the true measurements of people in the images. For example, \Figref{fig:error vs clothes} shows how clothing can impact the guesses: while the height guesses do not depend on clothing, the weight of fully\hyp dressed people is consistently estimated lower than the weight of similar (in terms of weight) partly\hyp naked people. Exploring this and other possible factors goes beyond the scope of this paper, and in the following we mainly focus on the bias caused by contraction towards the reference value.


% ...

\subsection{Dependencies between height and weight estimation}
\label{sec:Dependencies between height and weight estimation}

\begin{figure}[t]
\centering  
\begin{subfigure}{.48\hsize}
    \centering
    \includegraphics[width=\hsize]{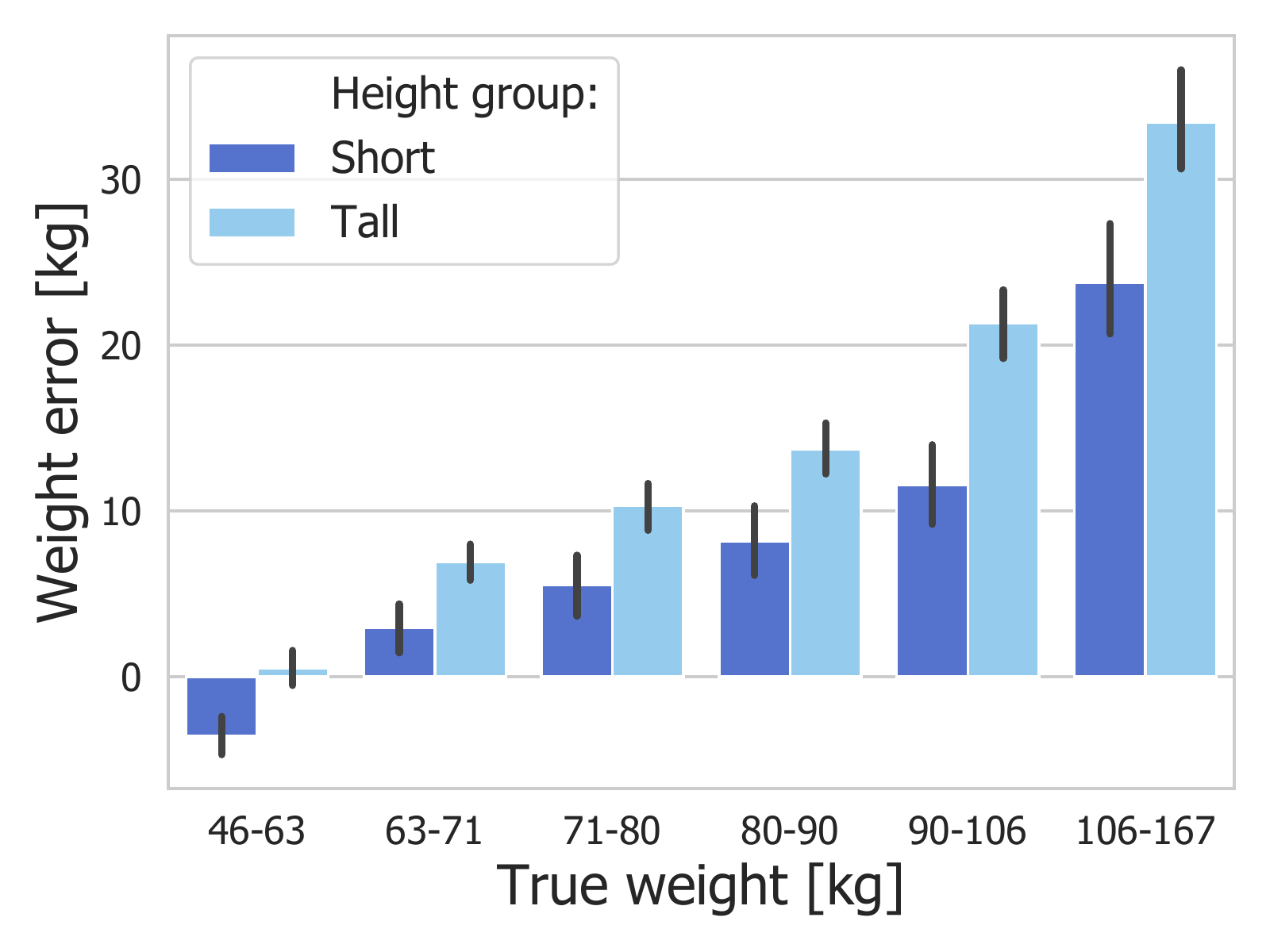}
    \caption{} \label{fig:weight error vs height type}
\end{subfigure}
\begin{subfigure}{.48\hsize}
    \centering
    \includegraphics[width=\hsize]{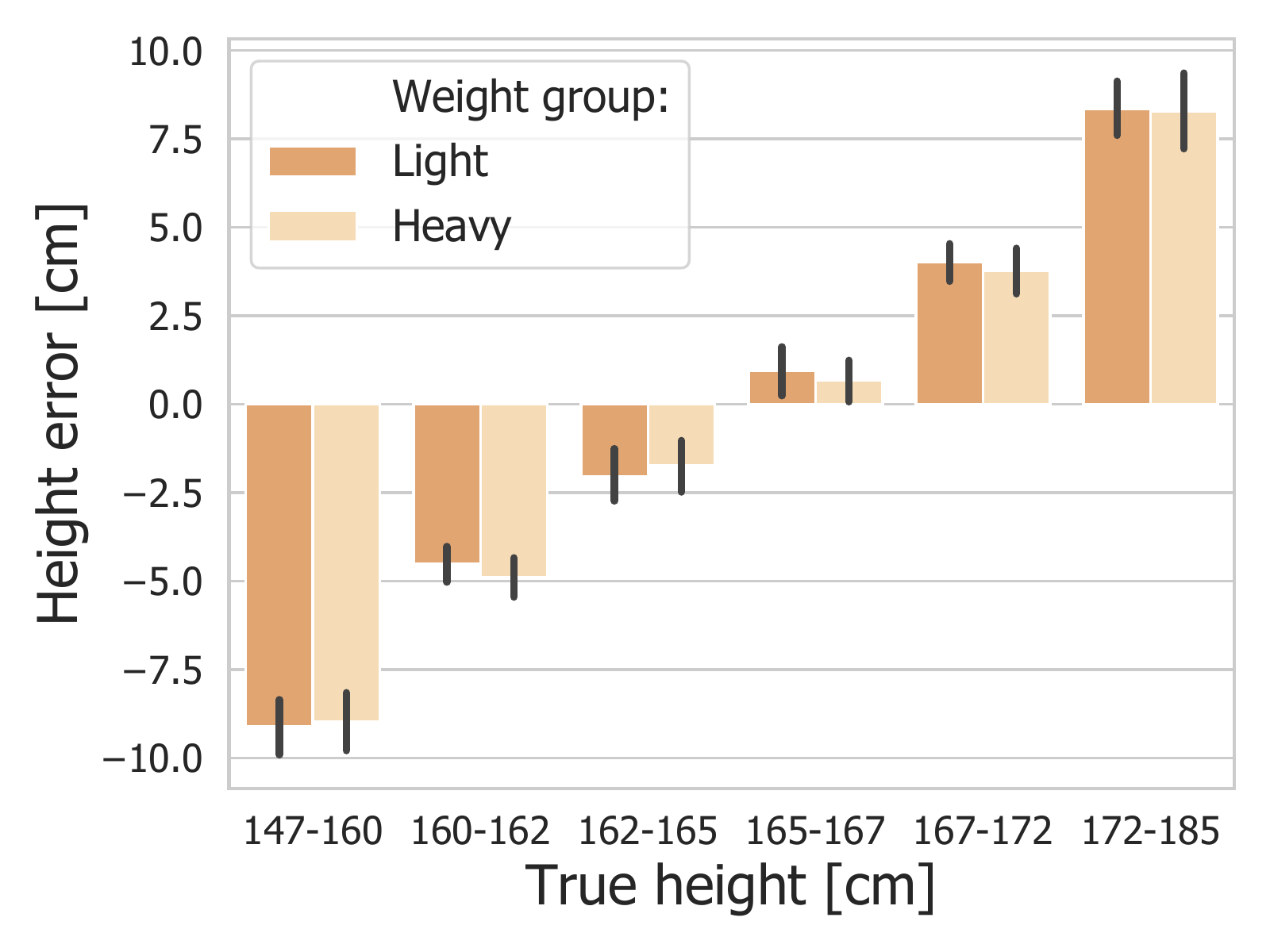}
    \caption{} \label{fig:height error vs weight type}
\end{subfigure}
\captionsetup{width=0.9\hsize} \caption{Dependence of the errors on the true measurements for female images. The black bars represent the 95\% confidence intervals.}
\label{fig:weight height dependence}
\end{figure}

In our experiment, workers estimated both the weight and height of people shown in the images. This setup allows us to gain further insights into how people approach this task.
In particular, we are interested in understanding whether a worker's weight estimate of a given person is also influenced by that person's height, rather than only by their weight (and analogously when swapping weight and height).
Since weight and height are correlated, we need to stratify the data in order to answer this question, as follows.
We first partition the set of all images into groups based on the quantiles of the weight distribution and further split each group into two subgroups: short and tall (i.e., height below \vs above group median).
We then plot the errors of the aggregated estimates for each subgroup of images (\Figref{fig:weight error vs height type}).%
\footnote{Since males and females differ with respect to their weight and height distributions, we perform this analysis separately for each gender. \Figref{fig:weight error vs height type} pertains to females. Similar results are obtained for males, but we omit them for space reasons.}
With a similar procedure, but partitioning first on height and then on the weight, we obtain \Figref{fig:height error vs weight type}.

\begin{figure}[t]
\centering  
    \centering
    \includegraphics[width=0.25\hsize]{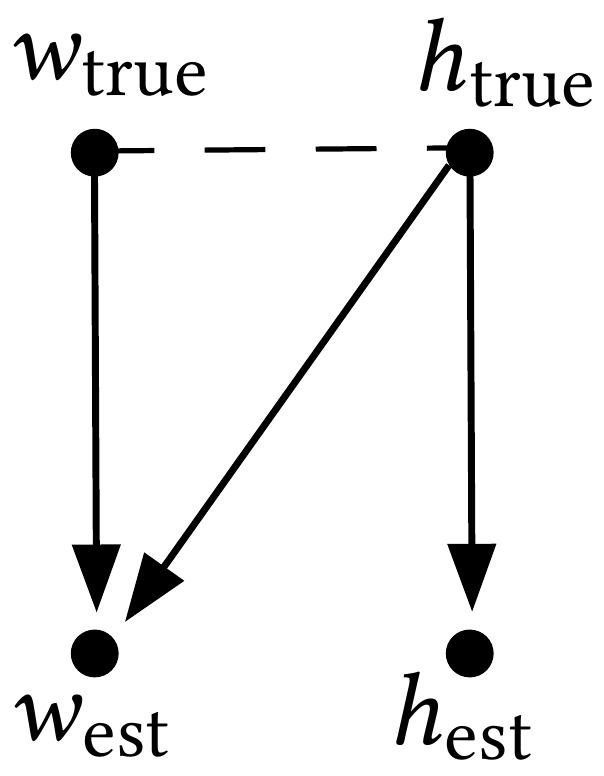}
\captionsetup{width=0.9\hsize} 
\caption{
Causal diagram \cite{pearl2009causality} capturing the dependence of weight and height estimates on true weight and height. The dashed line between true weight and height signifies that the two are correlated.
}
\label{fig:causal_diagrams}
\end{figure}

\Figref{fig:height error vs weight type} shows that, while the height error grows linearly with the true height, there is no significant difference between the two weight groups within each height group.%
\footnote{
Student's $t$-tests yield the following $p$-values for the null hypothesis of no difference in mean height errors between the light and heavy weight groups for fixed true height (from left to right in \Figref{fig:height error vs weight type}): 0.81, 0.34, 0.56, 0.55, 0.56, 0.90.
}
The pattern in \Figref{fig:weight error vs height type} is markedly different: in each weight group, the weight of tall people (whose height is larger than the group's median) is significantly more underestimated compared to the shorter half of the group.%
\footnote{
Student's $t$-tests yield the following $p$-values for the null hypothesis of no difference in mean weight errors between the short and tall height groups for fixed true weight (from left to right in \Figref{fig:weight error vs height type}): $2.1 \times 10^{-6}$,
$5.9 \times 10^{-5}$,
$7.7 \times 10^{-5}$,
$7.0 \times 10^{-5}$,
$1.3 \times 10^{-8}$,
$8.9 \times 10^{-5}$.
}
This implies that height estimates are conditionally independent of the true weight, given the true height (\Figref{fig:height error vs weight type}),
whereas weight estimates depend on the true height even conditioned on the true weight (\Figref{fig:weight error vs height type}).
For clarity, \Figref{fig:causal_diagrams} depicts this dependence structure as a causal diagram \cite{pearl2009causality}.

This finding has practical as well as theoretical implications.
In practical terms, we shall later (\Secref{sec:known height}) attempt to exploit the dependence of weight estimates on ground-truth height in order to improve weight estimates by supplying the ground-truth height (which is more stable and thus easier to obtain than the true weight) to workers at guessing time.
In theoretical terms, the unidirectional dependence can help us hypothesize about the mental models at work during weight and height estimation. We will return to this point in \Secref{sec:Relation to prior work}.

\section{Effect of worker characteristics}
\label{sec:worker spec}

So far, our discussion has focused on the results obtained by aggregating multiple guesses and did not take into account worker-specific characteristics. We now proceed to using the information collected from the workers in order to evaluate the impact of gender, age, country of residence, and, most important, the worker's own measurements on the produced guesses.

\subsection{General observations} \label{sec:worker spec general}
\setcounter{subfigure}{0}

\begin{figure}[t]
\centering  
\begin{subfigure}{.49\hsize}
    \centering
    \includegraphics[width=\hsize]{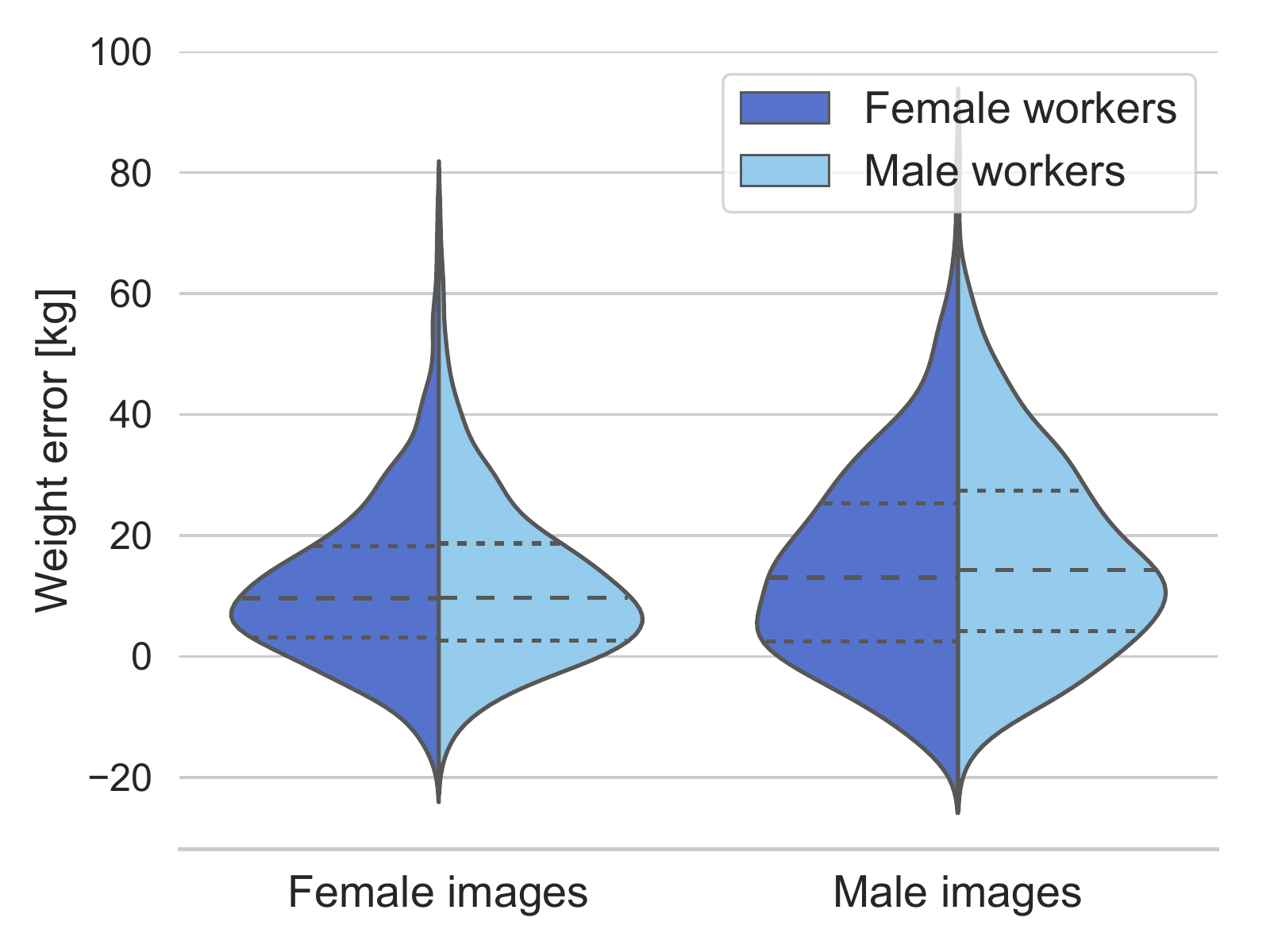}
    \caption{Weight error}
    \label{fig:weight error vs gender}
\end{subfigure}
\begin{subfigure}{.49\hsize}
    \centering
    \includegraphics[width=\hsize]{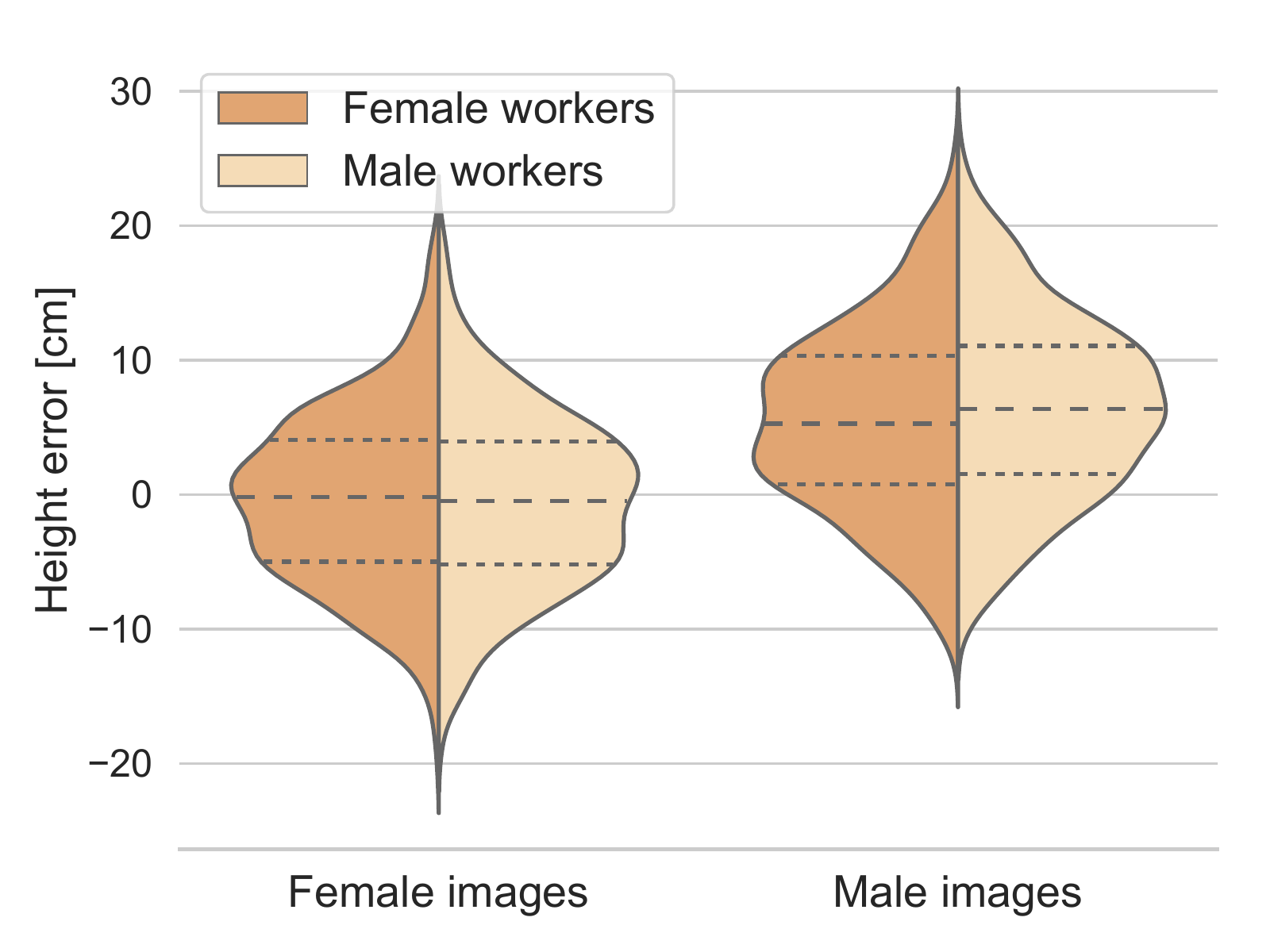}
    \caption{Height error}
    \label{fig:height error vs gender}
\end{subfigure}%
\captionsetup{width=0.9\hsize} \caption{Dependence of estimation errors on gender of workers and images.}
\label{fig:gender}
\end{figure}

We begin our discussion by looking at how the gender of workers and of people in the images influences the estimates. 
One might assume that workers would guess the measurements of people of their own gender more accurately. But, as \Figref{fig:gender} shows, accuracy is essentially identical for male and female workers: after separate aggregation of guesses from male and female workers, the weight and height errors made by male workers follow similar distributions as those made by female workers, for both male and female images.
(Student's $t$-tests yield $p=0.023$ for weight errors and $p=0.12$ for height errors; \ie, although the differences are minuscule, they may still be real, due to the large sample size.)

On the other hand, we can discern a clear difference in errors for male \vs\ female \textit{images,} with considerably larger mean height and weight errors for male than for female images ($p=1.7 \times 10^{-7}$ for weight errors and $p=1.0 \times 10^{-79}$ for height errors, according to Student's $t$-tests),
which is unexpected in the sense that the body-mass index (BMI) distributions are similar for males and females in our image data.
It might, however, be explained by the fact that the height and weight is larger for male images, which are therefore more underestimated (\cf\ \Figref{fig:error vs true value}).
Also, the weight error distribution (\Figref{fig:gender}a) has a larger variance for male images, which may be explained by the larger variance of the true weight of male images in the dataset (standard deviation 25.6~kg for men, \vs\ 21.7~kg for women).

\begin{figure}[t]
\centering  
\begin{subfigure}{.49\hsize}
    \centering
    \includegraphics[width=\hsize]{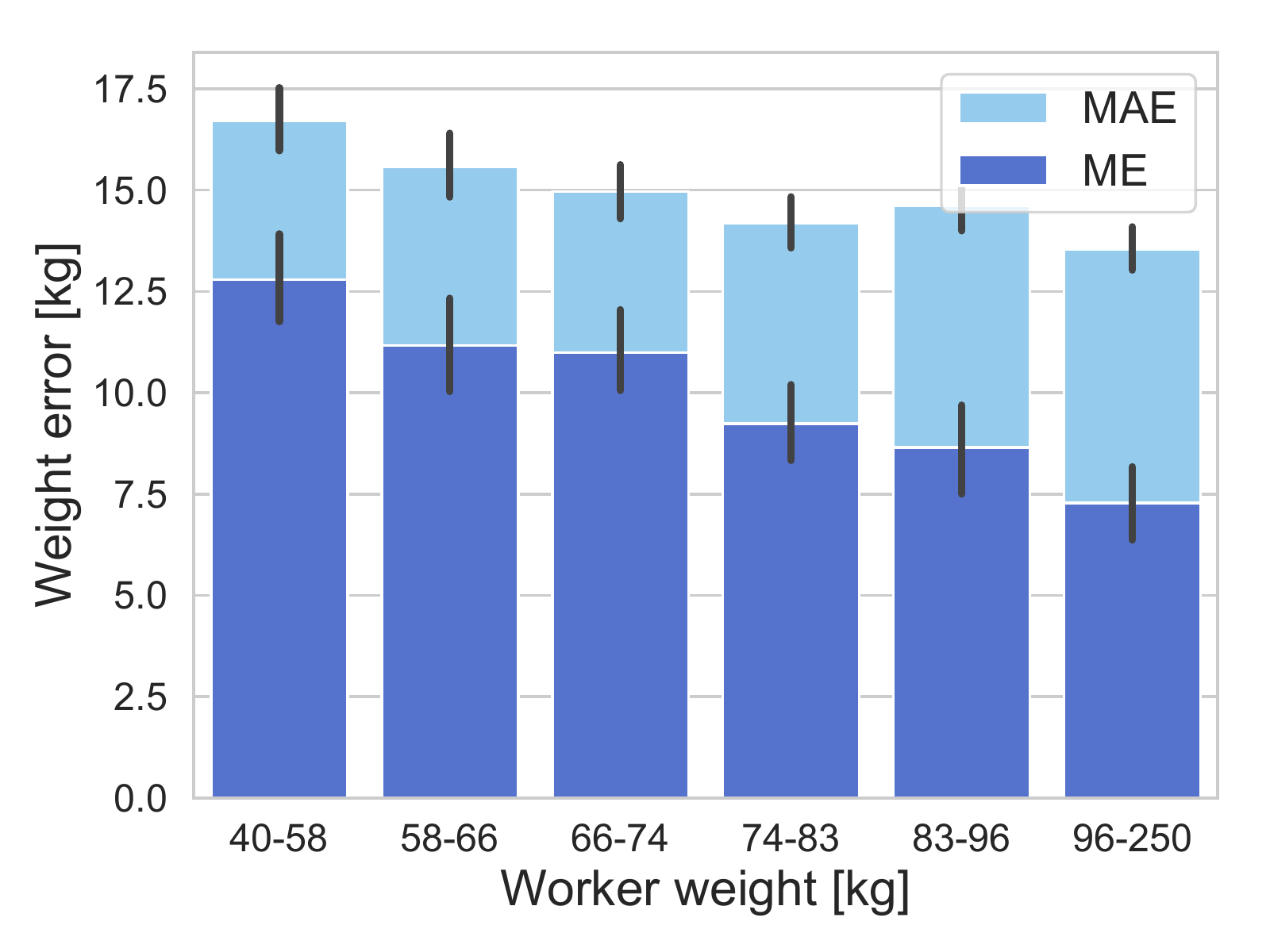}
\end{subfigure}
\begin{subfigure}{.49\hsize}
    \centering
    \includegraphics[width=\hsize]{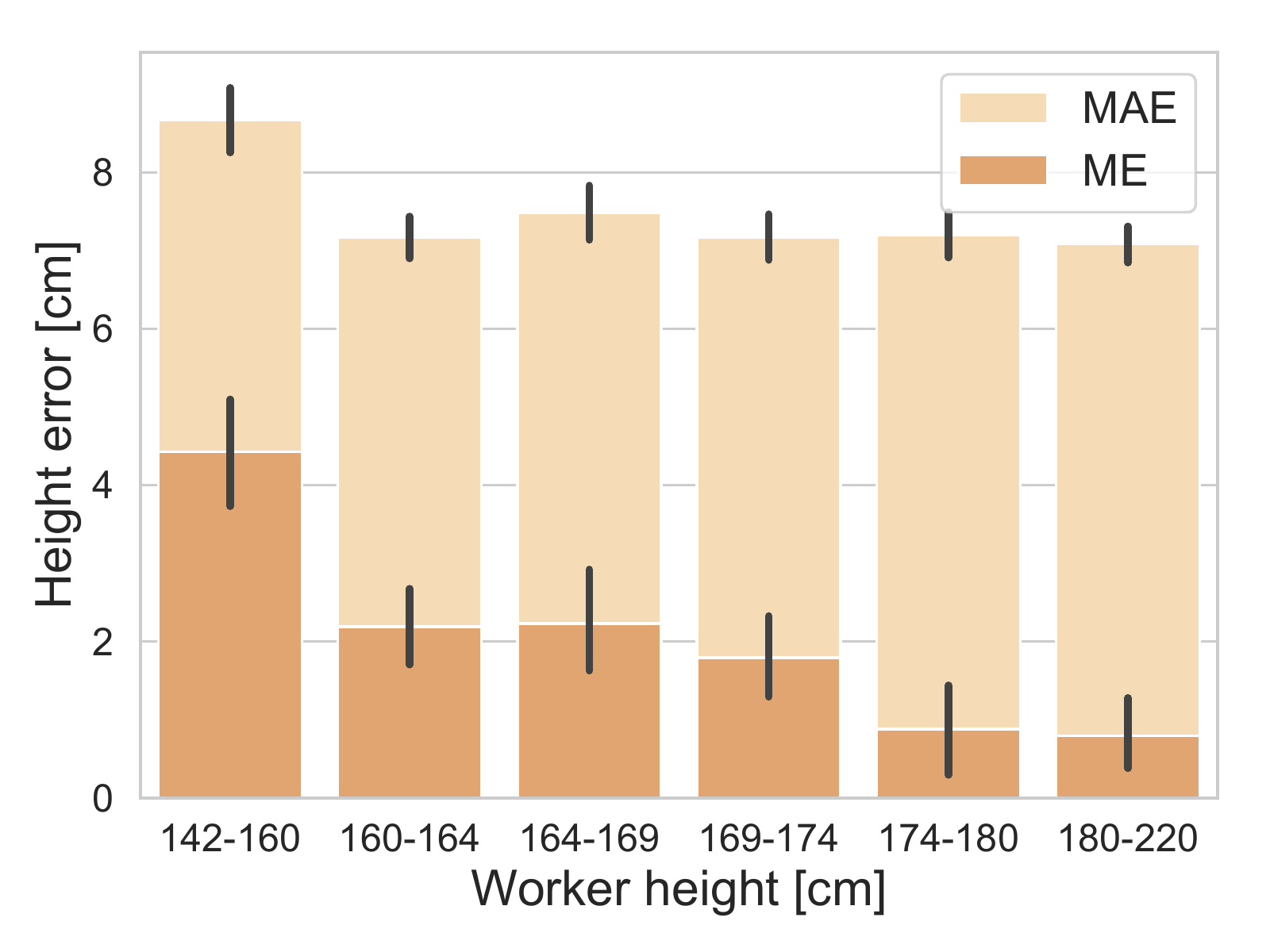}
\end{subfigure}%
    \captionsetup{width=0.9\hsize} \caption{Dependence of weight\slash height mean error (ME) and mean absolute error (MAE) of guesses on worker's weight\slash height.}
    \label{fig:error vs worker's weight}
\end{figure}

We next look at the dependence of the guesses on the workers' own body measurements. For this purpose, we split the workers into six bins based on their weight or height quantiles, and compute the mean error and the mean absolute error of all guesses provided by the workers from each bin. 
\Figref{fig:error vs worker's weight} shows the relationships between the workers' weight\slash height and their errors.
The fact that all mean errors (ME, dark bars) are positive implies that, regardless of their own weight and height, workers tend to underestimate weight and height in images.
Additionally, we can see a clear monotonic decrease of the mean error as the worker's weight or height, respectively, decreases.
This means that taller (heavier) workers tend to guess larger values for height (weight), leading to lower positive errors (\ie, less underestimation). 

\begin{figure}[t]
\centering  
\begin{subfigure}{.49\hsize}
    \centering
    \includegraphics[width=\hsize]{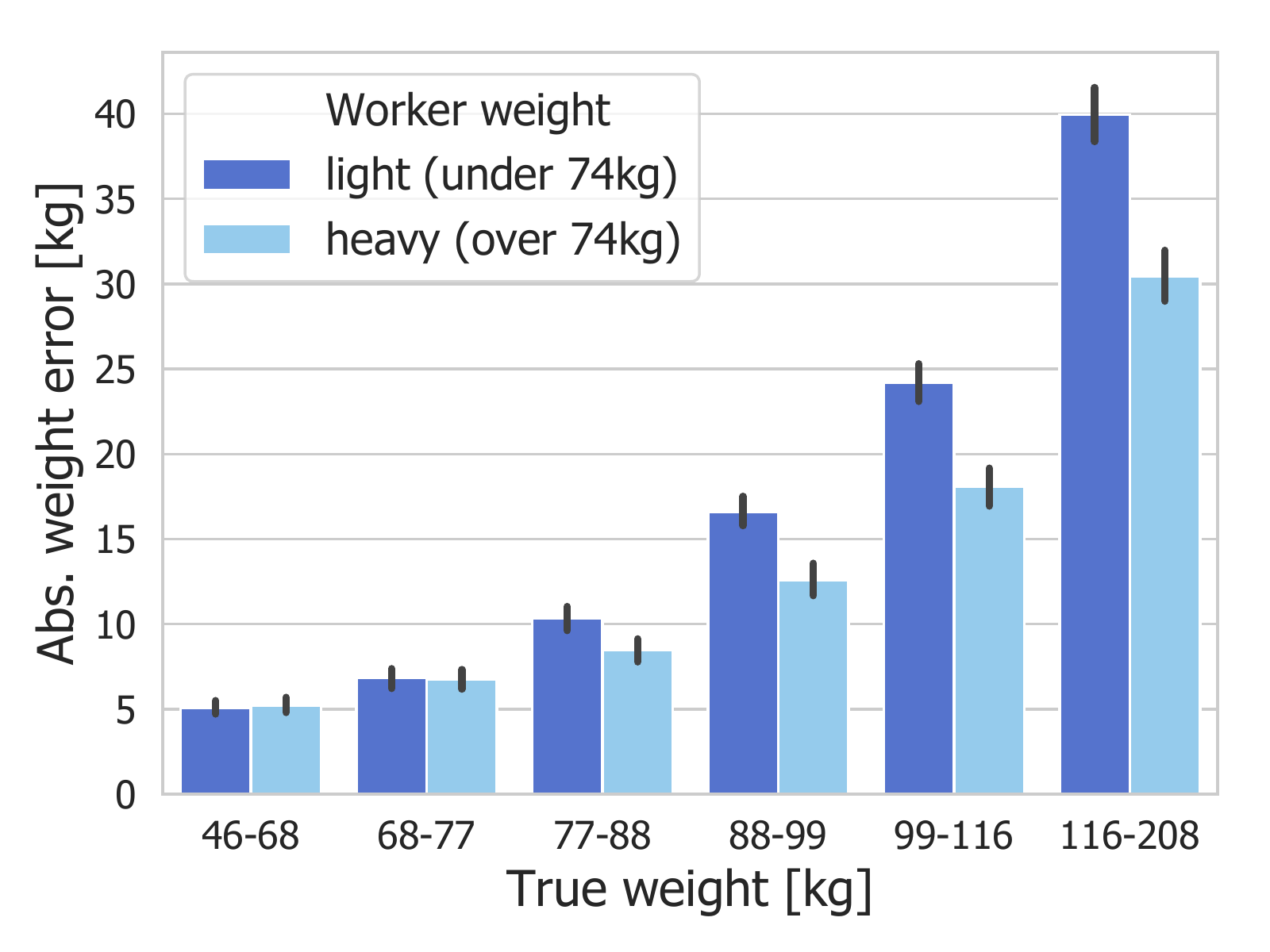}
\end{subfigure}
\begin{subfigure}{.49\hsize}
    \centering
    \includegraphics[width=\hsize]{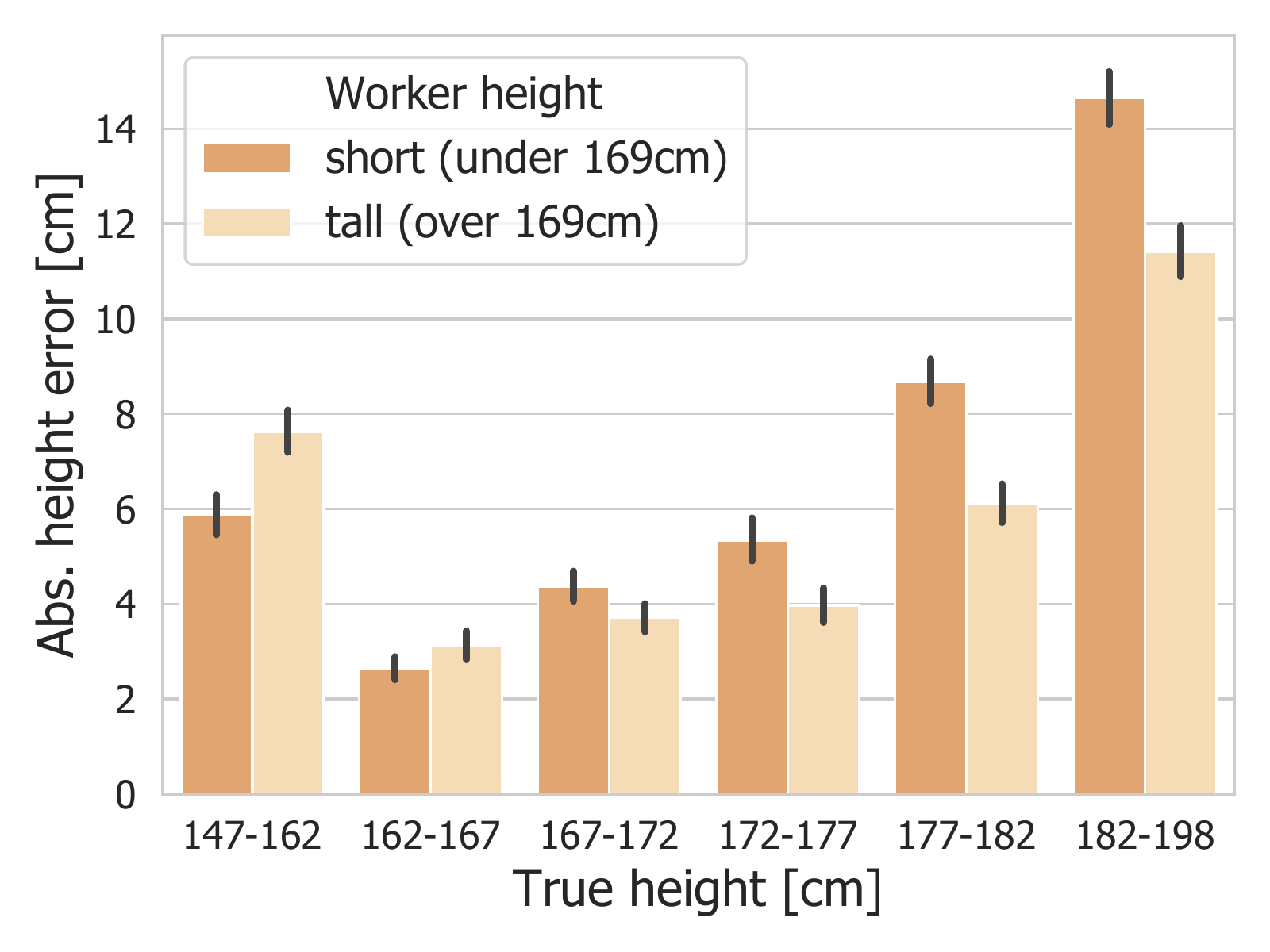}
\end{subfigure}%
    \captionsetup{width=0.9\hsize} \caption{Dependence of weight\slash height mean absolute error (MAE) of guesses on true weight\slash height of the person in the image ($x$-axis) and worker's weight\slash height.}
    \label{fig:error categorical vs worker's weight}
\end{figure}

\Figref{fig:error vs worker's weight} also shows a decrease in mean \textit{absolute} error (MAE, light bars).
\textit{A priori,} this could be caused by heavier and taller workers being more accurate in general.
However, this conclusion is invalidated by \Figref{fig:error categorical vs worker's weight}, which shows that heavier (taller) workers are more accurate than lighter (shorter) workers specifically on images of heavier (taller) people.
We therefore conclude that the decrease of errors with increasing worker weight (\Figref{fig:error vs worker's weight}) is simply an artefact of the overrepresentation of heavy images in our dataset.
(Similar conclusions hold for height, although the performance difference between short and tall workers is less drastic.)

\subsection{Model for inferring reference values}
\label{sec:ref values derivation}

Based on the results from \Secref{sec:worker spec}, in combination with the prior literature on reference values and contraction bias (\Secref{sec:relwork}), we hypothesize that heavier and taller workers have larger reference values.
To support this hypothesis, we now develop a simple generative model that describes the process of guessing and is based on the notions of reference values and contraction bias introduced in \Secref{sec:gen accuracy}. 
We describe our model for the case of weight; the case of height is fully analogous.

In our model, we assume that there are two factors that influence a worker's guess: the worker's personal reference value and the true weight of the person in the image.
More specifically, a worker $j$ produces a weight guess $w_j^i$ for an image $i$ with true weight $w^i_\true$ according to the following equation:
\begin{equation}
	w_j^i = \alpha_j w_j^{\reference} + (1-\alpha_j) w^i_\true + \epsilon,
	\label{eqn:estimate generation}
\end{equation}
where $w_j^{\reference}$ represents a worker\hyp specific reference value towards which the worker's guesses are skewed, $\alpha_j \in [0,1]$ represents the worker\hyp specific contraction coefficient and describes the power of the contraction effect (larger $\alpha_j$ implies stronger averaging due to contraction bias), and $\epsilon$ denotes the error and accounts for all remaining complexity of the visual perception task, such as various view angles, body positions, etc.
As the worker $j$ is fixed in the following derivations, we drop the worker index $j$ from here on for readability.

In order to infer the weight reference value $w^\reference$ and contraction coefficient $\alpha$ for a given worker,
we use Bayes' rule (\Eqnref{eq:bayes1}) and leverage two independence assumptions:
first, that a given worker's reference value and contraction coefficient are independent of the images shown to the worker
(\Eqnref{eq:bayes2}; this independence is implied by modeling the parameters as fixed for each worker);
and second, that the reference value and the contraction coefficient are independent of one another (\Eqnref{eqn:bayes}):
\begin{eqnarray}
	p(w^{\reference}, \alpha | w, w_\true) &=& \frac{p(w | w_\true, w^{\reference}, \alpha) \ p(w^{\reference}, \alpha | w_\true)}{p(w | w_\true)} \label{eq:bayes1} \\
	&=& \frac{p(w | w_\true, w^{\reference}, \alpha) \ p(w^{\reference}, \alpha)}{p(w | w_\true)} \label{eq:bayes2} \\
	&\propto& p(w | w_\true, w^{\reference}, \alpha) \ p(w^{\reference}, \alpha) \label{eq:bayes3} \\
	&=& p(w | w_\true, w^{\reference}, \alpha) \ p(w^{\reference}) \ p(\alpha), \label{eqn:bayes}
\end{eqnarray}
where $w$ is a vector that contain all guesses collected from the given worker, and $w_\true$ is a vector of the corresponding ground-truth labels. 

We assume wide priors for reference values:
$w^{\reference} \sim \mathcal{N}(\mu, \sigma)$, where $\mu = 70$~kg is an estimate of average human weight (taking into account the workers' countries of residence), and where $\sigma\approx 25$~kg is a large standard deviation, such that that all reasonable values of reference weight lie within one standard deviation from $\mu$ and more influence is given to the data evidence.
As the power of the contraction effect is hard to estimate \textit{a priori,} a uniform distribution on $[0,1]$ is used:
$\alpha \sim \mathcal{U} (0,1)$.
Finally, it is assumed that the errors follow a normal distribution:
$\epsilon \sim \mathcal{N}(0, \sigma_\epsilon)$, where $\sigma_\epsilon \approx 15$~kg is the empirical standard deviation of weight errors among the crowdsourced weight estimates.%
\footnote{For height, rather than weight, we use $\mu = 170$~cm, $\sigma=15$~cm, $\sigma_\epsilon=10$~cm.}

With these assumptions, the following distributions from \Eqnref{eqn:bayes} can be written explicitly:
\begin{eqnarray}
	p(w | w_\true, w^{\reference}, \alpha) &=& \prod_i  \frac{1}{\sqrt{2 \pi} \sigma_\epsilon} \exp\left(-\frac{\left(w^i - \alpha w^{\reference} - (1-\alpha) w^i_\true\right)^2}{2 \sigma_\epsilon^2}\right) \label{eqn:bayes data}\\
	p(w^{\reference}) &=&
\frac{1}{\sqrt{2 \pi} \sigma} \exp\left(-\frac{(w^{\reference} - \mu)^2}{2 \sigma^2}\right) \label{eqn:bayes prior} \\
	p(\alpha) &=&
\begin{cases}
1 & \text{ if } 0 \leq \alpha \leq 1,\\
0 & \text{ otherwise.}
\end{cases}
\end{eqnarray}

To simplify the notation, from now on we assume that only the meaningful range $0 \leq \alpha \leq 1$ is considered. Inserting \Eqnref{eqn:bayes data} and \ref{eqn:bayes prior} into the main \Eqnref{eqn:bayes}, we have
\begin{equation}
    \label{eqn:bayes posterior}
    p(w^{\reference},\alpha | w, w_\true) \propto \exp\left(-\frac{(w^{\reference} - \mu)^2}{2 \sigma^2}\right) \; \prod_i \exp\left(-\frac{\left(w^i - \alpha w^{\reference} - (1-\alpha) w^i_\true\right)^2}{2 \sigma_\epsilon^2}\right).
\end{equation}

For computing the maximum \textit{a-posteriori} (MAP) estimates of $w^{\reference}$ and $\alpha$, we take logarithms and multiply with $-2\sigma_\epsilon^2$, thus obtaining the loss function
\begin{equation}
	L(w^{\reference}, \alpha) = \frac{\sigma_\epsilon^2}{\sigma^2} \left(w^{\reference} - \mu\right)^2 + \sum_i \left(w^i - \alpha w^{\reference} - (1-\alpha) w^i_\true\right)^2.
\end{equation}
The MAP estimates of $w^{\reference}$ and $\alpha$ are then obtained by minimizing $L$ via gradient descent under the constraint $0 \leq \alpha \leq 1$.%
\footnote{
As a technical detail, we note that, although $L$ is convex for each variable separately, it is not convex in $(w^\reference, \alpha)$ jointly, due to the product $\alpha w^{\reference}$.
Gradient descent is hence not guaranteed to find a global, but merely a local, minimum.
As a sanity check, we hence also minimized $L$ by splitting the $\alpha$ interval $[0,1]$ into 100 pieces and performing constrained gradient descent on each piece separately, starting from 5 random initializations per piece.
Comparing to the results from vanilla gradient descent on the full interval $\alpha \in [0,1]$, the loss improved for only 1 out of about 300 workers, so we conclude that vanilla gradient descent is good enough for this setting.
}

We emphasize that reference values $w^{\reference}$ and contraction factors $\alpha$ are computed solely based on the respective worker's guesses and the ground-truth weight of the images they provide guesses on, and not based on the worker's own weight.
Next, we will show that, nonetheless, reference values correlate naturally with the worker's own weight.

\subsection{Analysis of reference values} 
\label{sec:ref values analysis}

In this section, we analyze the reference values and contraction coefficients inferred in \Secref{sec:ref values derivation} in order to gain further insights into the crowd workers' biases during weight and height estimation.

\begin{figure}[t]
\centering
    \begin{subfigure}[t]{.49\hsize}
        \centering
        \includegraphics[width=\hsize]{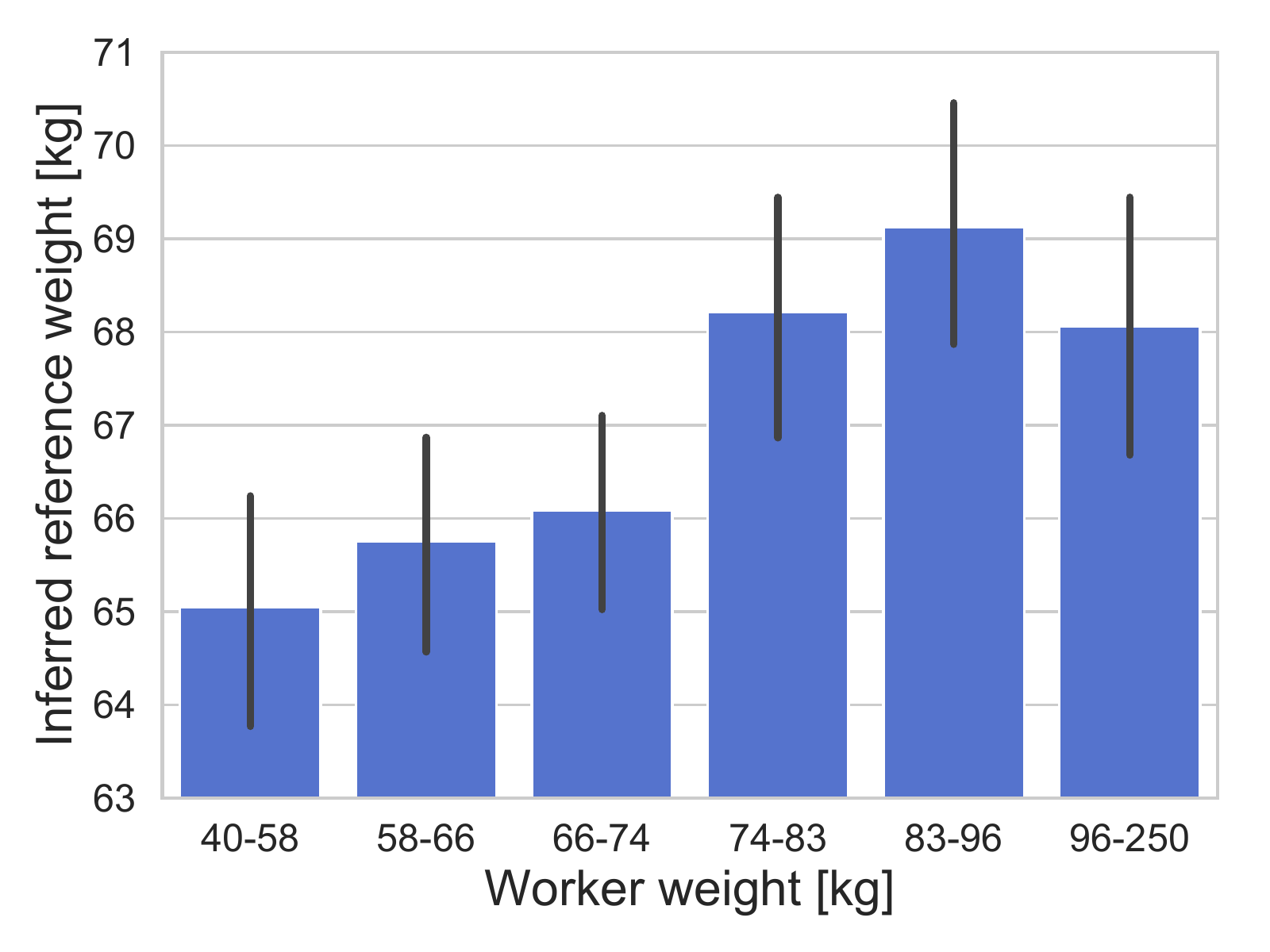}
        \caption{} \label{fig:ref values w}
    \end{subfigure}
    \begin{subfigure}[t]{.49\hsize}
        \centering
        \includegraphics[width=\hsize]{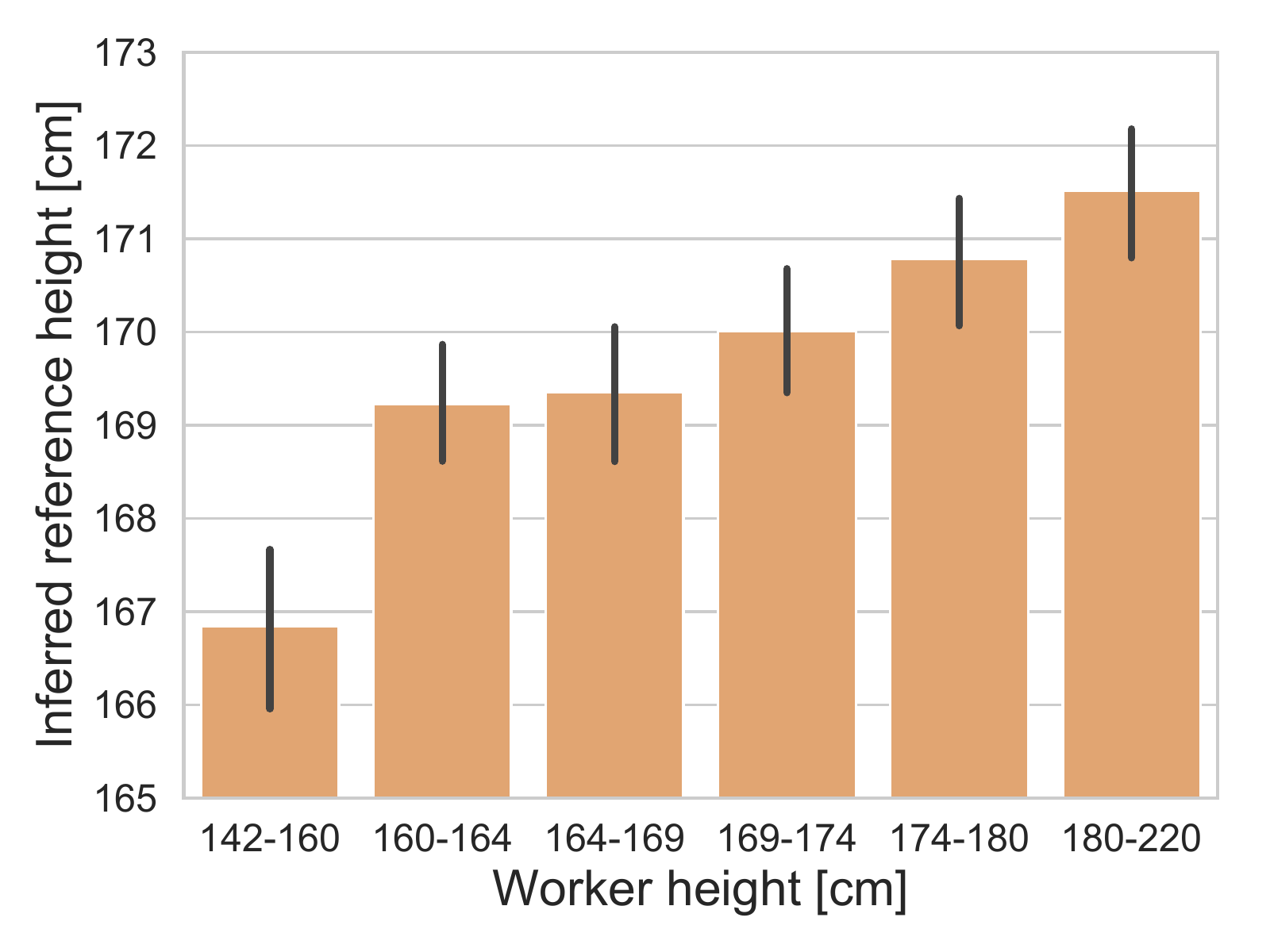}
        \caption{} \label{fig:ref values h}
    \end{subfigure}%
\captionsetup{width=0.9\hsize} \caption{Dependence of weight/height reference value on worker's weight/height.}
\label{fig:ref values}
\end{figure}

The dependence of the reference values on the worker's own weight and height is shown in \Figref{fig:ref values}. The plots support our previous hypothesis that the reference values grow with the worker's own weight and height. 
Note how the dependence for weight breaks with the last bin in \Figref{fig:ref values w}. It is possible that workers from this group provided wrong information (whether due to typos or on purpose) about their own weight, as we received suspiciously high values (up to 250~kg) from several workers.

\begin{figure}[t]
    \begin{subfigure}{.49\hsize}
        \centering
        \includegraphics[width=\hsize]{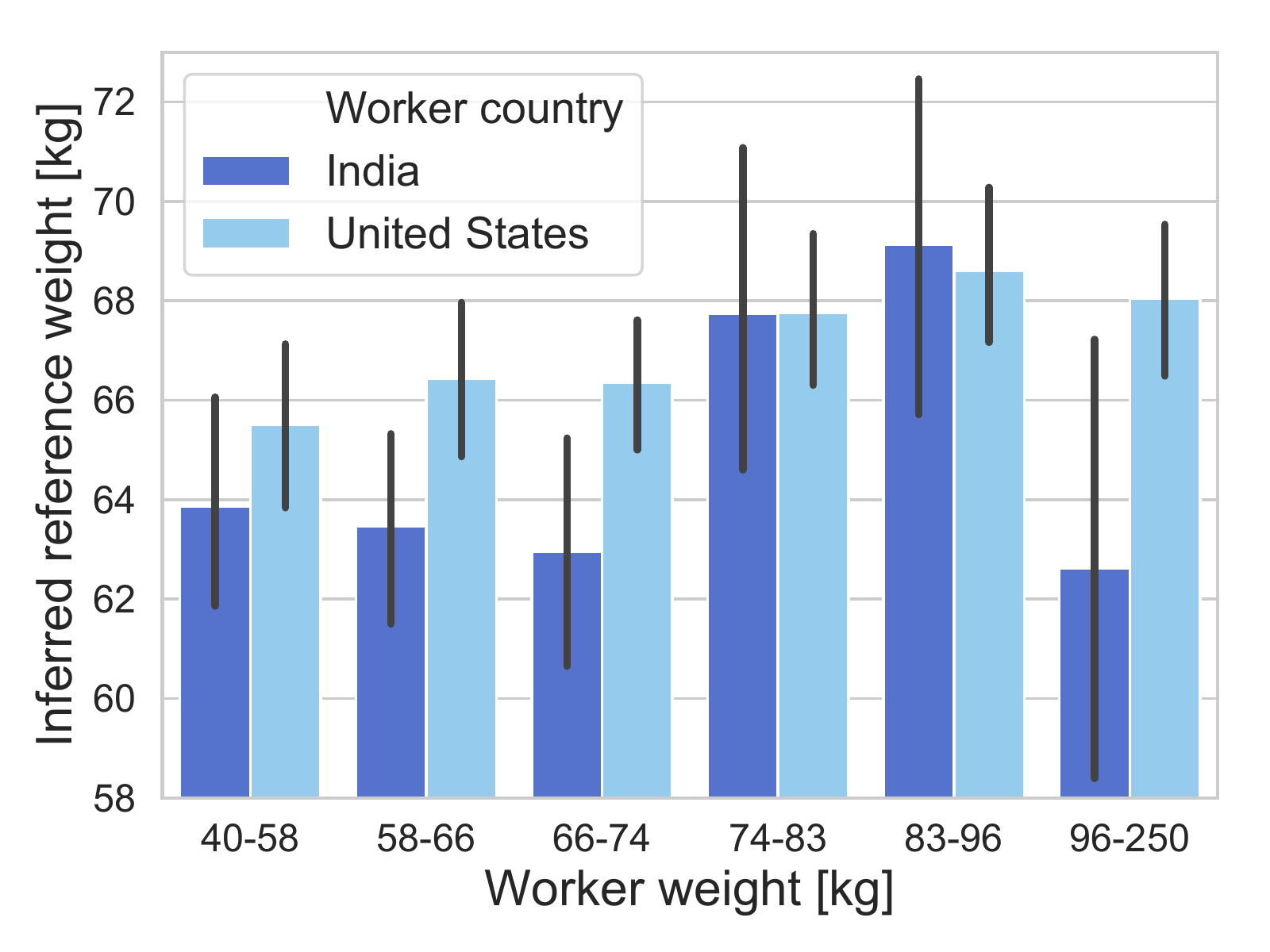}
    \end{subfigure}
    \hfill
    \begin{subfigure}{.49\hsize}
        \centering
        \includegraphics[width=\hsize]{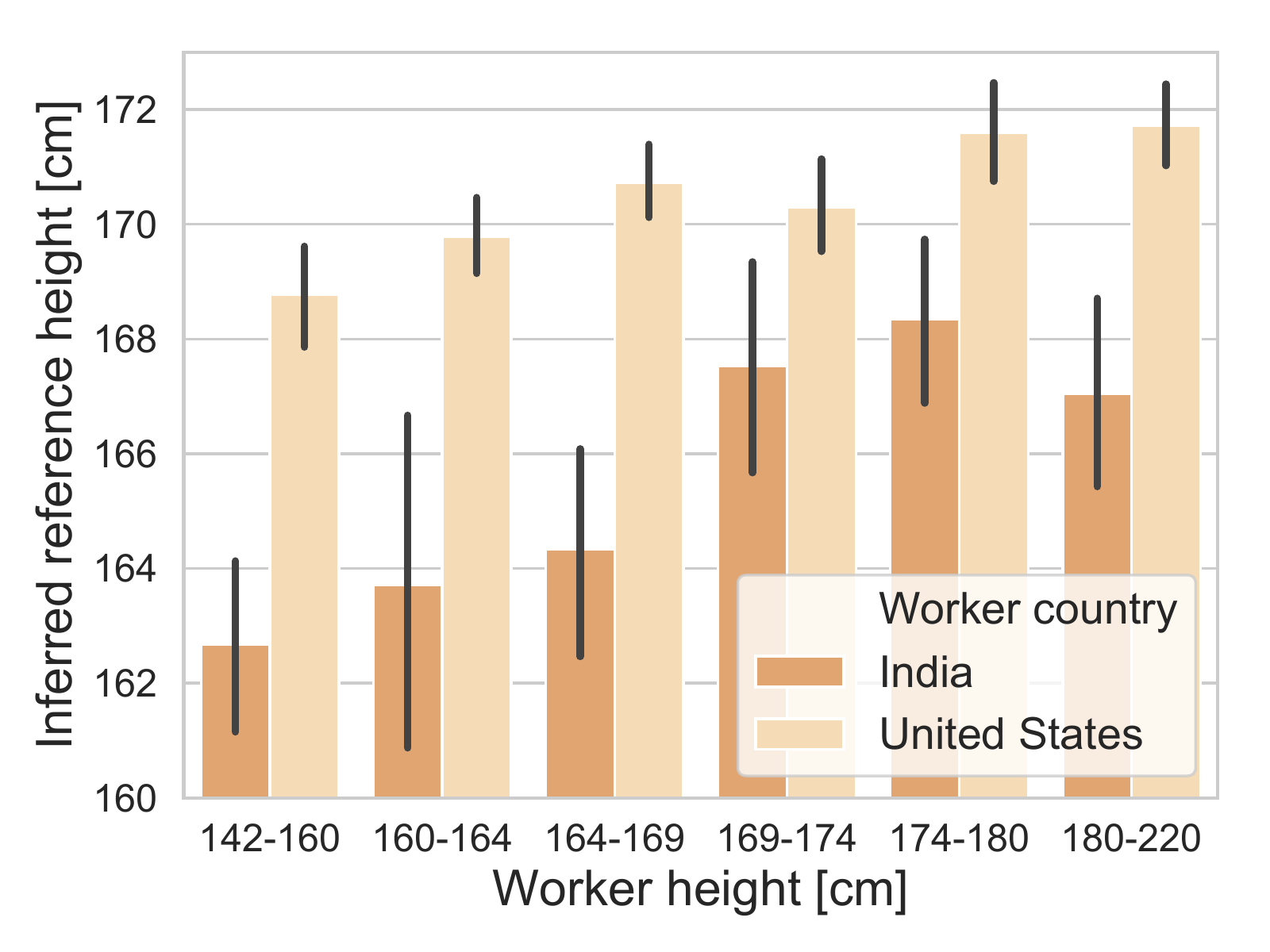}
    \end{subfigure} 
    \captionsetup{width=0.9\hsize} \caption{Dependence of weight/height reference value on worker's weight/height and country.}
    \label{fig:ref values vs country}
\end{figure}

Next, we look at differences in reference values depending on where the crowd workers are from.
(Since about 90\% of all our workers are from India or the U.S., we focus on these two countries.)
As indicated by \Figref{fig:ref values vs country}, the reference values for weight and especially height of workers from India are considerably smaller than those of workers from the U.S.
This fact may be explained by the large difference in height between residents of both countries: the average height of both men and women from India is about 10~cm smaller compared to the U.S.%
\footnote{\url{https://en.wikipedia.org/wiki/List_of_average_human_height_worldwide}} 
Thus, \Figref{fig:ref values vs country} shows how both the worker's own characteristics and the standards of their society impact the reference values. 

\begin{figure}[t]
    \begin{subfigure}{.49\hsize}
        \centering
        \includegraphics[width=\hsize]{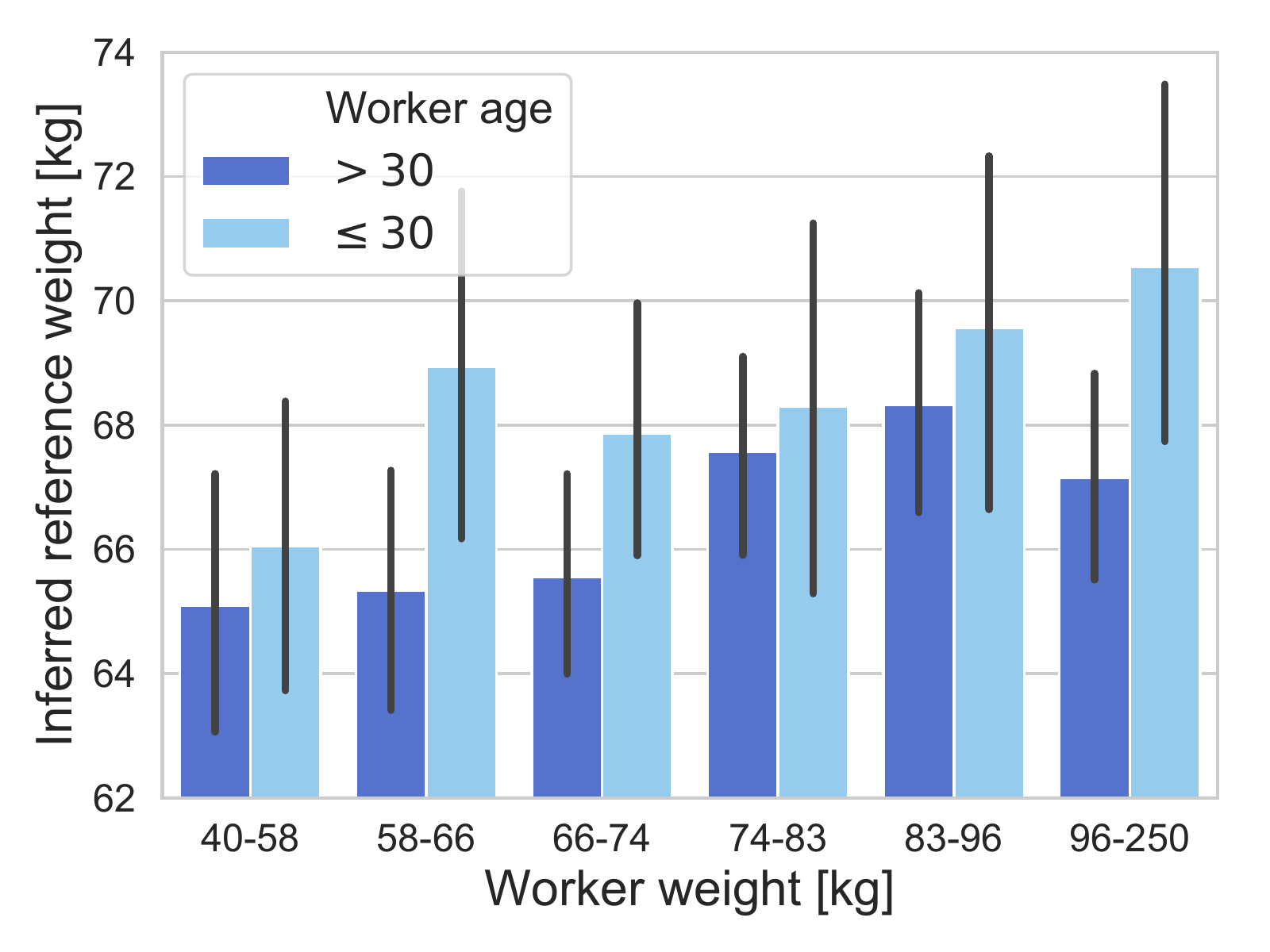}
    \end{subfigure}
    \hfill
    \begin{subfigure}{.49\hsize}
        \centering
        \includegraphics[width=\hsize]{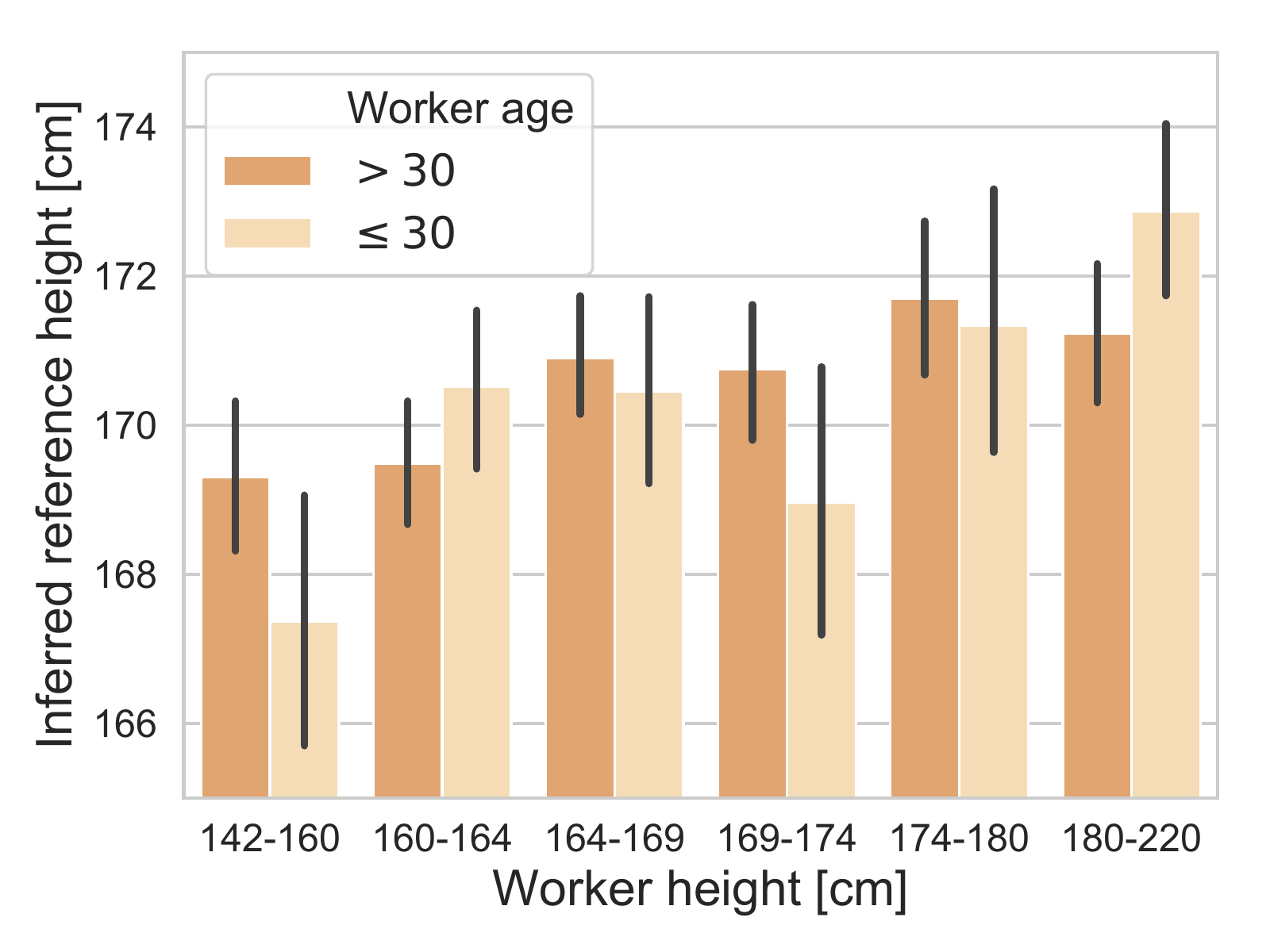}
    \end{subfigure}
    \captionsetup{width=0.9\hsize} \caption{Dependence of reference values on measurements and age for workers from the U.S.}
    \label{fig:ref values vs age}
\end{figure}

The reference values vary not only with geography, but also with the worker's age. \Figref{fig:ref values vs age} demonstrates this dependence for workers from the U.S. While the height values vary rather randomly among both groups, there is a clear trend for the weight values: younger workers have higher reference values in all weight groups. 
A possible explanation for this trend could be the increasing obesity rates in the U.S.\ \cite{mitchell2011,hansen2014generational}, 
due to which young people build their reference values while growing up in a society where obesity becomes more common, thus shifting their perception of normal weight towards higher values.
The lack of such a trend for height, rather than weight, may be explained by the fact that, across time, height distributions are much more stable than weight distributions \cite{ncd2016,burke2007social,hammond2014model}.

\begin{figure}[t]
\centering  
    \includegraphics[width=0.5\hsize]{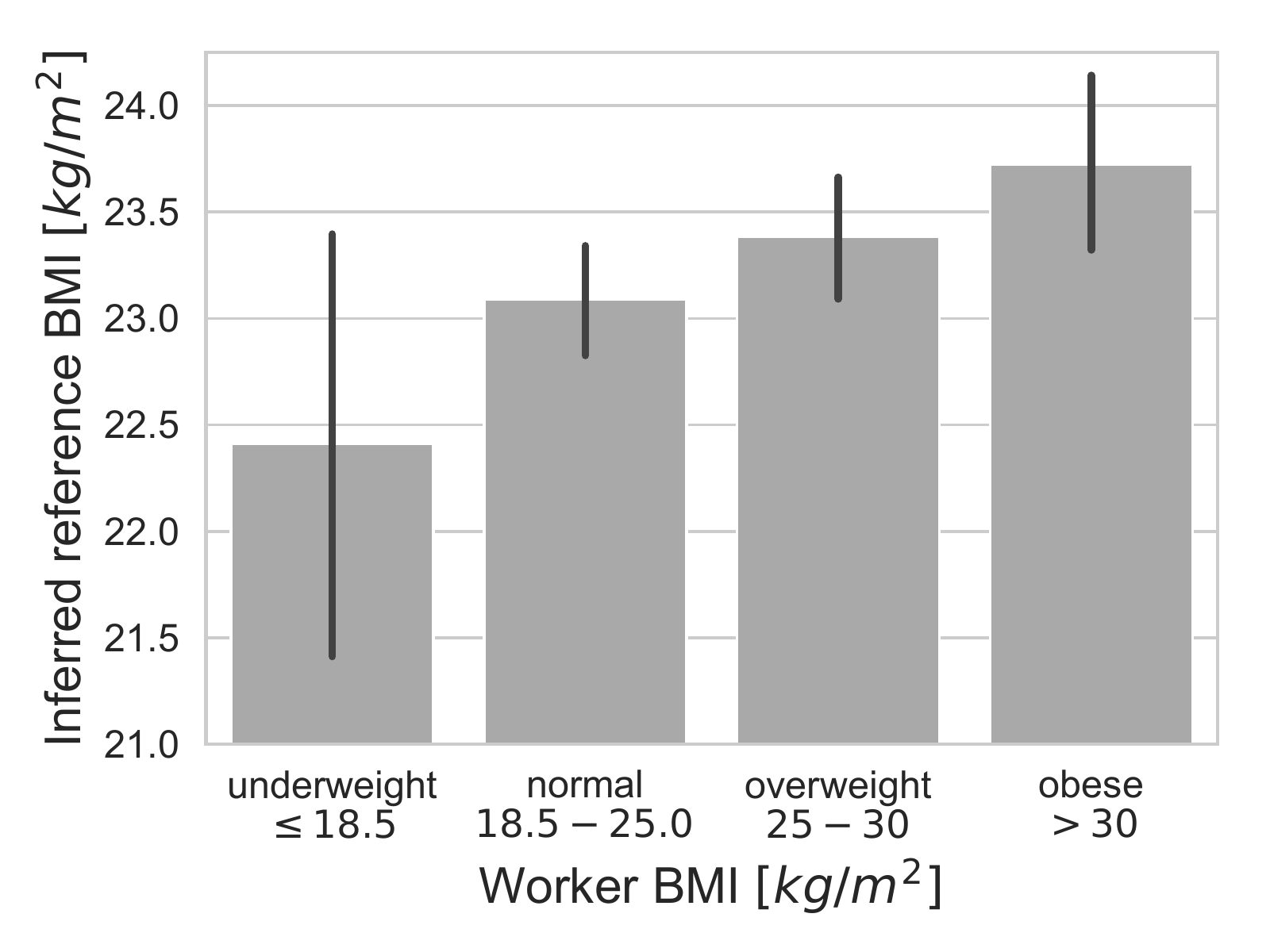}
\captionsetup{width=0.9\hsize} \caption{
Dependence of BMI reference value on worker's BMI.
}
\label{fig:ref bmi}
\end{figure}

Finally, as the dependence of the reference values on workers' own measurements holds for both weight and height, it is also reflected in the BMI: \Figref{fig:ref bmi} shows that the reference BMI increases together with the worker's own BMI. 

\begin{figure}[t]
    \begin{subfigure}[b]{.49\hsize}
        \centering
        \includegraphics[width=\hsize]{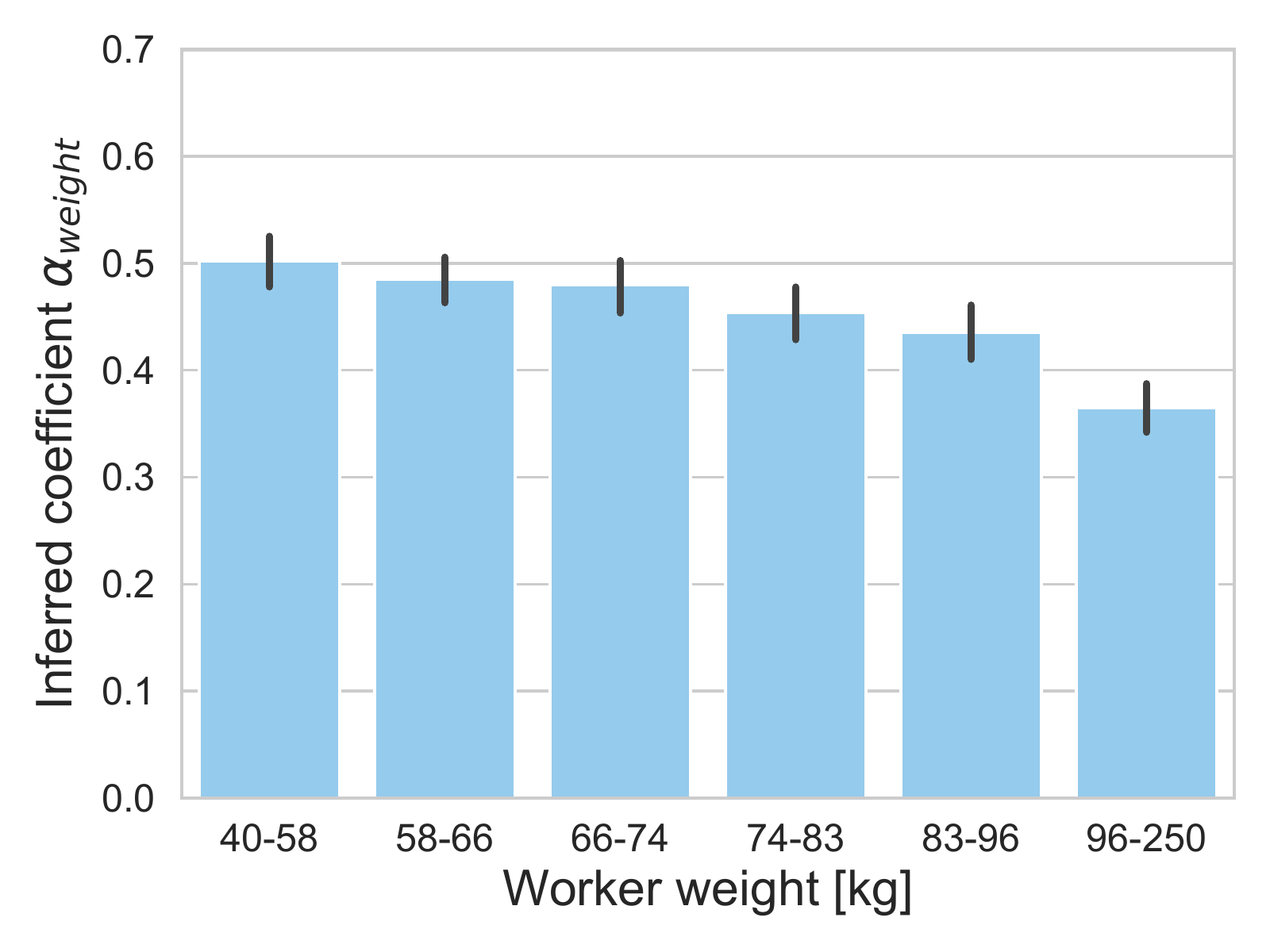}
        \caption{} \label{fig:ref alphaw}
    \end{subfigure}
    \begin{subfigure}[b]{.49\hsize}
        \centering
        \includegraphics[width=\hsize]{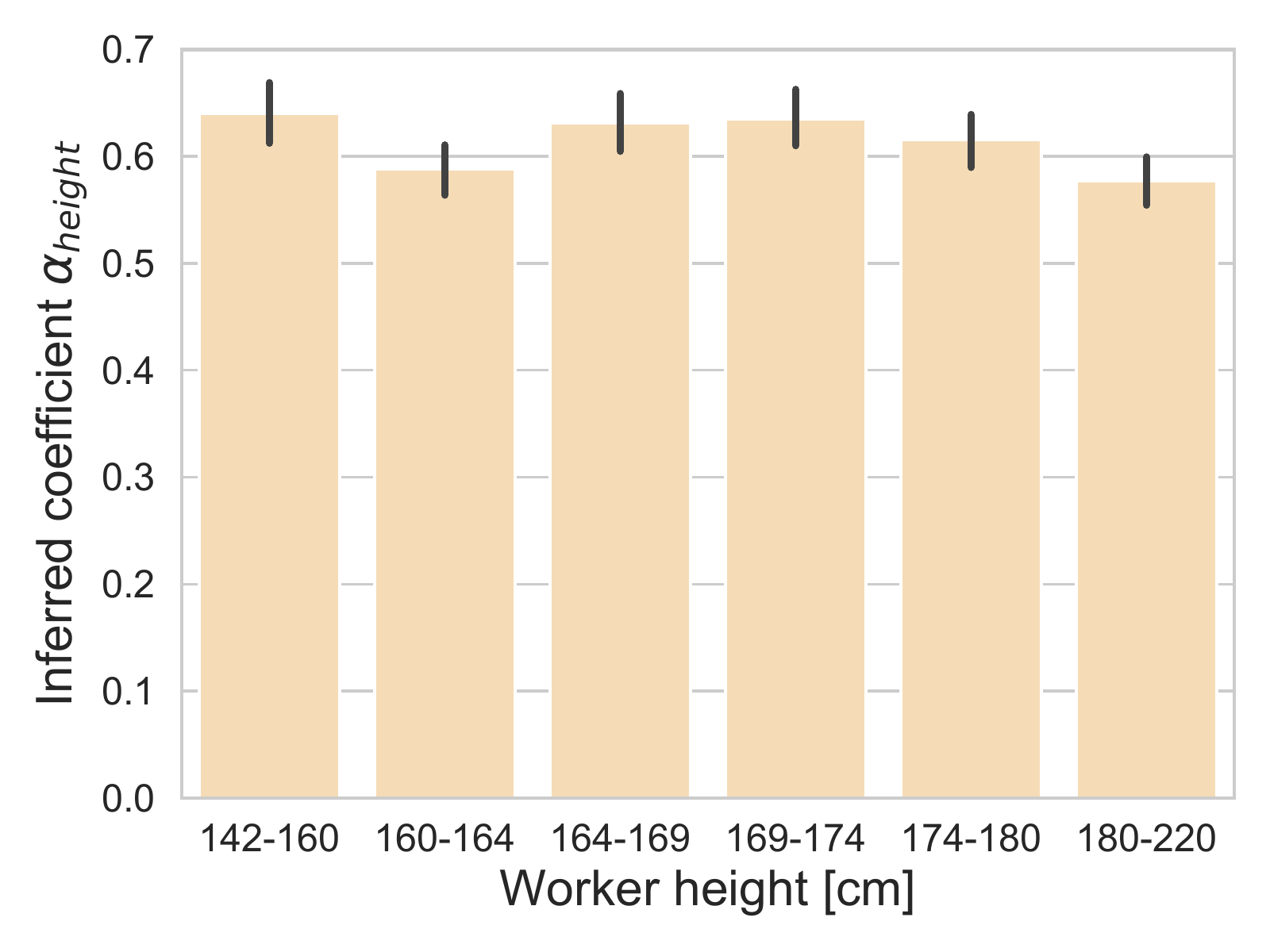}
        \caption{} \label{fig:ref alphah}
    \end{subfigure}
\captionsetup{width=0.9\hsize} \caption{Dependence of weight/height contraction coefficients on worker's weight/height.}
    \label{fig:ref alpha}
\end{figure}

Apart from the reference values, the model from \Secref{sec:ref values derivation} also contains the contraction coefficient $\alpha$. \Figref{fig:ref alpha} shows the coefficients for various worker weight and height groups. The contraction coefficient relates to the quality of the estimation: an $\alpha$ close to $1$ mean that the worker tends to guess their reference weight\slash height most of the times, whereas a small $\alpha$ indicates that the guesses closely follow the true labels. From \Figref{fig:ref alphah} we can observe that $\alpha_\mathrm{height}$ stays roughly constant across groups, whereas \Figref{fig:ref alphaw} shows that $\alpha_\mathrm{weight}$ decreases with increasing worker weight (and therefore with increasing reference weight). As previously discussed (\Secref{sec:data main}, \Figref{fig:men worker char} and \ref{fig:women worker char}), this is mainly caused by the specific dataset, which contains a lot of images of obese people and thus favors high reference values for weight. When the data is downsampled to represent the distribution of weight in the general population, the differences between groups become negligible. Another observation is that $\alpha_\mathrm{height}$ is in general larger than $\alpha_\mathrm{weight}$, which means that the averaging effect is stronger for height estimation. 

\section{Towards more accurate crowdsourcing}
\label{sec:Towards more accurate crowdsourcing}

The main contribution of this paper is to study human biases in height and weight estimation using crowdsourced data.
Moving beyond this scientific goal, if crowdsourced height and weight labels were accurate, they could potentially also be harnessed for engineering practically useful applications.
For instance, for monitoring obesity levels across time and space, it would be tremendously useful to have access to low-cost, low-latency estimates of population height and weight.
Such ``sensors'' could be built by feeding representative images sampled from a population (\eg, via social media) to a crowdsourcing system.
The data thus collected could further be used to train fully automated height and weight models based on machine learning~\cite{weber2016crowdsourcing}.

As we saw in the previous sections, the labels collected using our simple crowdsourcing method are rather inaccurate.
The present section explores simple ways of improving the accuracy, with the goal of establishing whether crowdsourcing is a promising candidate not only for measuring human biases (the main contribution of this paper), but also for collecting high\hyp quality labels for downstream tasks.

We proceed in two directions:
first, via statistical corrections to remove human biases (\Secref{sec:correction}), and
second, by giving crowd workers access to more information when guessing (\Secref{sec:known height} and \ref{sec:feedback}).
\trackedChange{
Neither of these two directions was found to increase crowd workers' accuracy sufficiently in order to directly enable a crowd\hyp powered obesity monitoring system as sketched above.
We shall reflect further on this negative result in \Secref{sec:Implications for crowdsourcing and machine learning}, where we discuss lessons learned from the present research for potential future crowdsourcing and machine learning applications.
}

\subsection{Statistical correction models}
\label{sec:correction}


We are interested in the performance of crowdsourcing and correction models on images representing the general population, whereas the dataset contains a disproportional number of images of obese people.
Hence, we downsample the data to 500 images in such a way that the distribution of weight among workers and samples becomes similar. 
Furthermore, for performance evaluation, the set is split into 400 training and 100 validation images.

\xhdr{Correction via reference values}
First, we try to use the previously derived reference values for correction, by rewriting \Eqnref{eqn:estimate generation} as
\begin{equation}
    w^i_\true = \frac{w^i_j - \alpha_j w^{\reference}_j - \epsilon}{1 - \alpha_j}. \label{eqn:estimate correction}
\end{equation}
Under the assumption that the noise $\epsilon$ follows a zero-centered normal distribution, we obtain a maximum \textit{a-posteriori} (MAP) estimate of the ground-truth label $w^i_\true$ by evaluating \Eqnref{eqn:estimate correction} with $\epsilon=0$. This MAP estimate represents a corrected weight label. 
Tables~\ref{tab:correction models} and~\ref{tab:correction models height} show that this approach leads to poor results.
The main problem is the poor performance of several workers whose $\alpha_j \approx 1$ 
makes the denominator of \Eqnref{eqn:estimate correction} very small, which in turn amplifies the noise in the guess $w_j^i$.

\xhdr{Linear regression}
Next, we introduce several linear regression models that differ with respect to the features they use.
The output of each model is a corrected weight or height estimate.
The following models are considered:   
\begin{squishlist}
    \item Global: a single model fitted after aggregating all individual guesses for the same image via the mean. The features we consider include the mean of all collected weight and height guesses (MW and MH, respectively) and the gender of the person in the image (GN).
    \item Per-worker: a separate model for each worker, taking as inputs individual guesses (i.e., each model learns different worker-specific parameters). The features we consider include the estimated weight and height (EW and EH, respectively) and the gender of the person in the image (GN). The outputs of all worker-specific models are averaged to get a single corrected weight label for each image.
    \item Mixed: a global correction model for individual guesses that also takes into account worker parameters. The inputs are individual weight and height estimates, the gender of the person in the image, as well as the worker's weight, height, and gender. Again, the average of all collected guesses is used as a final corrected estimate for each image.
\end{squishlist}

\begin{table}
    \caption{Correction models for weight estimation.}
    \begin{tabular}{ l | c | c}
        \hline
  		Model type & \makecell{MAE per image \\ train\slash{}test [kg]} & \makecell{MAE per guess \\ train\slash{}test [kg]} \\
		\hline 
		raw guesses & 8.83 \slash{} 8.65 & 11.39 \slash{} 11.38 \\
		ref. val. correction & 10.04 \slash{} 10.21 & 14.14 \slash{} 15.40 \\
		global: MW & 8.44 \slash{} 7.43 & --- \\
		global: MW, MH, GN & 7.98 \slash{} 7.02 & --- \\ 
		worker: EW & 7.46 \slash{} 7.41 & 9.15 \slash{} 10.07 \\
		worker: EW, EH, GN & 6.38 \slash{} 7.58 & 8.19 \slash{} 10.94 \\
		mixed & 8.81 \slash{} 7.99 & 10.61 \slash{} 10.16 \\
	    \hline
	\end{tabular}
    \label{tab:correction models}
\end{table}

\begin{table}
    \caption{Correction models for height estimation.}
    \begin{tabular}{ l | c | c}
        \hline
  		Model type & \makecell{MAE per image \\ train\slash{}test [cm]} & \makecell{MAE per guess \\ train\slash{}test [cm]} \\
		\hline 
		raw guesses & 5.28 \slash{} 4.91 & 7.01 \slash{} 6.82 \\
		ref. val. correction & 6.99 \slash{} 6.42 & 12.05 \slash{} 12.21 \\
		global: MH & 5.24 \slash{} 4.70 & --- \\
		global: MH, MW, GN & 4.91 \slash{} 4.64 & --- \\ 
		worker: EH & 4.93 \slash{} 5.14 & 5.50 \slash{} 5.93 \\
		worker: EH, EW, GN & 4.05 \slash{} 4.93 & 4.40 \slash{} 5.81 \\
		mixed & 5.21 \slash{} 5.09 & 5.24 \slash{} 5.13 \\
	    \hline
	\end{tabular}
    \label{tab:correction models height}
\end{table}	

The performance of raw crowdsourcing and different correction models is summarized in Tables \ref{tab:correction models} and \ref{tab:correction models height}.%
\footnote{Errors are lower than in \Tabref{tab:mean vs median} because the dataset used here was downsampled to match the worker population and thus contains fewer overweight images.}
For space reasons, we focus on weight (\Tabref{tab:correction models}) in our discussion.
The best performance on the test set is obtained by the global correction models. We believe that these models can better leverage the wisdom of crowds and are thus more suited for crowdsourcing tasks, because the global models work on the aggregated guesses of multiple workers, whereas the worker\hyp specific models try to correct single estimates that are more noisy by their nature.
In particular, the worker-specific models fit both the relevant signal and the noise of separate guesses, and do not generalize well (as can be seen when comparing training and testing errors: the difference between the two is smaller for the worker\hyp specific models).
An interesting effect is observed for the mixed models: while the error of single estimates decreases compared to the raw guesses, this improvement is lost once the corrected guesses are aggregated. For example, on the training set, the MAE of single estimates drops from 11.39 to 10.61 after correction (\Tabref{tab:correction models}), but the MAE for images remains almost the same.



\xhdr{Limits of correction}
The simple correction models discussed so far demonstrate that correction of the collected estimates poses a challenging task. In this section, we provide evidence that more sophisticated statistical models would also have limitations resulting from the collected data.

The initial dataset contains pairs of images (``before'' and ``after'') and weight labels (\cf\ \Secref{sec:data main}). Each of the two weight labels in a pair needs to be assigned to one of the two images in the pair. An obvious solution, discussed in \Secref{sec:gen accuracy}, is to aggregate the guesses (\eg, by averaging, as we did throughout the paper), and to assign the larger ground-truth label to the image with the higher aggregated value.

Though the resulting assignment is correct for almost all samples, it is wrong in some rare cases. 
\Figref{fig:sample1} contains an example of a wrong assignment, for an image pair capturing a particularly large weight transformation. It shows how after a major weight loss of almost 19~kg the person is labeled as slightly heavier than before: the mean of ``before'' guesses is 79.4~kg, whereas the mean of ``after'' guesses is 81~kg. Furthermore, we observed that the guesses collected for both images follow similar distributions, and even come from similar pools of workers (in terms of the workers' own weight and height distributions). Under such conditions, no statistical model can correct the results. The important difference here is purely visual: the images are taken from different angles and under very different circumstances (most importantly, fully dressed \vs\ bare-chested), which makes the task unequally hard across images.




\subsection{Crowdsourcing variation 1: Guessing weight for known height}
\label{sec:known height}

Above, we showed that correcting the estimates \textit{post hoc} is challenging. Next, we suggest two different setups that could simplify the task for crowd workers and lead to more accurate results.
The experiments in these modified setups are again performed on the reduced set of 500 images that were chosen to represent the general population, as discussed in \Secref{sec:correction}.

Given that height does not change significantly during adulthood, whereas weight may change widely and might therefore have to be frequently re\hyp estimated for the same person, one might provide workers with the true height labels for the images and collect only weight guesses.

For this setup, we apply an additional filter based on workers' countries of residence: as previously shown, reference values of height for people in Asian countries do not match the dataset at hand. Thus, providing the actual height of the person in the image can further confuse such workers (indeed, the quality of their guesses slightly decreases when the true height is revealed). Hence, we focus on the performance of crowd workers from Europe and the U.S. The results in \Tabref{tab:error basic vs known height} summarize the error distributions for both scenarios (\ie, with and without height given),
indicating only small improvements of the mean error (ME) and mean absolute error (MAE) for the new setup. Moreover, the improvements are not significant: both a $t$-test for the mean and a Bartlett test for the variance show no significant difference between the distributions of errors for two scenarios ($p=0.66$ and $0.72$ respectively). 

\begin{table}
    \caption{Comparison of weight-guessing results when showing \vs\ not showing true height labels to workers, with bootstrapped 95\% confidence intervals (only workers from Europe and U.S.).}
    \begin{tabular}{ l | c | c}
        \hline
  		 & Height not given & Height given \\
		\hline 
		ME [kg] & 4.54 (3.54, 5.57)  & 4.22 (3.24, 5.23) \\
		MAE [kg] & 8.70 (7.96, 9.46) & 8.47 (7.73, 9.21)\\
		\hline
	\end{tabular}
    \label{tab:error basic vs known height}
\end{table}	

\begin{figure}[h]
\centering  
     \includegraphics[width=0.5\hsize]{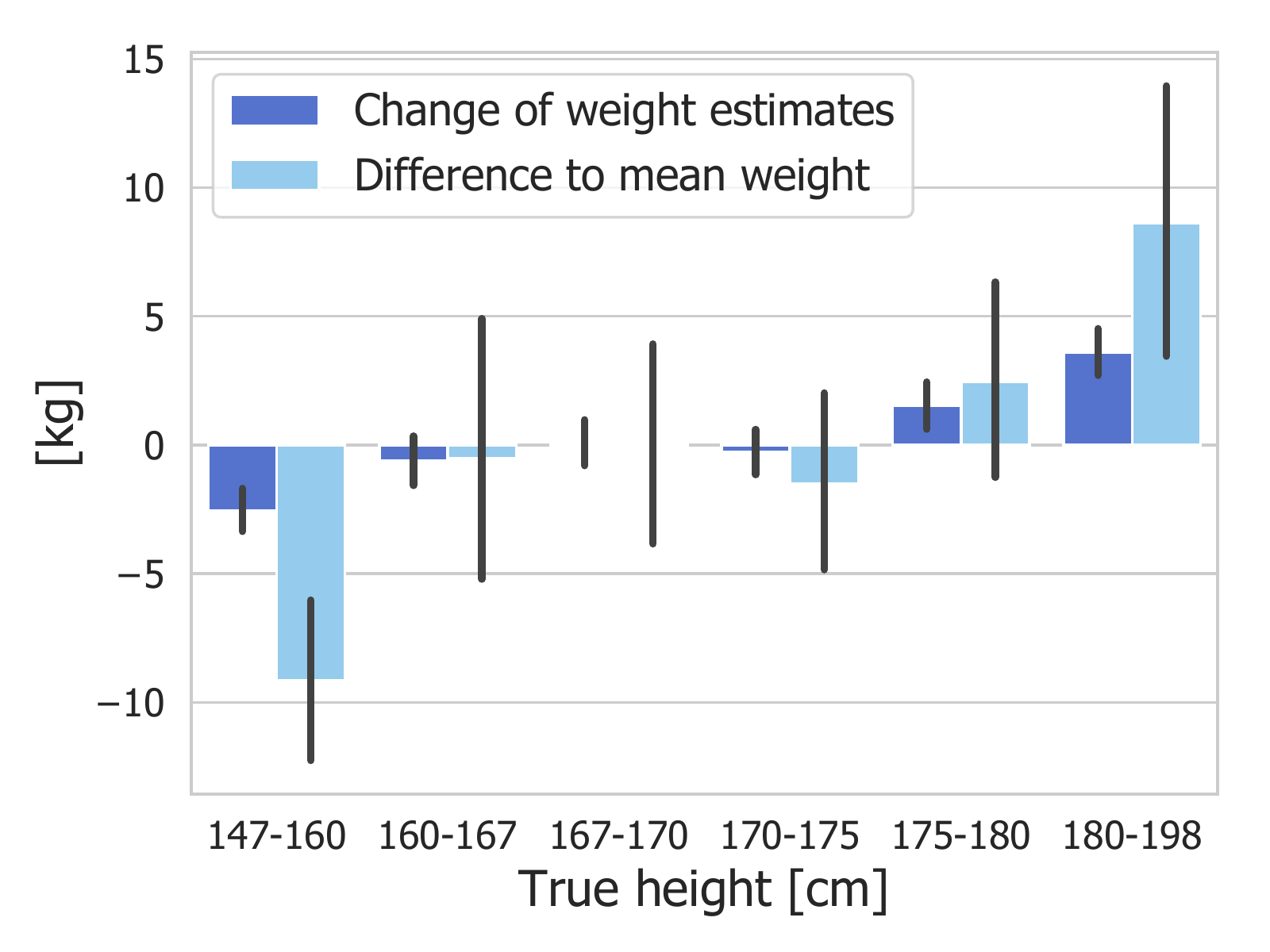}
     \captionsetup{width=0.9\hsize} \caption{
     Change of weight guesses for various height groups, when shown true height of person in image.
     }
     \label{fig:weight guesses vs known height}
\end{figure}

The comparison in \Figref{fig:weight guesses vs known height} shows how the weight estimates are affected by the known true height (dark bars) and how the true weight differs from the overall average of the true weight (across all images in the dataset; light bars) as a function of the true height. The figure provides a possible explanation for the small performance improvement. First, while workers generally shift their estimates into the right direction (add or subtract a few kilograms based on the actual height), the magnitude of this shift seems to be too small, in particular for the shortest and tallest images. Furthermore, the bars that show difference to the mean weight have wide confidence intervals and, thus indicate large variations of weight within each group. Indeed, the growth of weight as a function of height, as depicted in \Figref{fig:weight guesses vs known height}, appears only after aggregating a large number of images. The real relationship between weight and height is complicated and nonlinear~\cite{bmi2005}. This is further supported by statistics of the dataset at hand: the correlation between weight and height is weak according to both Pearson's (0.27) and Spearman's (0.28) correlation coefficient.



\subsection{Crowdsourcing variation 2: Guessing weight and height with feedback}
\label{sec:feedback}

Next, we consider a second variation: instead of making the task itself simpler, we suggest a design that gives workers an opportunity to learn, by providing immediate feedback to the workers after each guess. The feedback contains the true weight and height of the person in the image, as well as the errors of their guess.
Of course, if we already knew the ground truth for every image, this would defeat the purpose of collecting it via crowdsourcing.
We might, however, have a few labeled images that could be used in this setup for teaching workers early on.

\begin{figure}[t]
    \begin{subfigure}{.49\hsize}
        \centering
        \includegraphics[width=\hsize]{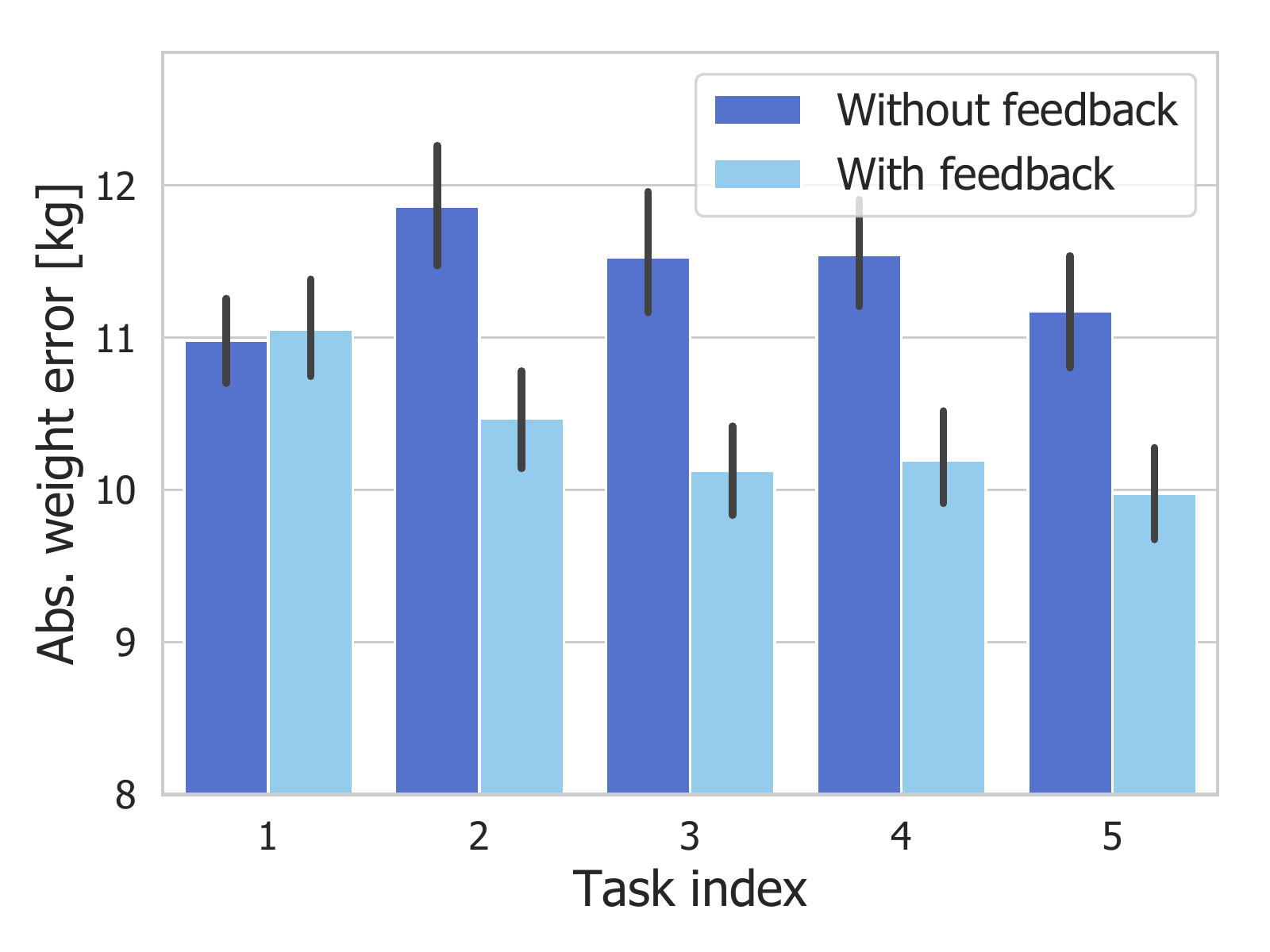}
    \end{subfigure}
    \begin{subfigure}{.49\hsize}
        \centering
        \includegraphics[width=\hsize]{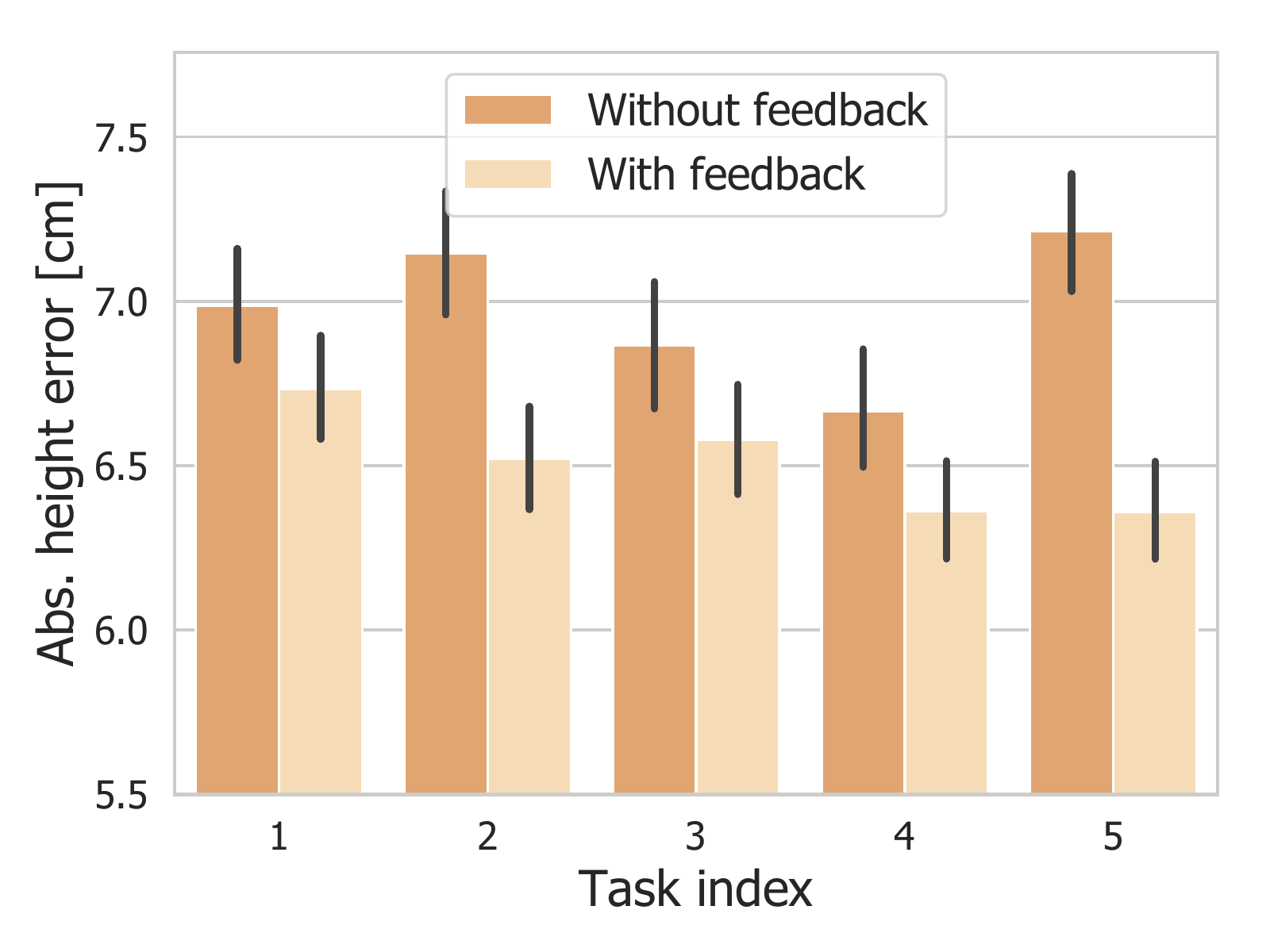}
    \end{subfigure}
    \captionsetup{width=0.9\hsize} \caption{Dependence of the mean absolute error of estimates on the number of previously seen images.}
    \label{fig:learning effect}
\end{figure}

The results in \Tabref{tab:error basic vs feedback} show that the accuracy slightly improves in the setup where workers receive feedback. The values in \Tabref{tab:error basic vs feedback} were computed after aggregating all guesses per image, and the improvements appear to be minor. However, the quality of individual guesses (before aggregating per image) improved significantly with feedback: the mean absolute error for weight decreased from 11.4~kg in the basic setup to 10.4~kg in the feedback setup, and similarly for height, where it decreased from 7.0~cm to 6.5~cm ($p < 10^{-17}$ according to Student's $t$-tests, for both errors and absolute errors).
Further evidence of the increase in accuracy due to training effects can be seen in \Figref{fig:learning effect}. Here, single estimates are split into bins according to the number of preceding guesses (and, thus, instances of feedback received). In this way, the estimates in later bins are produced  by more ``experienced'' workers who have previously received more feedback. 
The plots in \Figref{fig:learning effect} show how the MAE of the weight and height guesses decreases with progressing bin indices in the experiment with feedback, while no improvement is observed for the initial setup. 

\begin{table}
    \caption{Comparison of weight- and height\hyp guessing results with \vs\ without feedback on the previous guess, with bootstrapped 95\% confidence intervals (only workers from Europe and U.S.).}
    \begin{tabular}{ l | c | c}
        \hline
  		 & Without feedback & With feedback \\
		\hline 
		Weight ME [kg] & 4.90 (3.90, 5.95) & 3.25 (2.33, 4.24) \\
		Weight MAE [kg] & 8.80 (8.05, 9.57) & 8.11 (7.48, 8.82)\\ 
		\hline
        Height ME [cm] & 0.47 ($-0.12$, 1.07) & $-0.26$ ($-0.8$, 0.34) \\
    	Height MAE [cm] & 5.20 (4.86, 5.55) & 5.10 (4.76, 5.45)\\
    	\hline
	\end{tabular}
    \label{tab:error basic vs feedback}
\end{table}

The benefits of training can also be observed when we look at the change of the reference values. For this purpose, it is interesting to consider crowd workers from Asian countries (mostly India). As previously discussed, their reference values tend to deviate from the reference values of Europeans and Americans, and thus hinder their performance on the given dataset. \Figref{fig:ref values basic vs feedback, Asia} shows how the reference values for both weight and height are changing in the experiment with feedback. Since the mean weight and height of the images from the dataset (shown with a dashed line) exceeds all initial reference values, crowd workers shift their references values towards higher weights and heights as they learn these new standards and adapt to the dataset. On the other hand, we did not observe any considerable change of the reference values for European or American workers. 

\begin{figure}[t]
    \begin{subfigure}{.49\hsize}
        \centering
        \includegraphics[width=\hsize]{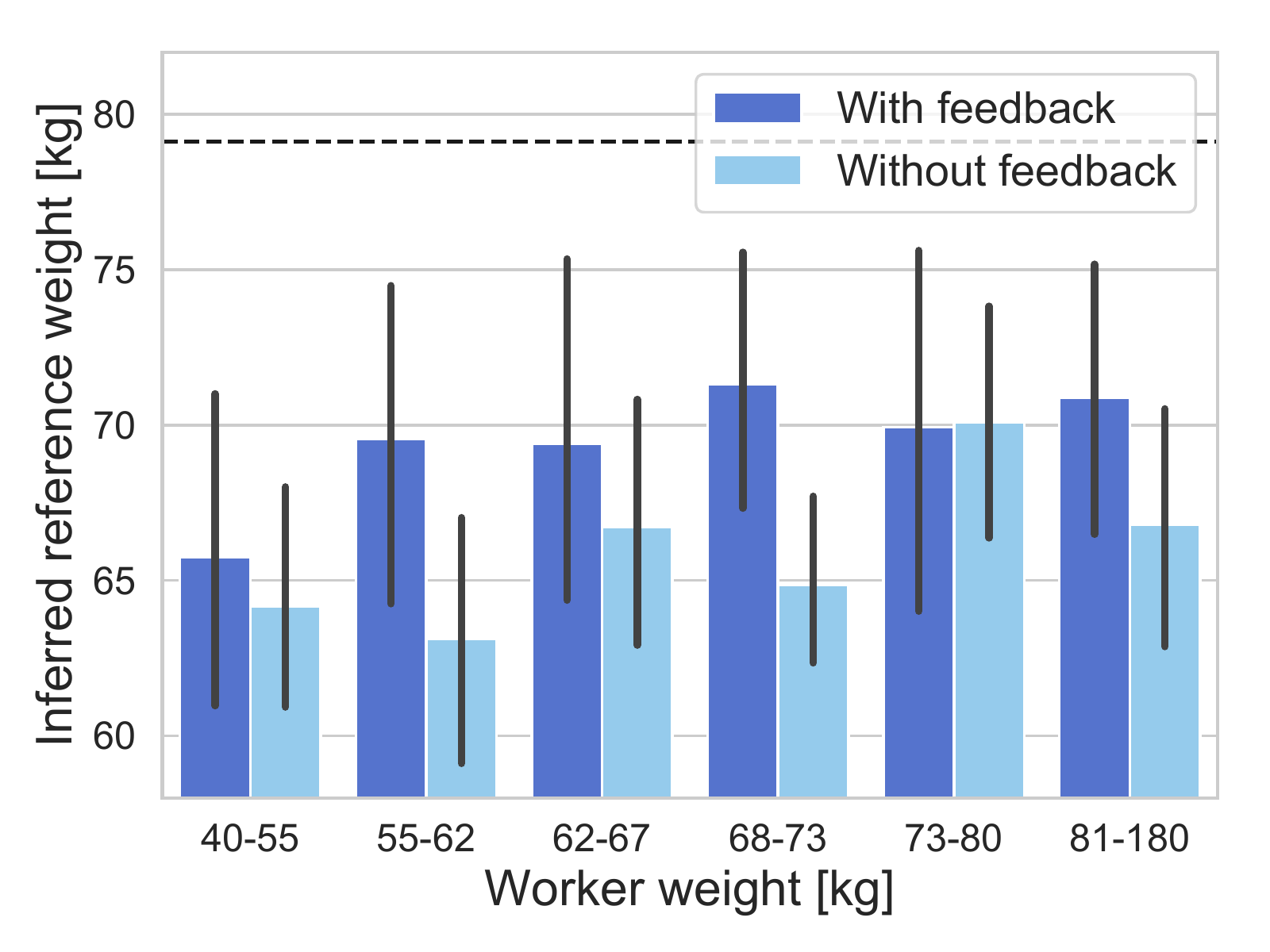}
    \end{subfigure}
    \begin{subfigure}{.49\hsize}
        \centering
        \includegraphics[width=\hsize]{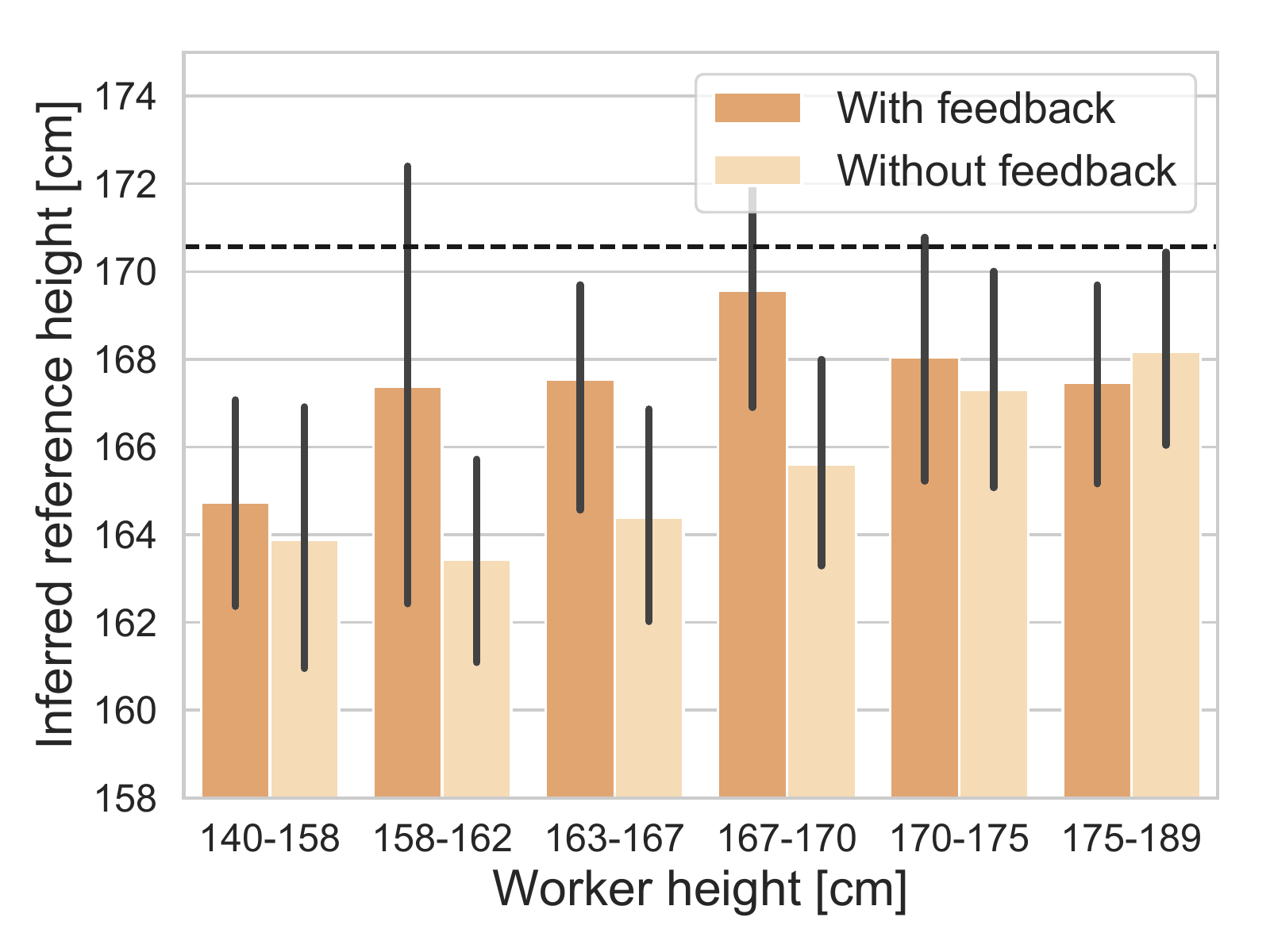}
    \end{subfigure}
    \captionsetup{width=0.9\hsize} \caption{Reference values inferred for the setups with and without feedback, for workers from Asia only. Dashed lines denote the true mean weight\slash height over all images.}
    \label{fig:ref values basic vs feedback, Asia}
\end{figure}

\begin{figure}[t]
    \begin{subfigure}{.49\hsize}
        \centering
        \includegraphics[width=\hsize]{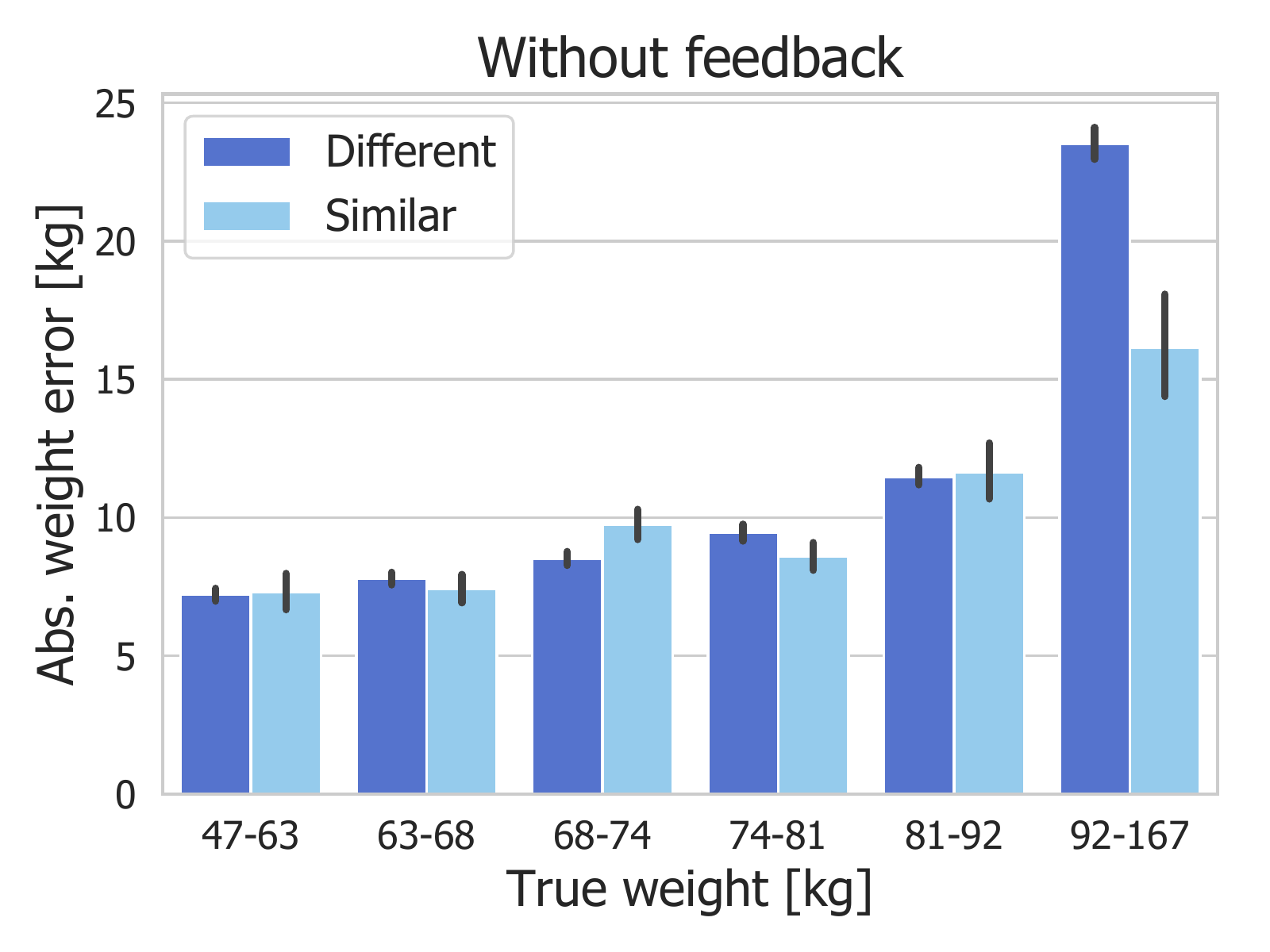}
    \end{subfigure}
    \begin{subfigure}{.49\hsize}
        \centering
        \includegraphics[width=\hsize]{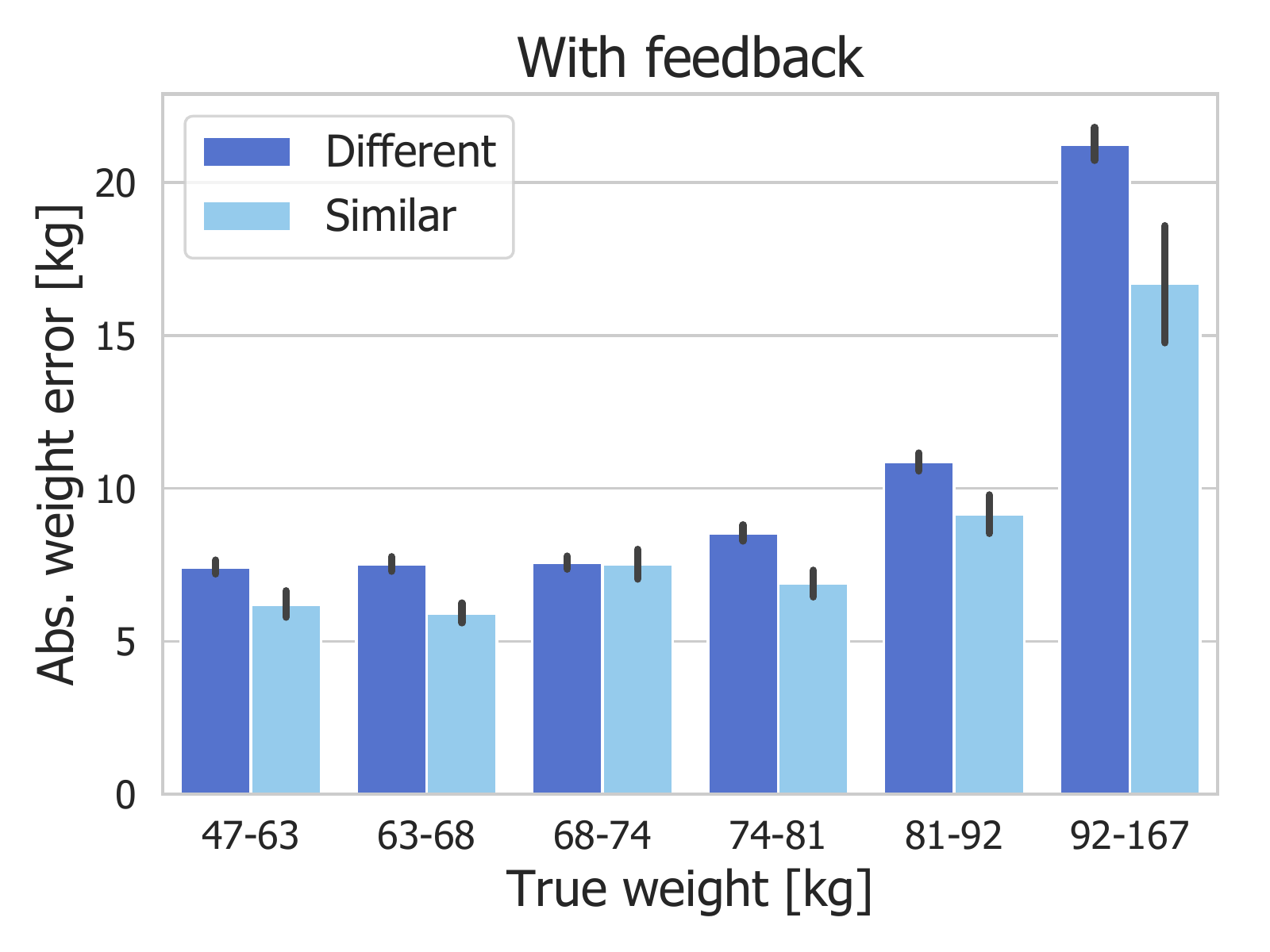}
    \end{subfigure}
    \begin{subfigure}{.49\hsize}
        \centering
        \includegraphics[width=\hsize]{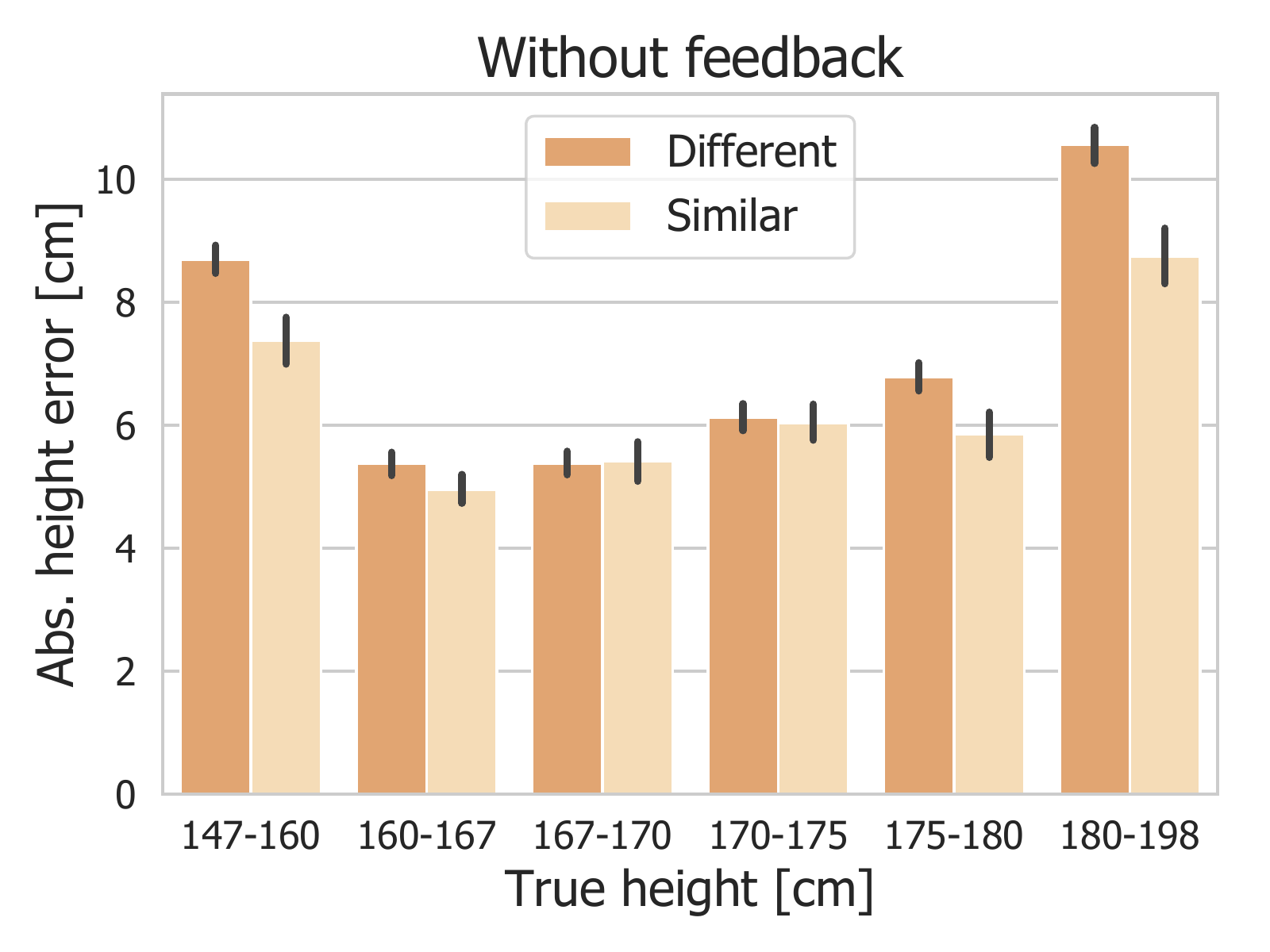}
    \end{subfigure}
    \begin{subfigure}{.49\hsize}
        \centering
        \includegraphics[width=\hsize]{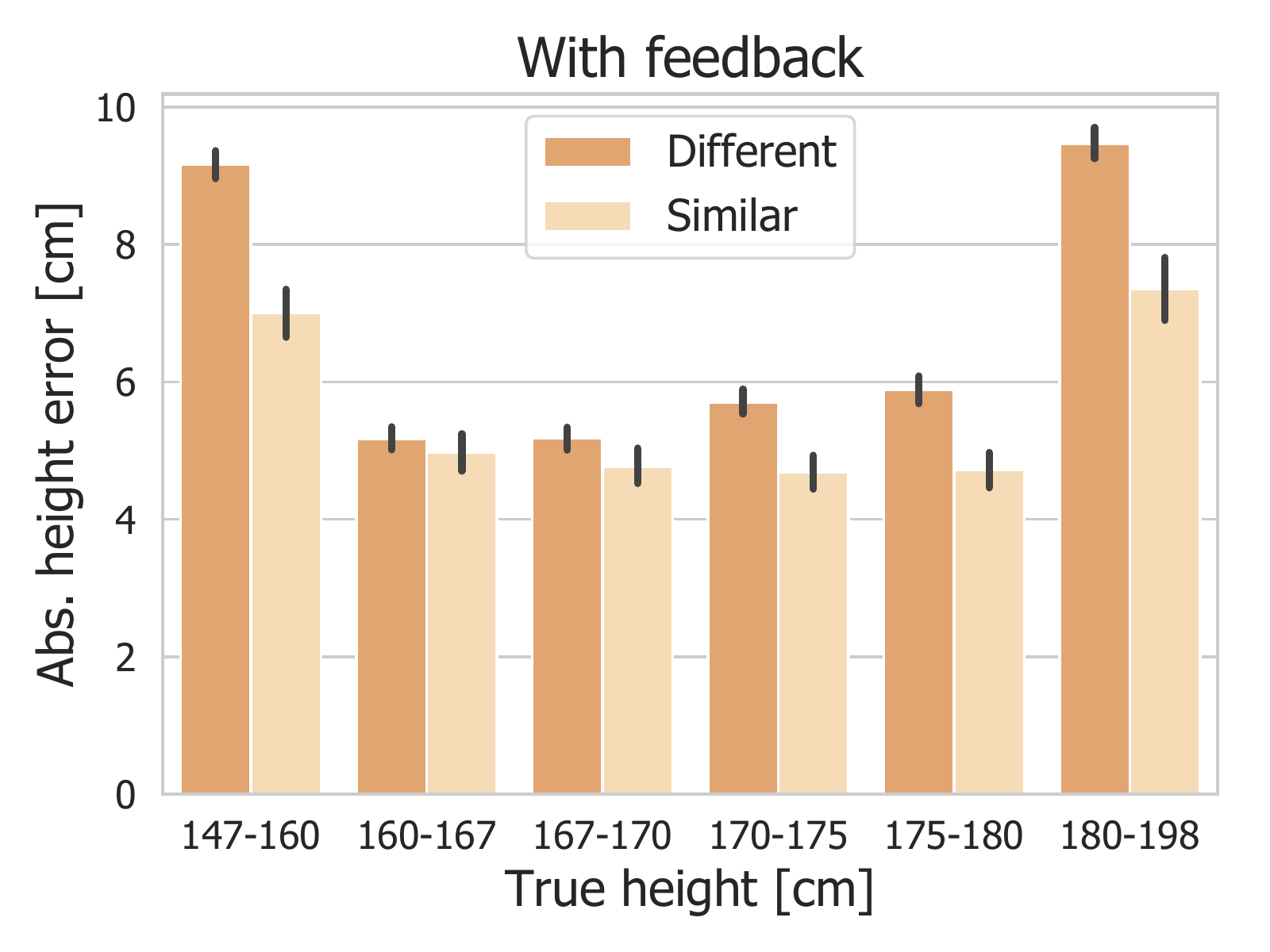}
    \end{subfigure}
    \captionsetup{width=0.9\hsize} \caption{Mean absolute error of estimates in dependence on the previous image for the setups without feedback (left) and with feedback (right).}
    \label{fig:dependence on prev}
\end{figure}

The previous discussion was centered around long-term training effects, i.e., performance improvement as a result of feedback on multiple previous guesses. Now, we focus on the short-term effects of getting feedback. \Figref{fig:dependence on prev} shows how the MAE depends on the previous image seen during a task (remember that each task consisted of 10 images shown one after the other). In particular, we distinguish between two cases: 
\begin{squishlist}
    \item Similar: the previous image shows a person of the same gender, whose weight (height) differs from the weight of the person in the current image by at most 7.5~kg (7.5~cm).
    \item Different: the previous image does not satisfy the above condition.
\end{squishlist}

\Figref{fig:dependence on prev} shows how the MAE depends on the previous image across weight and height groups, and compares the two versions of the experiment (with \vs\ without feedback). 
Clearly, the impact of the previous image is marginal when no feedback is shown: for most groups, the MAE remains roughly the same for similar and different images. 
The previous image, however, appears to be important when workers start to receive feedback: in almost all cases, the average accuracy of the guesses improved when the new image was similar to the previous one.



\section{Discussion}
\label{sec:discussion}

We conclude the paper by discussing relations to prior work (\Secref{sec:Relation to prior work};
also cf. \Secref{sec:relwork}), limitations of our methodology
(\Secref{sec:Limitations}),
and implications and best practices for crowdsourcing and machine learning
tasks involving body measurements (\Secref{sec:Implications for crowdsourcing and machine learning}).

\subsection{Relation to prior work}
\label{sec:Relation to prior work}

We observe that crowd estimates are skewed toward an intermediate value close to the population\hyp wide average, a phenomenon known as \textit{contraction bias}~\cite{poulton1989bias}.
This bias has been shown to in particular lead to the underestimation of overweight and obese bodies~\cite{cornelissen2016visual}, which may in turn compromise people's ability to recognize weight gain and undertake compensatory weight\hyp control behaviors.

Taking a closer look, we observe that the contraction does not, however, skew all estimates to a global constant, but rather to the typical height and weight of the environment in which the respective crowd worker lives (\Figref{fig:ref values vs country}), as well as to their own height and weight (\Figref{fig:ref values}).
We confirm this effect via a simple model, which elicits individual workers' reference values (\Secref{sec:ref values derivation}) from their estimates in combination with the ground-truth height and weight of the images they judge.
The fitted model parameters indicate, \eg, that workers from India have lower reference heights and weights than workers from the U.S., reflecting lower average measurements in India.
Similarly, older U.S.\ workers have lower reference weights than younger U.S.\ workers, potentially caused by an increasing American average weight.
This echoes the claims of \textit{visual normalization theory} \cite{robinson2015visual,robinson2017overweight,robinson2014he}, which states that weight status is judged relative to body\hyp size norms prevalent in society, leading to the systematic misestimation of certain body shapes.

We also provide evidence that users anchor their weight estimates in height estimates, but less so in the reverse direction (\Figref{fig:weight height dependence}).
Regarding height, prior research has shown that estimates are also heavily anchored in facial features \cite{burton2013judgments,coetzee2009facial} and head\hyp to\hyp shoulder ratio~\cite{mather2010head}.


In his seminal 1907 paper about human performance at guessing the weight of an ox \cite{galton1907vox}, Francis Galton wrote:
``I have not sufficient knowledge of the mental methods followed by those who judge weights [\dots].''
Here Galton raises the question of mental models.
Although our work is not primarily concerned with eliciting mental models, our findings can lead the way to a deeper scrutiny of this aspect.
For instance, in \Secref{sec:Dependencies between height and weight estimation} we showed that height estimates are independent of true weight for a fixed true height, whereas weight estimates still depend on true height even for a fixed true weight (\cf\ \Figref{fig:causal_diagrams}).
This asymmetry might hint at a mental process where weight and height estimation are performed sequentially, rather than simultaneously.
Future work should investigate the question of mental models further, \eg, with experiments for discerning whether height and weight estimation indeed happen sequentially.


\subsection{Limitations}
\label{sec:Limitations}

Our study is limited in several ways.
First, as our dataset of weight- and height\hyp labeled images was collected from an online forum that is primarily concerned with weight loss, the data is biased towards heavier people.
Additionally, the crowd workers who provided estimates came primarily from the United States and India, leading to a biased sample of the global population.
Future work should therefore validate our findings on further datasets and with further worker demographics.

The above two sources of bias, stemming from images and workers, respectively, are distinct from one another.
Where required, we circumvented this issue by subsampling images to reflect the demographics of the worker population.

A final limitation stems from the fact that users posting their images and measurements on Reddit, as well as crowd workers, may in principle have indicated their own weight and height incorrectly, \eg, due to social desirability, lacking information on true measurements, or sheer malice.
While we cannot rule out such behavior, neither users labeled in images nor crowd workers had an obvious incentive to wilfully act this way.
Nevertheless, our findings could be strengthened by follow-up studies with images and estimates collected in a more controlled environment, \eg, where researchers themselves measure the height and weight of both estimated and estimating participants.

\subsection{\trackedChange{Implications for crowdsourcing and machine learning}}
\label{sec:Implications for crowdsourcing and machine learning}


Based on the studies of this paper, numerous lessons can be drawn that can inform the design of future crowdsourcing and machine learning systems for height and weight estimation.
We conclude this section by summarizing these lessons.

The cause for the rather low crowd accuracy (\Secref{sec:gen accuracy}) is not that workers' estimates are widely dispersed; in fact 20 to 30 independent estimates suffice for the mean to converge (\Figref{fig:convergence}).
The issue is rather that estimates are systematically off due to the contraction bias toward workers' reference values.
Hence, if the images to be annotated are known to be drawn from a certain subpopulation, workers should be recruited to match that subpopulation.
If, on the contrary, the images depict bodies from a wide variety of backgrounds, one should strive to assemble a worker pool reflecting that variety.

To raise the accuracy, we explored the option of facilitating weight estimation by providing workers with known height labels, finding, however, that this makes guessing the weight only mildly easier (\Tabref{tab:error basic vs known height}), presumably because the relationship between height and weight is nonlinear and complex \cite{bmi2005}, such that workers cannot accurately incorporate the height information (\Figref{fig:weight guesses vs known height}).

What does help, on the other hand, is giving workers feedback on their previous guess.
We therefore conclude that it is advisable to train workers on a set of ground-truth images if available.
Our results also indicate that workers are slightly more accurate when subsequent images depict people of similar measurements.
The latter finding points to a promising direction:
instead of simply sampling a random image at every step, future work should design more intelligent crowd algorithms for cleverly assigning images to the workers most likely to make accurate guesses, given their personal background and task history.
Starting from random assignments, such algorithms could learn more about workers with every guess and could later on route tasks more intelligently \cite{shahaf2010generalized,zhang2012task}.

Orthogonally, we explored ways to correct the contraction bias in a postprocessing step (\Secref{sec:correction}).
In line with previous work~\cite{brettschneider2015development,perez2015measuring}, accounting for worker height and weight allows us to significantly reduce the error in individual workers' estimates, but interestingly, these gains are lost once we aggregate all workers' estimates, such that simply averaging raw guesses and shifting and scaling the result by global constants achieves better results than first correcting on a per-worker basis and then averaging (Tables~\ref{tab:correction models} and \ref{tab:correction models height}).
We conclude that aggregating many potentially diverse guesses (``wisdom of crowds'') is more essential than correcting individual guesses.
Overall, bias correction remains challenging, with even our best models lowering the error only marginally, from 8.7~kg to 7.0~kg (\Tabref{tab:correction models}).

\trackedChange{
This negative result has important practical implications.
}
Given the large errors, we expect a crowd\hyp labeled image set to be too noisy for training accurate machine learning models, even with correction.
Human performance is, of course, not necessarily an upper bound for machine performance. It is well conceivable that machine learning algorithms trained on ground-truth, rather than crowdsourced, height and weight labels may surpass human performance.

That said, noisy, crowdsourced labels might still become useful via transfer learning \cite{oquab2014cvpr}: deep learning needs large amounts of training data, so one may start by training a weak model with many noisy labels and then fine-tune it with more accurate training data.
We hope that future work will build on our insights to develop better health monitoring solutions based on crowdsourcing and machine learning.


\section{Conclusion}

We investigated human performance at estimating body measurements by using a novel large-scale dataset of height- and weight\hyp labeled images from the Web as input to an estimation task deployed to a diverse set of human guessers via crowdsourcing.
We find that human estimates are overall of low accuracy, with mean absolute errors of 15.5~kg for weight and 6.3~cm for height (8.8~kg and 5.2~cm, respectively, when subsampling images to represent the height and weight distributions among participating crowd workers).
Estimates are biased in distinct ways, such that errors vary systematically with properties of both the estimated and the estimating person.
Future work should extend this research by shedding further light on the mental mechanisms at play during estimation and by building tools for improving people's weight awareness.


\begin{backmatter}

\section*{Availability of data and material}
  The datasets used and/or analysed during the current study are available from the corresponding author on reasonable request.

\section*{Funding}
This project was partly funded by a grant from the Swiss Data Science Center.

\section*{Competing interests}
  The authors declare that they have no competing interests.

\section*{Author's contributions} 
 KM conducted all the analysis. KM and KG collected the data. KM, KG, and
RW designed the analysis and experiments and wrote the manuscript.
KM and KG performed the work mostly while at EPFL.
All authors read and approved the final manuscript.

\section*{Acknowledgements}
We thank Klaus Sch\"onenberger, Jean-Philippe Thiran, Fernando P\'erez Cruz, and their teams, as well as Kayla de la Haye, Siddharth Suri, Jake Hofman, and Manoel Horta Ribeiro,
for valuable discussions and input.
We also thank Ingmar Weber and Ferda Olfi for giving us access to their image data.
RW is grateful to Google, Microsoft, and Facebook for generous gifts to his lab.
  

\bibliographystyle{bmc-mathphys} 
\bibliography{biblio.bib}     




\end{backmatter}
\end{document}